\documentclass[aps,pre,reprint,twocolumn,superscriptaddress,longbibliography,showpacs,floatfix]{revtex4-2}

\usepackage{graphicx}
\usepackage{amsmath}
\usepackage{amssymb}
\usepackage{epsf}
\usepackage{color}
\usepackage{mathtools}
\usepackage{bm}
\usepackage[colorlinks=true, citecolor=blue, linkcolor=blue, urlcolor=blue]{hyperref}
\usepackage{bibunits}

\begin{document}

\title{An information-theoretic perspective on feed-forward loop abundances in transcriptional networks}

\author{Mintu Nandi}
\email{mintunandi@ubi.s.u-tokyo.ac.jp}
\affiliation{Universal Biology Institute, The University of Tokyo, 7-3-1 Hongo, Bunkyo-ku, Tokyo 113-0033, Japan}

\author{Sudip Chattopadhyay}
\email{sudip@chem.iiests.ac.in}
\affiliation{Department of Chemistry, Indian Institute of Engineering Science and Technology, Shibpur, Howrah 711103, India.}

\author{Suman K Banik}
\email{skbanik@jcbose.ac.in}
\affiliation{Department of Chemical Sciences, Bose Institute, EN 80, Sector V, Bidhan Nagar, Kolkata 700091, India.}

\begin{abstract}

In the transcriptional networks of \textit{Escherichia coli} and \textit{Saccharomyces cerevisiae}, the eight feed-forward loop (FFL) motifs occur at markedly different frequencies. Although previous studies have linked the abundant C1- and I1-FFLs to specific dynamical functions, a common account of the broader pattern is lacking. To address this gap, we adopt an information-theoretic approach. An FFL transmits information through two paths that share an input and converge on an output, suggesting that their interaction may contribute to differences among motifs. To investigate this, we decompose input-output mutual information (MI) into pathway MI and interference MI (IMI). Here, the pathway MI collects the contributions of individual pathways, whereas IMI identifies the contribution that arises from their interference. We find that within a representative parameter regime, IMI values of FFLs produce a hierarchy that closely resembles the abundance ordering in \textit{E. coli}, whereas total and pathway MIs do not. Motivated by this observation, we perform constrained optimization that identifies parameter regimes compatible with abundance-like targets for both \textit{S. cerevisiae} and \textit{E. coli}. We relate the optimized IMI hierarchy to pathway interference strength and local pathway sensitivities to understand the underlying biophysical origin. Our framework thus probes the correspondence between pathway interference, information transmission, and motif-abundance hierarchies.

\end{abstract}


\maketitle

\begin{bibunit}


\section{Introduction}

Gene regulatory networks (GRNs) perform a multitude of cellular functions, from orchestrating developmental programs to maintaining homeostasis and enabling adaptation to environmental signals \cite{Davidson2006, Alon2006}. GRNs are composed of multiple interacting transcription factors (TFs) that control gene expression. Some common subcircuits, called motifs, often recur within GRNs and serve as the elemental building blocks that dictate the mechanisms of complex biological processes, e.g., development, cell signaling, and metabolism \cite{Milo2002, Shen-Orr2002, Alon2006, Kholodenko2012, Lim2013}.

\begin{figure*}[!t]
\includegraphics[width=2.0\columnwidth,angle=0]{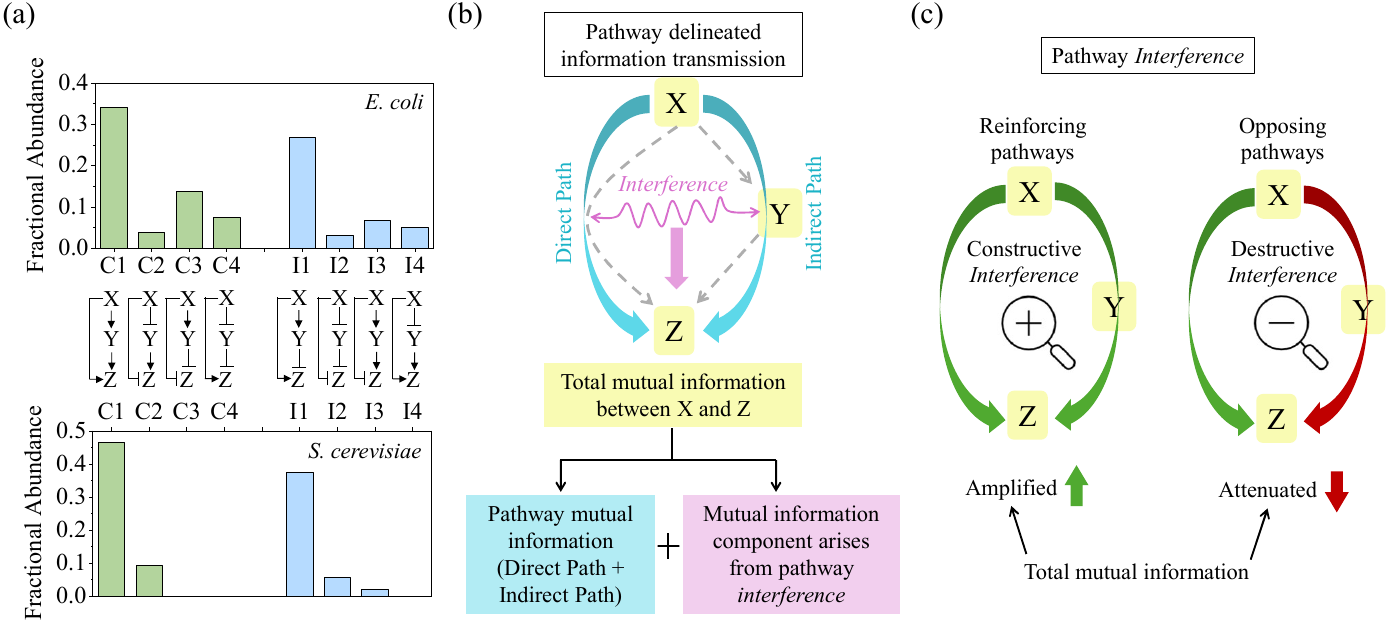}
\caption {\textbf{Abundance pattern and mutual information decomposition.}
(a) Relative abundances of the eight coherent and incoherent FFL types in \textit{E. coli} and \textit{S. cerevisiae}, extracted from the data reported by Mangan \textit{et al.}~\cite{Mangan2006}.
(b) Schematic of the pathway-delineated mutual information decomposition in a generic FFL. Here, $\dashrightarrow$ represents a general notation for activation ($\rightarrow$) and repression ($\dashv$). 
(c) Schematics of the nature of the pathway interference giving rise to amplification and attenuation in total MI. Here, coherent FFLs exhibit reinforcing pathways, whereas incoherent FFLs contain opposing pathways.
}
\label{f1}
\end{figure*}

Feed-forward loops (FFLs) are a canonical class of gene regulatory motifs that occur frequently in the transcriptional networks of \textit{Escherichia coli} and \textit{Saccharomyces cerevisiae} \cite{Shen-Orr2002, Mangan2003, Ma2004, Mangan2006}. In an FFL, an upstream TF $X$ regulates a target TF $Z$ in two ways: directly and indirectly through an intermediate TF $Y$ (see Fig.~\ref{f1}) \cite{Milo2002, Alon2006}. The regulatory edges connecting the TFs can be either activating or repressing, giving rise to eight distinct FFL types \cite{Mangan2003}. These are grouped into four coherent FFLs, C1--C4, and four incoherent FFLs, I1--I4, depending on whether the direct and indirect regulatory paths have the same or opposite overall signs.


The relative abundances of these eight FFLs in \textit{E. coli} and \textit{S. cerevisiae} show distinct non-uniform patterns (Fig.~\ref{f1}a), as reported previously \cite{Mangan2006, Alon2006}. In particular, C1- and I1-FFLs occur more frequently than the other types. Earlier studies have related these dominant motifs to several dynamical functions, including persistence detection, response acceleration, pulse generation, and functional plasticity \cite{Mangan2003, Mangan2003a, Kalir2005, Mangan2006, Alon2006, Murugan2012}. Although these studies establish important dynamical functions of particular FFLs, they do not by themselves explain the relative frequencies of all eight types \cite{Mangan2003, Ingram2006}. Non-adaptive network evolution may also contribute to motif abundance \cite{Widder2012}. This raises a broader question: whether a property of information transmission can be related to the abundance patterns across FFL types, and which biochemical conditions support such a relationship.

To address this question, we consider how the direct and indirect pathways of an FFL jointly influence information transmission. An FFL is not simply the sum of two independent regulatory paths. The same upstream regulator drives both the direct and indirect paths, and the two contributions converge on a common output. Their joint effect can therefore reinforce or oppose the information transmitted through the individual paths. This motivates us to investigate how regulatory architecture shapes information transmission, which requires examining not only the individual routes, but also their joint action. Shannon's formalism of information theory provides a means to make this distinction using the stochastic dynamics of gene expression  \cite{Shannon1948, Shannon1963, Cheong2011, Tkacik2011, Tkacik2016, Tkacik2025}. Building on our earlier analysis of direct–indirect pathway coupling at the fluctuation level \cite{Nandi2026}, we extend the framework to explore how pathway interference contributes to information transmission across the eight FFL types.

Here, we develop a pathway-resolved description of steady-state mutual information (MI) between the upstream regulator $X$ and the output $Z$ (Fig.~\ref{f1}b). The decomposition separates a contribution due to individual pathways from an interference contribution. We term the former as pathway MI and the latter as interference MI (IMI). We show that reinforcing pathways can enhance information transmission through constructive pathway interference, whereas opposing pathways can attenuate it through destructive pathway interference (see Fig.~\ref{f1}c). These variations, relative to the pathway MI, are quantified by positive and negative IMI. We find that within a representative parameter regime, IMI values of different FFLs closely produce an \textit{E. coli} abundance-like hierarchy, whereas the total and pathway MIs do not. Motivated by this observation, we use the cross-entropy method (CEM) to identify the parameter regimes in which the IMI hierarchies approximately follow the distinct bacterial and yeast abundance patterns. We then relate the resulting IMI hierarchies to pathway interference strength and local regulatory sensitivities to understand the underlying biophysical origin. Together, these analyses connect an abundance-like information hierarchy to the biochemical conditions under which the two regulatory paths interact. We note that this pathway interference framework may also extend to other multi-path regulatory architectures in GRNs \cite{Lipshtat2008, Bleris2011, Burda2011, Gutierrez2021}.

\begin{figure*}[!t]
\includegraphics[width=2.0\columnwidth,angle=0]{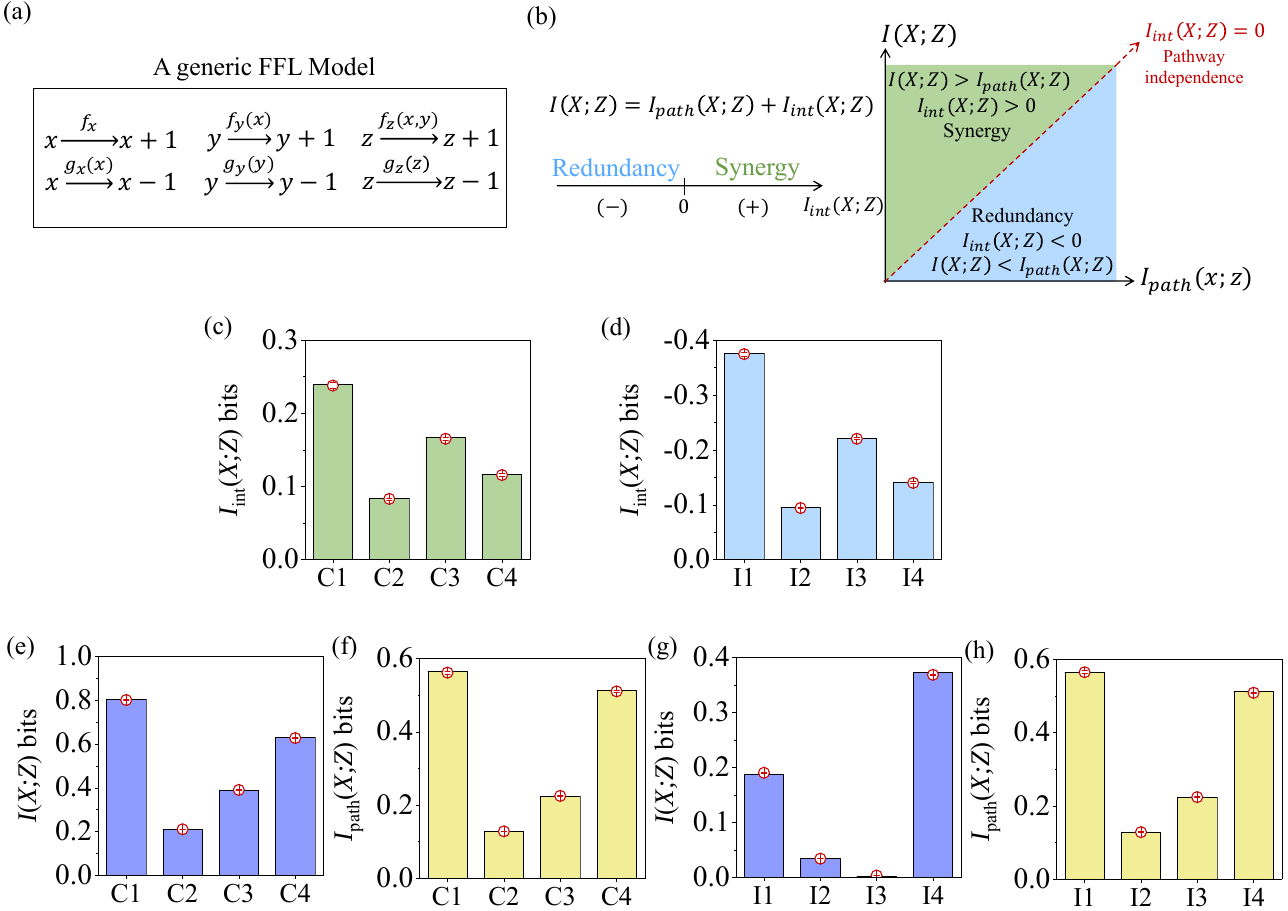}
\caption{\textbf{FFL model and interference mutual information.}
(a) Kinetic scheme of a generic FFL motif.
(b) Decomposition of MI, $I(X;Z)$, into pathway MI, $I_{\rm path}(X;Z)$, and IMI, $I_{\rm int}(X;Z)$. Positive, negative, and zero IMI correspond to synergistic pathway interference, redundant pathway interference, and pathway independence, respectively. Pathway independence is exhibited along the diagonal of the representative total MI versus Pathway MI plot. 
(c,d) IMI magnitudes for coherent and incoherent FFLs with an AND gate, respectively. These patterns closely resemble the abundance pattern of \textit{E. coli} in Fig.~\ref{f1}a.
(e,f) Total MI and pathway MI for coherent FFLs.
(g,h) Corresponding total MI and pathway MI for incoherent FFLs.
The total and pathway MI magnitudes for both FFL types substantially differ from any of the abundance hierarchies. In (c--h), bars denote analytical predictions under the LNA, whereas open symbols with error bars denote numerical values generated from stochastic simulation using the Gillespie algorithm. Simulation protocols are described in Sec.~\ref{sec4C}. The parameters used to generate the plots are: $\beta_X=0.1$ min$^{-1}$, $\beta_Y=1.0$ min$^{-1}$, $\beta_Z=10.0$ min$^{-1}$, $K_{XY}=100$ molecules/cell, $K_{XZ}=100$ molecules/cell, $K_{YZ}=100$ molecules/cell, $\langle x\rangle=40$ molecules/cell, $\langle y\rangle=100$ molecules/cell, and $\langle z\rangle=100$ molecules/cell. The other parameters, such as the synthesis rate constants of each TF, can be obtained from the steady-state expressions of $\langle x\rangle$, $\langle y\rangle$, and $\langle z\rangle$. Corresponding OR-gate results are shown in \textit{SM Fig.~S2}, which qualitatively preserves the IMI hierarchy.
}
\label{f2}
\end{figure*}

\section{Results}

\subsection{The information-theoretic framework} 

To understand how the two regulatory paths contribute to information transmission, we first quantify the steady-state covariance matrix between the upstream regulator $X$ and the output $Z$. Conventionally, MI asks how much of the output's fluctuations become predictable once the input is known. Under the Gaussian channel approximation, the MI between $X$ and $Z$ can thus be written as \cite{Nandi2024},
\begin{equation}
    I(X;Z)=\frac{1}{2} \log_2 \left[\frac{\eta_Z^2}{\eta_{Z|X}^2}\right],
    \label{eq1}
\end{equation}

\noindent where $\eta_Z^2$ and $\eta_{Z|X}^2$ refer to the steady-state output fluctuation and normalized conditional output variance, respectively. Therefore, the predictability is controlled entirely by the fluctuations in molecular abundances. Here, $\eta_Z^2:=\sigma_Z^2/\langle z\rangle^2$ and $\eta_{Z|X}^2:=\eta_Z^2-(\zeta_{XZ}^2/\eta_X^2)$, with $\eta_X^2:=\sigma_X^2/\langle x\rangle^2$ quantifying input fluctuations and $\zeta_{XZ}:=\sigma_{XZ}/(\langle x\rangle \langle z\rangle)$ quantifying the normalized covariance between $X$ and $Z$ \cite{Nandi2024}. Note that $\sigma_X^2$ and $\sigma_Z^2$ denote the variances of $X$ and $Z$, whereas $\sigma_{XZ}$ represents the covariance between them. Moreover, $\langle x\rangle$ and $\langle z\rangle$ denote the steady-state mean copy numbers of $X$ and $Z$.

Under the linear noise approximation (LNA), the total fluctuation of $Z$ is simply the superposition of the fluctuations arriving via the direct and indirect paths. As a result, we can decompose the normalized covariance as $\zeta_{XZ} = \zeta_{XZ,d} + \zeta_{XZ,ind}$, where $\zeta_{XZ,d}$ and $\zeta_{XZ,ind}$ measure the covariance due to direct and indirect paths, respectively. Moreover, due to correlation of fluctuations of $Z$ with itself, the normalized output variance can separate as $\eta_Z^2 = \eta_{Z,path}^2 + \eta_{Z,int}^2$ \cite{Nandi2026}. Here, $\eta_{Z,path}^2$ quantifies the pathway contribution and $\eta_{Z,int}^2$ is the contribution arising from interference of the two path-mediated fluctuations at $Z$. The expressions of these statistical quantities can be obtained under LNA of the governing chemical master equations for the kinetics, shown in Fig.~\ref{f2}a, of a generic FFL motif. see Sec.~\ref{sec4A} and \textit{SM Sec.~S1} for details. 

Based on the decomposition of $\zeta_{XZ}$ and $\eta_Z^2$, $\eta_{Z|X}^2$ can be separated into a pathway part and an interference part as $\eta_{Z|X}^2 = \eta_{Z|X,path}^2 + \eta_{Z|X,int}^2$. Here the pathway conditional fluctuation is $\eta_{Z|X,path}^2=\eta_{Z,path}^2-[(\zeta_{XZ,d}^2+\zeta_{XZ,ind}^2)/\eta_X^2]$ and the interference conditional fluctuation is
\begin{equation}
    \eta_{Z|X,int}^2=\eta_{Z,int}^2-\frac{2\zeta_{XZ,d}\zeta_{XZ,ind}}{\eta_X^2}.
    \label{eq1a}
\end{equation}

\noindent In the pathway term, the direct and indirect covariance contributions enter additively through the squared terms. This signifies that the pathway term measures the residual output fluctuation after conditioning on $X$ transmitted separately through the two pathways. As a consequence, $\eta_{Z|X,path}^2 \le \eta_{Z,path}^2$ always holds. In contrast, the interference conditional term contains the product of direct and indirect covariance contributions and therefore measures the residual part of the direct-indirect pathway interference after conditioning on $X$.

To decompose $I(X;Z)$, we define the relative interference contributions as $\tilde{\eta}_{Z,int}^2=\eta_{Z,int}^2/\eta_{Z,path}^2$ and $\tilde{\eta}_{Z|X,int}^2=\eta_{Z|X,int}^2/\eta_{Z|X,path}^2$. Using these quantities, the MI $I(X;Z)$ separates into
\begin{equation}
    I(X;Z) = I_{path}(X;Z) + I_{int}(X;Z),
    \label{eq2}
\end{equation}

\noindent where 
\begin{eqnarray}
    I_{path}(X;Z) &=& \frac{1}{2}\log_2\left[\frac{\eta_{Z,path}^2}{\eta_{Z|X,path}^2}\right],
    \label{eq3} \\
    I_{int}(X;Z) &=& \frac{1}{2}\log_2\left[\frac{1+\tilde{\eta}_{Z,int}^2}{1+\tilde{\eta}_{Z|X,int}^2}\right].
    \label{eq4}
\end{eqnarray}

\noindent Here, $I_{path}(X;Z)$ is the pathway MI representing the information contribution associated with the two regulatory paths, excluding their non-additive interference. This contribution is non-negative because conditioning on $X$ does not increase the pathway uncertainty of $Z$. In contrast, the IMI $I_{int}(X;Z)$ quantifies the signed contribution arising from interference between the direct- and indirect-path-mediated fluctuations. Positive IMI corresponds to constructive interference, increasing the total MI above $I_{path}(X;Z)$, whereas negative IMI corresponds to destructive interference, reducing it below $I_{path}(X;Z)$. Thus, the amplification and attenuation illustrated in Fig.~\ref{f1}c are measured relative to the pathway MI. Further details are provided in \textit{SM Sec.~S2}.

\begin{figure*}[!t]
\includegraphics[width=2.0\columnwidth,angle=0]{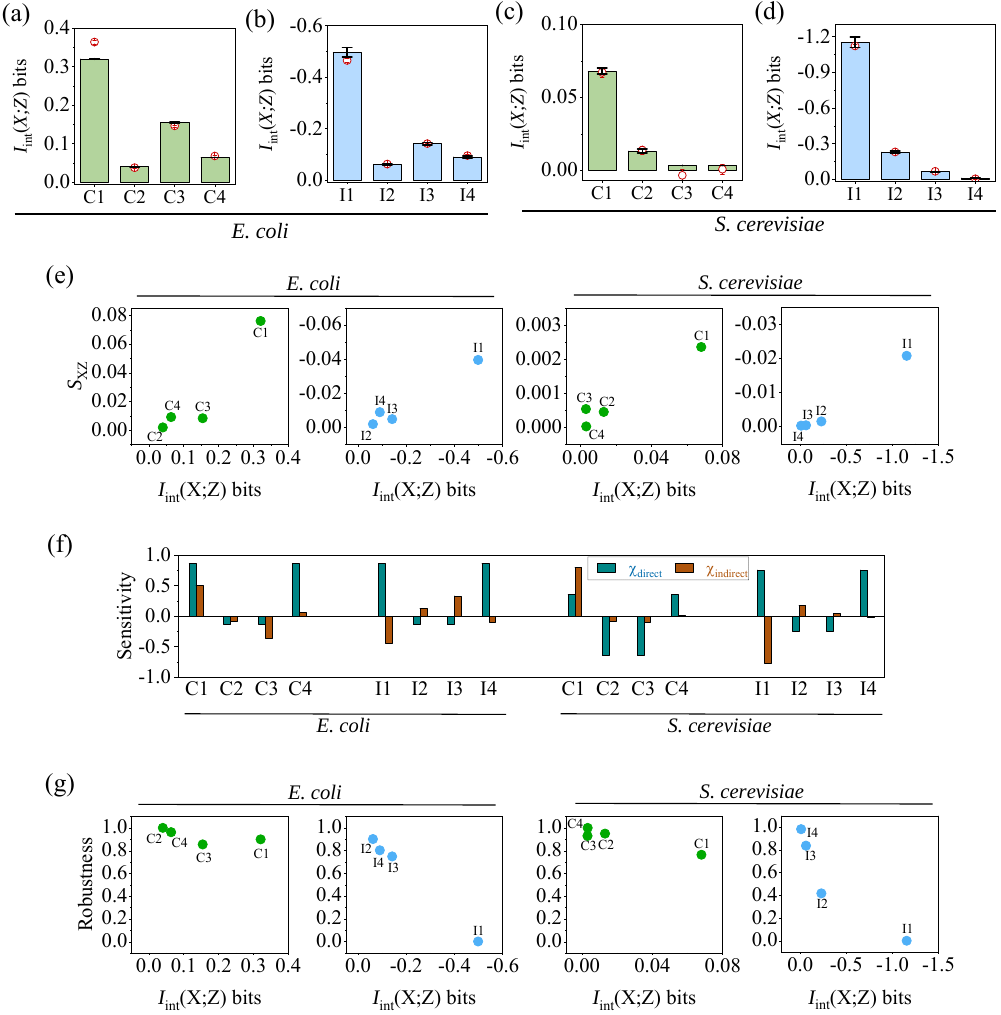}
\caption{\textbf{Optimized IMI patterns and their biophysical origin.}
(a-d) Optimized IMI values for AND-gate FFLs in \textit{E. coli} and \textit{S. cerevisiae}. The resulting IMI hierarchy closely resembles the empirical abundance hierarchy shown in Fig.~\ref{f1}a. Here, bars denote optimized mean IMI values, with error bars indicating standard deviations across independent CEM runs. Open symbols denote mean IMI values obtained from stochastic simulations using the Gillespie algorithm, with error bars indicating standard errors of the mean across simulation replicates. Simulation protocols are described in Sec.~\ref{sec4C}. Corresponding OR-gate results are shown in \textit{SM Fig.~S8}, which qualitatively preserves the IMI hierarchy.
(e) Optimized IMI $I_{int}(X;Z)$ plotted against pathway interference strength $S_{XZ}$.
(f) Direct- and indirect-path sensitivities, $\chi_{_{direct}}$ and $\chi_{_{indirect}}$ for different FFL types.
(g) Optimized IMI plotted against robustness to parameter perturbations, $\mathcal{R}_m$. Here, we use $\varepsilon=0.01$. The results in (e-g) are for AND-gate FFLs. Corresponding OR-gate results are shown in \textit{SM Fig.~S10}, which show a similar observation.
In generating the plots, we use the optimized parameter values listed in \textit{SM Table~S5}.}
\label{f3}
\end{figure*}

\subsection{Pathway interference in information transmission} 

The sign of IMI follows directly from the relative interference terms in Eq.~(\ref{eq4}). If the relative interference contribution is larger before conditioning than after conditioning, i.e., $\tilde{\eta}_{Z,int}^2 \ge \tilde{\eta}_{Z|X,int}^2$, then $I_{int}(X;Z) \ge 0$. If it is smaller before conditioning than after conditioning, i.e., $\tilde{\eta}_{Z,int}^2 \le \tilde{\eta}_{Z|X,int}^2$, then $I_{int}(X;Z) \le 0$. This is directly related to the sign structure of the two FFL paths.

In coherent FFLs, the direct and indirect paths have the same overall sign, and hence $\zeta_{XZ,d}\zeta_{XZ,ind} > 0$. The two path-mediated fluctuations, therefore, interfere constructively at the output (see Fig.~\ref{f1}c). Conditioning on $X$, thus, reduces this shared input-mediated interference, giving $\eta_{Z,int}^2>\eta_{Z|X,int}^2$ [see Eq.~(\ref{eq1a})]. Both interference contributions remain positive in the present model, as follows from the expressions in \textit{SM Sec.~S1}. However, conditioning also reduces the pathway fluctuations, so the reduction in interference alone does not determine the sign of IMI. We consider operating regimes satisfying $(\eta_{Z,int}^2/\eta_{Z|X,int}^2) > (\eta_{Z,path}^2/\eta_{Z|X,path}^2)$. This condition means that conditioning removes a larger fraction of the interference contribution than of the pathway contribution. It is equivalent to $\tilde{\eta}_{Z,int}^2 > \tilde{\eta}_{Z|X,int}^2$ and therefore gives $I_{int}(X;Z) > 0$. This positive IMI corresponds to synergistic pathway interference, where the two paths jointly enhance input-output information transmission beyond their separate pathway contributions (see Fig.~\ref{f2}b). 

In incoherent FFLs, the direct and indirect paths have opposite overall signs, giving $\zeta_{XZ,d}\zeta_{XZ,ind} < 0$. The two path-mediated fluctuations therefore interfere destructively at the output (see Fig.~\ref{f1}c). Conditioning on $X$ makes this shared input-mediated interference contribution less negative. In the present model, both interference contributions remain negative, giving
$\eta_{Z,int}^2<\eta_{Z|X,int}^2<0$ [Eq.~(\ref{eq1a}) and \textit{SM Sec.~S1}]. However, conditioning also reduces the pathway fluctuations, so this ordering alone does not determine the sign of IMI. We consider operating regimes satisfying $(\eta_{Z,int}^2/\eta_{Z|X,int}^2) > (\eta_{Z,path}^2/\eta_{Z|X,path}^2)$. This condition means that conditioning removes a larger fraction of the magnitude of the interference contribution than of the pathway contribution. Since both interference contributions are negative, it is equivalent to $\tilde{\eta}_{Z,int}^2 < \tilde{\eta}_{Z|X,int}^2$ and therefore gives $I_{int}(X;Z) < 0$. This negative IMI indicates redundancy or compensation between the two paths. This signifies that the direct and indirect routes carry input-dependent fluctuations with opposite signs, so their joint effect reduces the net information gain relative to the pathway contribution (see Fig.~\ref{f2}b).

In addition to these two cases, we define a limiting case termed ``pathway independence", where $I_{int}(X;Z)=0$ (see Fig.~\ref{f2}b), which means that the two paths neither synergistically enhance nor redundantly reduce the overall information transmission. From Eq.~(\ref{eq4}), this condition is obtained when $\tilde{\eta}_{Z,int}^2 = \tilde{\eta}_{Z|X,int}^2$, indicating that the relative interference contribution is unchanged after conditioning on $X$. This situation may arise when the direct-indirect interference is very weak, i.e., $\zeta_{XZ,d}\zeta_{XZ,ind} \simeq 0$. However, it can also appear in an open-loop system, which is an equivalent reference system of the corresponding FFL. In the open-loop system, the two routes are driven by statistically equivalent but independent inputs, $X$ and $\tilde{X}$, rather than by a common upstream regulator (see \textit{SM Fig.~S1}). We further consider that the open-loop equivalent (OLE) systems keep the direct and indirect routes parameter-matched to the corresponding FFL. Under these assumptions, we find that the pathway MI in the FFL matches the total information transmission in the corresponding OLE of that FFL, i.e., $I_{path}(X;Z) = I_{OLE}(X,\tilde{X};Z)$, as discussed in the \textit{SM Sec.~S3}. As a consequence, we may write the IMI as the difference between the MI content in FFL [$I_{FFL}(X;Z)$] and in the corresponding OLE [$I_{OLE}(X,\tilde{X};Z)$], i.e.,
\begin{equation}
    I_{int}(X;Z) = I_{FFL}(X;Z) - I_{OLE}(X,\tilde{X};Z).
    \label{eq40}
\end{equation}

\noindent This identity is exact under the stated assumptions. The above comparison thus clarifies that $I_{path}(X;Z)$ represents the information carried by the two routes exclusively, irrespective of how they are coupled, whereas $I_{int}(X;Z)$ captures this coupling and thereby quantifies the topological effect on information transmission. We use Eq.~(\ref{eq40}) to estimate the IMI numerically using the Gillespie algorithm (see Sec.~\ref{sec4C}) \cite{Gillespie1976, Gillespie1977}.

\subsection{An initial IMI hierarchy resembles \textit{E. coli} abundance pattern} 

Since IMI encodes the architectural signatures of FFL in information transmission, we ask whether it can exhibit a pattern similar to the empirical abundance patterns across FFL types in \textit{E. coli} and \textit{S. cerevisiae} (see Fig.~\ref{f1}a). To examine this, we first compare the IMI values of eight FFL types using a representative parameter set. The parameter choice (see Fig.~\ref{f2} caption) follows the timescale separation considered in our earlier studies \cite{Nandi2024, Nandi2024_a, Nandi2026} and related analyses of stochastic signal transduction \cite{Maity2015}. Specifically, the degradation rates satisfy $\beta_X<\beta_Y<\beta_Z$, corresponding to intrinsic relaxation times $\tau_X>\tau_Y>\tau_Z$, where $\tau_\mathcal{J}=1/\beta_\mathcal{J}$. This timescale separation favors downstream components responding to input fluctuations before they decorrelate. The prescribed expression levels satisfy $\langle x\rangle<\langle y\rangle=\langle z\rangle$, while the regulator-to-threshold ratios are $\langle x\rangle/K_{XY}=\langle x\rangle/K_{XZ}=0.4$ and $\langle y\rangle/K_{YZ}=1$. Thus, the $X$-dependent regulatory interactions operate below their half-response thresholds, whereas the $Y$-dependent interaction operates at half-maximal response. 

Within this parameter regime, the IMI patterns for both coherent and incoherent FFLs closely match the \textit{E. coli} abundance hierarchy (Fig.~\ref{f2}c,d). Crucially, the high abundance of C1- and I1-FFLs matches their strong synergistic and redundant IMI profiles. In contrast, total MI and pathway MI yield entirely different structural orderings under the same parameter settings (Fig.~\ref{f2}e-h). This observation suggests that pathway interference distinguishes features of the abundance pattern that are not captured by the total information alone. However, it does not establish that the illustrative parameters describe natural \textit{E. coli} circuits. Rather, it provides a reason to examine the relationship between information transmission and natural abundances systematically. We note that the different yeast ordering is not recovered in the initial manual exploration, motivating a constrained parameter search through optimization.

\subsection{Constrained optimization produces abundance-like IMI hierarchies for both organisms} 

The fixed-parameter result raises the question of whether the same information measure can reproduce the different abundance patterns of bacteria and yeast. To test this possibility, we optimize the kinetic rate parameters of each FFL class using the cross-entropy method (CEM). The optimization is restricted to biologically plausible parameter ranges (see \textit{SM Table~S3}) chosen from previous studies \cite{Bintu2005, Belle2006, Garcia2011, Hintsche2013, Christiano2014, Li2014, Hansen2015, Milo2015, Gupta2024}. For each organism, FFL class, and output gate, the same degradation rates, regulatory thresholds, and mean copy numbers are applied to all four FFL types, preventing independent motif-by-motif fitting. Moreover, the observed abundance ratios are used to define broad target ranges for the relative IMI magnitudes, as described in \textit{SM Sec. S4} and \textit{Table S4}. We refer to Sec.~\ref{sec4B} and \textit{SM Sec.~S4} for finer details of the optimization procedure.

The optimized IMI values show motif-specific hierarchies that follow the empirical abundance patterns within the coherent and incoherent FFL classes in both organisms (see Fig.~\ref{f1}a and Fig.~\ref{f3}a-d). Specifically, C1-FFL exhibits the highest synergistic IMI among coherent FFLs, whereas I1-FFL exhibits the largest redundant IMI among incoherent FFLs. This observation is consistent with their highest empirical frequencies in \textit{E. coli} and \textit{S. cerevisiae}. The remaining FFL types exhibit weaker IMI values, consistent with the lower-frequency part of the abundance hierarchy. 

The IMI-abundance correspondence suggests that the coherent-incoherent classification determines the sign of pathway interference, whereas the non-uniform abundance pattern within each class is associated with the strength of that interference. This strength is dictated by how activation and repression are distributed over the three regulatory edges. This changes the local regulatory sensitivities, the balance between direct and indirect paths, and the strength of their interference at the output. As a result, the magnitude of IMI varies across motif types even when its sign is selected by coherence or incoherence. In this sense, the enrichment of C1- and I1-FFLs is associated with their ability to generate the strongest pathway interference information within their respective classes. Thus, IMI provides a possible information-processing basis for the non-uniform enrichment of FFL types in the transcriptional networks.

A possible concern is that optimization could force any information measure to follow the abundance pattern. We therefore repeat the complete procedure using the total MI, $I(X;Z)$, and the pathway MI, $I_{path}(X;Z)$. The CEM optimization is performed for total MI and pathway MI under similar protocols as used for IMI. Unlike IMI, these quantities do not consistently satisfy the abundance-like target hierarchy across organisms and FFL classes, as discussed in the \textit{SM Sec.~S4 and Fig.~S9}. This indicates that the observed correspondence is not a generic outcome of optimizing an arbitrary information measure. Among the measures tested, it depends specifically on the interference contribution.

\subsection{Biophysical origin of the IMI hierarchy} 

With IMI showing a clear connection to the abundance pattern where total and pathway MI do not, we next ask which biophysical quantities determine the interference contribution. To find out the relevant quantities, we seek answers to the following three questions: (1) How strongly do the two paths interfere? (2) How does pathway sensitivity relate to this strength? (3) How robust is the resulting IMI hierarchy to perturbations of biochemical parameters? These questions motivate the three measures, namely, pathway interference strength, direct and indirect pathway sensitivities, and robustness to parameter perturbations. These quantities may help to identify the biophysical origin of the abundance-like IMI hierarchy.

We quantify pathway interference strength as $S_{XZ}=2\zeta_{XZ,d}\zeta_{XZ,ind}/\eta_X^2$. It measures the extent to which the direct and indirect paths interfere through their shared dependence on the input $X$. Its sign distinguishes the nature of the interference, for instance, $S_{XZ}>0$ for coherent FFLs, where the two paths interfere constructively, and $S_{XZ}<0$ for incoherent FFLs, where the two paths interfere destructively. We find that the hierarchy of $S_{XZ}$ closely follows the optimized IMI hierarchy (see Fig.~\ref{f3}e). In particular, C1-FFL exhibits the largest positive $S_{XZ}$, corresponding to the strongest synergistic IMI. In contrast, I1-FFL exhibits the largest negative $S_{XZ}$ in magnitude, corresponding to the strongest redundant or compensatory IMI. Thus, the non-uniform IMI hierarchy is directly linked to the strength of pathway interference encoded in the covariance structure. 

We next examine how this interference strength relates to the sensitivities of the direct and indirect paths, denoted by $\chi_{_{direct}}$ and $\chi_{_{indirect}}$, respectively. Their definitions are given in Sec.~\ref{sec4D}. The sign of $\chi_{_{direct}}$ reflects the direct-path sign, while the sign of $\chi_{_{indirect}}$ reflects the effective sign of the indirect path. We find that different FFL types display distinct balances between $\chi_{_{direct}}$ and $\chi_{_{indirect}}$ (see Fig.~\ref{f3}f). Motifs with stronger IMI generally show larger and more comparable magnitudes of the direct and indirect sensitivities, irrespective of the sign. This is expected because IMI is generated by the interference between the two paths; if one path is locally weak, the path-interference contribution is limited, whereas strong, comparable sensitivities in both paths allow their interference at the output to become more pronounced. Thus, local pathway sensitivity provides another biophysical origin for the IMI hierarchy.

We now examine whether the optimized IMI hierarchy is robust to perturbations of biochemical parameters. To address this, we define a robustness measure $\mathcal{R}_m$ for motif $m$, based on the sensitivity of IMI to parameter perturbations. See Sec.~\ref{sec4E} for details. Here, $\mathcal{R}_m \simeq 1$ indicates that IMI is weakly affected by parameter perturbations, whereas $\mathcal{R}_m \simeq 0$ indicates stronger parameter dependence. We find that a larger IMI magnitude is generally associated with reduced robustness (see Fig.~\ref{f3}g), indicating that strong pathway interference information requires tighter control of biochemical parameters. This decrease in robustness is more prominent for incoherent FFLs than for coherent FFLs. This trade-off between IMI and the robustness measure suggests that FFL prevalence may reflect a balance between pathway interference strength and tolerance to biochemical perturbations. Thus, the robustness analysis provides a plausible information-cost principle for maintaining abundance-like IMI patterns.

These results place the abundance-like IMI patterns on a biophysical footing. The motifs with the strongest IMI also exhibit the strongest pathway interference, as evidenced by stronger, more balanced local path sensitivities. Moreover, a larger IMI is accompanied by reduced parameter robustness, indicating a cost for maintaining strong pathway interference and, hence, IMI. Thus, IMI links motif-specific information processing to measurable biophysical properties, providing a plausible physical basis for the non-uniform abundance-like patterns.

\section{Discussion}

FFLs are enriched regulatory architectures in transcriptional networks, but the eight FFL types are not equally represented in \textit{E. coli} and \textit{S. cerevisiae}. This unequal abundance has remained an open problem since the early identification of network motifs. Previous studies have explained important dynamical functions of the dominant C1- and I1-FFLs, but a common description of the broader abundance pattern has been lacking. Here, we identify pathway interference as a candidate organizing principle.

By decomposing input-output MI into pathway MI and IMI, we first observe that IMI magnitudes closely follow the qualitative abundance ordering of \textit{E. coli} in a representative parameter regime. In contrast, total MI and pathway MI yield different orderings at the same operating points. This initial correspondence motivates a systematic investigation of whether the same information measure can mimic the distinct abundance patterns of bacteria and yeast. We therefore use the CEM to search for the parameter regimes within biologically motivated parameter bounds. The search identifies regimes in which the IMI magnitudes exhibit abundance-like hierarchies within both the coherent and incoherent FFL classes in each organism. However, the total MI and the pathway MI do not consistently satisfy the same target hierarchy across organisms and FFL classes. Thus, the abundance pattern is more closely associated with the information contribution generated by pathway interference than with the total amount of information transmitted by the circuit.

The biophysical analysis further explains why different FFL types within the same class exhibit different IMI magnitudes. We find that the magnitude of pathway interference strength follows the IMI hierarchy. Local pathway sensitivities provide the next layer of explanation, showing that stronger IMI generally exhibits stronger, more balanced direct- and indirect-path sensitivities. This indicates that a large IMI requires stronger pathway interference and highly responsive pathways. Moreover, the robustness analysis reveals an additional cost, where a larger IMI is generally associated with reduced robustness to biochemical parameter perturbations. This information-robustness trade-off follows directly from the interference of the two pathways.

The role of optimization in this study is to determine whether the observed hierarchy can be realized within biologically motivated operating regimes. The four motifs within each class are evaluated under the same degradation rates, regulatory thresholds, and expression levels, preventing independent motif-by-motif fitting. Moreover, IMI is not selected by the optimization itself. Total MI and pathway MI are also tested under a similar optimization protocol and numerical settings, but they do not consistently reach the feasible hierarchy. This comparison shows that the result depends on the interference contribution rather than on parameter optimization alone.

We note that motif prevalence in real transcriptional networks can also reflect evolutionary history, network-growth processes, mutational biases, organism-specific constraints, and selection for specific dynamical functions. Our study does not rule out these explanations, rather, it identifies an information-processing feature consistent with the observed abundance hierarchy. Our framework may therefore extend beyond FFLs to other multi-path architectures, including diamond motifs and bifans in transcriptional networks \cite{Burda2011, Lipshtat2008}, convergent pathways in signaling and neural circuits \cite{Gutierrez2021}, and multi-route designs in synthetic gene circuits \cite{Bleris2011}.

A limitation of our analysis is its reliance on Hill-type regulatory functions and LNA around the steady state. It therefore does not explicitly include transcriptional bursting, chromatin regulation, transient adaptive dynamics, growth-dependent effects, or context-dependent interactions with other network components. In addition, the input is represented by fluctuations in the abundance of $X$, while regulatory inputs can also arise through changes in transcription-factor activity or localization. The present analysis is thus focused on steady-state information transmission and is complementary to the established temporal functions of FFLs, including response delays, pulse generation, adaptation, and response acceleration. Despite these simplifications, the framework provides a tractable starting point for investigating how pathway interference shapes information transmission in more detailed, transient, and non-Gaussian biochemical systems.

Another related limitation is the empirical benchmark. The abundance data in Fig.~\ref{f1}a reflect network reconstructions available at the time of the original studies \cite{Mangan2003, Mangan2006}. We adopt this benchmark for comparability with prior theoretical work \cite{Murugan2012, Widder2012} rather than as a settled ground truth. However, C1- and I1-FFL remain dominant within their classes in a larger \textit{E. coli} network, though the coherent-incoherent balance differs markedly from the original study \cite{Ma2004}. Furthermore, a perturbation-based \textit{S. cerevisiae} reconstruction identifies additional overrepresented FFL types \cite{Kemmeren2014}. We therefore interpret our result as evidence of within-class dominance of C1- and I1-FFL rather than as a perfect fit to precise abundances. The reported frequencies may also reflect the overall balance of activating and repressing interactions and the way motif instances are counted.

More broadly, our results suggest that the non-uniform abundance of circuit motifs may reflect not only their dynamical functions but also how their architecture organizes information across convergent regulatory paths. The framework, thus, offers a lens for understanding architectural enrichment across regulatory networks and for guiding the design of synthetic circuits. An experimental test would compare synthetic C1–C4 or I1–I4 circuits under matched expression levels and timescales, where C1 and I1 are predicted to exhibit the strongest interference contributions within their respective classes. Because convergent regulatory paths are also common in developmental and signaling networks, the same framework may help examine how circuit architecture shapes the propagation of biological information in other systems.

\section{Methods}
\label{sec4}

\subsection{Model} 
\label{sec4A}

Using the kinetic scheme in Fig.~\ref{f2}a, which defines the elementary production and degradation reaction steps of the three TFs, we formulate the CME \cite{Kampen2007, Gardiner2009} as,
\begin{eqnarray}
    \frac{dP(\bm{n};t)}{dt} &=&
    \sum_{\mathcal{J} \in \{X,Y,Z\}} \left[ 
    \left( \mathbb{E}_\mathcal{J}^{+1}-1 \right) g_\mathcal{J}(\bm{n}) P(\bm{n};t) \right. \nonumber \\
    && \left. +
    \left( \mathbb{E}_\mathcal{J}^{-1}-1 \right) f_\mathcal{J}(\bm{n}) P(\bm{n};t)
    \right],
    \label{eq5}
\end{eqnarray}

\noindent where, $\bm{n} = (n_X,n_Y,n_Z)^\top \equiv (x,y,z)^\top$ is the copy-number vector of the TFs $X$, $Y$, and $Z$ and $P(\bm{n};t)$ is the corresponding probability of state $\bm{n}$ at time $t$. The operators $\mathbb{E}_\mathcal{J}^{\pm 1}$ are step operators that shift the copy number of TF $\mathcal{J}$ by $\pm1$. The functions $f_\mathcal{J}(\bm{n})$ and $g_\mathcal{J}(\bm{n})$ denote the production and degradation propensities of TF $\mathcal{J}$, respectively. Their explicit forms depend on the type of the FFL and are listed explicitly in \textit{SM Table~S1}.

We apply LNA on the CME to derive steady-state expressions of second moments, which are, in turn, used to derive the pathway-delineated decomposed terms in Eqs.~(\ref{eq3})~and~(\ref{eq4}). A detailed solution of the CME and component-wise derivation of the decomposed terms are provided in the \textit{SM Sec.~S1}.

\subsection{CEM optimization} 
\label{sec4B}

CEM is an adaptive population-based stochastic optimization approach for rare-event simulation and continuous nonconvex optimization \cite{Rubinstein1997, Rubinstein1999, deBoer2005}. In this work, the biochemical parameters are optimized using the continuous CEM following the standard adaptive-sampling framework \cite{Rubinstein1999, deBoer2005}. For each organism, FFL class, and output gate logic, the search is performed in log-parameter space to ensure positive kinetic parameters and efficient exploration over several orders of magnitude. At each CEM iteration, a population of candidate parameter vectors is sampled from a multivariate Gaussian distribution. For each candidate, the IMI is evaluated using its analytical expression in Eq.~(\ref{eq4}). Candidate parameter sets are ranked according to an objective function that favors larger IMI strength while penalizing violations of the abundance-like hierarchy constraints. The elite candidates are then used to update the sampling mean and covariance, and this procedure is repeated until the final CEM iteration. Independent CEM runs are performed for each condition, and the optimized values reported in Fig.~\ref{f3}a-d correspond to the mean and standard deviation across these runs. The same procedure is applied to $I(X;Z)$, and $I_{path}(X;Z)$ to compare whether each quantity could reproduce the abundance-like motif hierarchy. We note that CEM is suitable here because the optimization is derivative-free, nonconvex, constrained by motif-ratio penalties, and aimed at identifying feasible parameter regimes rather than a unique biochemical parameter set. A detailed procedure is provided in the \textit{SM Sec.~S4}.

\subsection{Stochastic simulation protocols}
\label{sec4C}

We simulate the FFLs and their corresponding OLEs using the Gillespie direct method \cite{Gillespie1976, Gillespie1977}. The reaction propensities for FFLs are specified by \textit{SM Table~S1}, with production-rate constants determined from the prescribed mean-field steady-state conditions. The propensities for the corresponding OLE can easily be inferred from the given propensities of FFLs (see \textit{SM Sec.~S3} for details). Each FFL comprised six production and degradation channels, whereas its OLE comprised eight channels following the independent-input construction described in \textit{SM Sec.~S3}.

For each system and parameter set, we generate $N_{\mathrm{rep}}=10$ independent trajectories using distinct, reproducible random seeds. Trajectories are initialized at the prescribed operating points and simulated for $T_{\mathrm{sim}}=10^5/\beta_X$, with the first $20/\beta_X$ discarded as an initial transient. Thus, simulation durations are scaled by the input correlation time, $1/\beta_X$, and are equal for each FFL--OLE pair. Means, variances, and covariances are calculated from residence-time-weighted first and second moments over the remaining trajectory. Normalized quantities are evaluated using the simulated means of the corresponding system and replicate, rather than the prescribed mean-field values.

For each FFL trajectory, the total MI, $I_{FFL}(X;Z)$, is estimated by substituting the simulated moments into Eq.~(\ref{eq1}). For comparison with the analytical pathway MI, we calculate the Gaussian joint-input information of the corresponding OLE, 
\begin{equation}
I_{\mathrm{OLE}}(X,\widetilde{X};Z)
=
\frac{1}{2}\log_2
\left[
\frac{\eta_{Z,\mathrm{OLE}}^2}
{\eta_{Z|X,\widetilde{X},\mathrm{OLE}}^2}
\right],
\label{eq6_2}
\end{equation}

\noindent where,
\begin{equation}
\eta_{Z|X,\widetilde{X},\mathrm{OLE}}^2
=
\eta_{Z,\mathrm{OLE}}^2
-
\frac{\left(\zeta_{XZ}^{\mathrm{OLE}}\right)^2}
{\eta_{X,\mathrm{OLE}}^2}
-
\frac{\left(\zeta_{\widetilde{X}Z}^{\mathrm{OLE}}\right)^2}
{\eta_{\widetilde{X},\mathrm{OLE}}^2}.
\label{eq6_3}
\end{equation}

\noindent See \textit{SM Sec.~S3} for further details on moment calculations. The input cross-covariance is set to its model value of zero, reflecting the independence of $X$ and $\widetilde{X}$. The simulation-based IMI, $I_{\mathrm{int}}^{\mathrm{sim}}(X;Z)$, is then obtained for each replicate from the FFL--OLE information difference in Eq.~(\ref{eq40}).

For each simulation-derived information quantity, we report the replicate mean and its standard error, calculated as the sample standard deviation divided by $\sqrt{N_{\mathrm{rep}}}$. For IMI, these statistics are calculated from the replicate-wise differences. We use $|\overline{I_{\mathrm{int}}^{\mathrm{sim}}}|>3\,\mathrm{SEM}$ as a numerical criterion for resolving its sign. These estimates retain the Gaussian information approximation while using moments obtained from the nonlinear stochastic dynamics. Their comparison with the analytical results therefore assesses the LNA predictions for moment-based information quantities, rather than independently validating the Gaussian approximation or estimating the full discrete MI.

We apply this protocol to all eight FFL types under AND and OR regulation in two settings: (1) the representative parameter regime used in Fig.~\ref{f2}c--h and its OR-gate counterpart in \textit{SM Fig.~S2}, and (2) the optimized parameter sets listed in \textit{SM Table~S5}, corresponding to Fig.~\ref{f3}a--d and the OR-gate results in \textit{SM Fig.~S8}. In each case, the simulations and analytical calculations use the same prescribed operating-point parameters.

\subsection{Pathway sensitivities} 
\label{sec4D}

The pathway sensitivities are defined from the logarithmic response of the regulatory Hill factors around the optimized steady state. We therefore quantify the local direct- and indirect-path sensitivities as,
\begin{eqnarray}
    \chi_{_{direct}} &=& \frac{\partial \ln h_{ZX}}{ \partial \ln x} \Bigg|_{\bm{n}=\langle \bm{n}\rangle},
    \label{eq6} \\
    \chi_{_{indirect}} &=& \frac{\partial \ln h_{YX}}{ \partial \ln x} \frac{\partial \ln h_{ZY}}{ \partial \ln y} \Bigg|_{\bm{n}=\langle \bm{n}\rangle},
    \label{eq7}
\end{eqnarray}

\noindent where $\bm{n} = (n_X,n_Y,n_Z)^\top \equiv (x,y,z)^\top$ and $\langle \bm{n}\rangle = (\langle n_X\rangle, \langle n_Y\rangle, \langle n_Z\rangle)^\top \equiv (\langle x\rangle, \langle y\rangle, \langle z\rangle)^\top$. In these equations, $h_{ZX}$ is the regulatory factor for the direct regulation $X \dashrightarrow Z$,  $h_{YX}$ is for $X \dashrightarrow Y$, and  $h_{ZY}$ is for $Y \dashrightarrow Z$, where $\dashrightarrow$ represents a general notation for activation ($\rightarrow$) and repression ($\dashv$). For C1-FFL, $h_{YX}=x/(K_{XY}+x)$, $h_{ZX}=x/(K_{XZ}+x)$, and $h_{ZY}=y/(K_{YZ}+y)$, which are obtained from the production propensities $f_Y$ and $f_Z$ listed in \textit{SM Table~S1}. The corresponding regulatory factors for the other FFL types are extracted in the same way.

\subsection{IMI robustness to parameter perturbations} 
\label{sec4E}

For each motif $m$, let $\bm{\theta}_m^\ast$ be the optimized parameter vector (see \textit{SM Table~S5} for the optimized values). A particular parameter $\theta_{m,i}^\ast$ within $\bm{\theta}_m^\ast$ is perturbed by a small relative amount $\varepsilon$ while keeping all other parameters fixed at their optimized values. Specifically, the perturbation is applied as $\theta_{m,i}^\ast \to \theta_{m,i}^\ast e^\varepsilon$ and $\theta_{m,i}^\ast \to \theta_{m,i}^\ast e^{-\varepsilon}$, giving two perturbed vectors $\bm{\theta}_m^{\ast(+i)}$ and $\bm{\theta}_m^{\ast(-i)}$, respectively. Using these two perturbed vectors, the corresponding IMI values are evaluated. The sensitivity is then estimated as
\begin{equation}
    \mathcal{S}_{m,i} = \left\lvert \frac{I_{int}^m(\bm{\theta}_m^{\ast(+i)})-I_{int}^m(\bm{\theta}_m^{\ast(-i)})}{2\varepsilon} \right\rvert.
    \label{eq8}
\end{equation}

\noindent A motif-level sensitivity is then obtained by averaging $\mathcal{S}_{m,i}$ over the perturbed parameters, $\bar{\mathcal{S}}_m=\langle \mathcal{S}_{m,i} \rangle_i$. Robustness for motif $m$ is then defined as
\begin{equation}
    \mathcal{R}_m = 1- \frac{\bar{\mathcal{S}}_m - \bar{\mathcal{S}}_{\rm min}}{\bar{\mathcal{S}}_{\rm max}-\bar{\mathcal{S}}_{\rm min}},
    \label{eq9}
\end{equation}

\noindent where $\bar{\mathcal{S}}_{\rm min}$ and $\bar{\mathcal{S}}_{\rm max}$ are the minimum and maximum mean sensitivities among the motifs being compared within the same organism.


%

\end{bibunit}

\clearpage


\onecolumngrid


\begin{center}

{\large\bfseries Supplementary Material}

\vspace{0.8em}

{\large\bfseries
An information-theoretic perspective on feed-forward loop abundances in transcriptional networks
}

\vspace{1.2em}

{\large
Mintu Nandi$^{1}$, Sudip Chattopadhyay$^{2}$, and Suman K Banik$^{3}$
}

\vspace{0.5em}

{\small
$^{1}$Universal Biology Institute, The University of Tokyo, 
7-3-1 Hongo, Bunkyo-ku, Tokyo 113-0033, Japan\\
$^{2}$Department of Chemistry, Indian Institute of Engineering Science and Technology, 
Shibpur, Howrah 711103, India\\
$^{3}$Department of Chemical Sciences, Bose Institute, 
EN 80, Sector V, Bidhan Nagar, Kolkata 700091, India
}

\end{center}



\setcounter{section}{0}
\setcounter{figure}{0}
\setcounter{table}{0}
\setcounter{equation}{0}

\renewcommand{\thesection}{S\arabic{section}} 
\renewcommand{\thetable}{S\arabic{table}}  
\renewcommand{\thefigure}{S\arabic{figure}} 
\renewcommand{\theequation}{S\arabic{equation}}


\begin{bibunit}

\section{Formulation of statistical moments}

The feed-forward loop contains two regulatory routes from the input $X$ to the output $Z$: a direct path $X\dashrightarrow Z$, and an indirect path $X\dashrightarrow Y\dashrightarrow Z$, where $\dashrightarrow$ represents a general notation for activation ($\rightarrow$) and repression ($\dashv$). Under the linear noise approximation (LNA) \cite{Elf2003, Hayot2004, Kampen2007, Gardiner2009}, fluctuations around the deterministic steady-state obey a linear stochastic dynamics. This linearity is important because it allows the principle of superposition: the total fluctuation in the output can be decomposed into the sum of the fluctuations generated through distinct paths. Thus, the fluctuation in $Z$ ($\delta z = z - \langle z\rangle$, with $\langle z\rangle$ being the steady-state copy number) can be written as
\begin{equation}
    \delta z = \delta z_0 + \delta z_d + \delta z_{ind},
    \label{eqs1}
\end{equation}

\noindent where, $\delta z_0$ denotes the intrinsic fluctuations generated locally at $Z$, $\delta z_d$ denotes the fluctuation propagated through the direct path, and $\delta z_{ind}$ denotes the fluctuation propagated through the indirect path. The input-output covariance and the output variance respond differently to this decomposition. The covariance between $X$ and $Z$ becomes,
\begin{equation}
    \sigma_{XZ} = \langle \delta x \delta z \rangle 
    = \langle \delta x (\delta z_0 + \delta z_d + \delta z_{ind}) \rangle.
    \label{eqs2}
\end{equation}

\noindent The intrinsic fluctuation $\delta z_0$ is independent of the upstream fluctuation $\delta x$, hence $\langle \delta x \delta z_0 \rangle =0$. Therefore, the input-output covariance separates into direct and indirect path contributions,
\begin{equation}
    \sigma_{XZ} = \langle \delta x \delta z_d \rangle + \langle \delta x \delta z_{ind} \rangle = \sigma_{XZ,d} + \sigma_{XZ,ind}.
    \label{eqs3}
\end{equation}

\noindent Upon normalization by the product of mean copy numbers of $X$ [$\langle x\rangle$] and $Z$ [$\langle z\rangle$], we obtain
\begin{equation}
    \zeta_{XZ} = \zeta_{XZ,d} + \zeta_{XZ,ind},
    \label{eqs4}
\end{equation}

\noindent where, $\zeta_{XZ}=\sigma_{XZ}/(\langle x\rangle \langle z\rangle)$, the direct path contribution is $\zeta_{XZ,d}=\sigma_{XZ,d}/(\langle x\rangle \langle z\rangle)$, and indirect path contribution is $\zeta_{XZ,ind}=\sigma_{XZ,ind}/(\langle x\rangle \langle z\rangle)$. \\

Similarly, the variance of $Z$ can be written as,
\begin{equation}
    \sigma_Z^2 = \langle (\delta z)^2 \rangle 
    = \langle (\delta z_0 + \delta z_d + \delta z_{ind})^2 \rangle.
    \label{eqs5}
\end{equation}

\noindent Expanding this expression gives
\begin{equation}
    \sigma_Z^2 = \langle \delta z_0^2 \rangle + \langle \delta z_d^2 \rangle + \langle \delta z_{ind}^2 \rangle + 2 \langle \delta z_d \delta z_{ind} \rangle,
    \label{eqs6}
\end{equation}

\noindent where the cross terms involving $\delta z_0$ vanish because the intrinsic fluctuation at Z is independent of the upstream path-mediated fluctuations. In Eq.~(\ref{eqs6}), the first term is the intrinsic fluctuations. The second and third terms are the fluctuations propagated through the direct and indirect paths, respectively. The final term, $2 \langle \delta z_d \delta z_{ind} \rangle$ is the interference contribution. It appears because the direct and indirect path-mediated fluctuations are not independent at $Z$, rather both are coupled to the upstream regulator $X$, and both converge on the same output node. We therefore write
\begin{equation}
    \sigma_Z^2 = \sigma_{Z,0}^2 + \sigma_{Z,d}^2 + \sigma_{Z,ind}^2 + \sigma_{Z,int}^2,
    \label{eqs7}
\end{equation}

\noindent with $\sigma_{Z,0}^2=\langle \delta z_0^2 \rangle$, $\sigma_{Z,d}^2=\langle \delta z_d^2 \rangle$, $\sigma_{Z,ind}^2=\langle \delta z_{ind}^2 \rangle$, and $\sigma_{Z,int}^2=2 \langle \delta z_d \delta z_{ind} \rangle$. Upon normalization by $\langle z\rangle^2$ yields the squared coefficient of variation (CV),
\begin{eqnarray}
    \eta_Z^2 = \frac{\sigma_Z^2}{\langle z\rangle^2}
    &=& \frac{\sigma_{Z,0}^2}{\langle z\rangle^2} + \frac{\sigma_{Z,d}^2}{\langle z\rangle^2} + \frac{\sigma_{Z,ind}^2}{\langle z\rangle^2} + \frac{\sigma_{Z,int}^2}{\langle z\rangle^2} \nonumber \\
    &=& \eta_{Z,0}^2 + \eta_{Z,d}^2 + \eta_{Z,ind}^2 + \eta_{Z,int}^2, \nonumber \\
    &=& \eta_{Z,path}^2 + \eta_{Z,int}^2,
    \label{eqs8}
\end{eqnarray}

\noindent where, $\eta_{Z,0}^2=1/\langle z\rangle$ due to Poisson nature of the birth-death processes. Here, we define the pathway contribution towards total relative fluctuations by $\eta_{Z,path}^2=\eta_{Z,0}^2 + \eta_{Z,d}^2 + \eta_{Z,ind}^2$. The interference contribution $\eta_{Z,int}^2$ quantifies the relative fluctuations propagation through both the pathways simultaneously. This decomposition establishes the physical origin of the sources of fluctuations in terms of squared CV. We next derive the analytical expressions of each decomposed term of Eqs.~(\ref{eqs4})~and~(\ref{eqs8}) using LNA. \\

To begin with, we rewrite the chemical master equation (CME) \cite{Kampen2007, Gardiner2009} governing the dynamics of the FFL as [see Eq.~(7) in the main text],
\begin{eqnarray}
    \frac{dP(\bm{n};t)}{dt} &=&
    \sum_{\mathcal{J} \in \{X,Y,Z\}} \left[ 
    \left( \mathbb{E}_\mathcal{J}^{+1}-1 \right) g_\mathcal{J}(\bm{n}) P(\bm{n};t) 
    +
    \left( \mathbb{E}_\mathcal{J}^{-1}-1 \right) f_\mathcal{J}(\bm{n}) P(\bm{n};t)
    \right]
    \label{eqs9}
\end{eqnarray}

\noindent where, $\mathcal{J}$ denote the TF index with the corresponding copy-number vector (state space) $\bm{n} = (n_X,n_Y,n_Z)^\top = (x,y,z)^\top \in \mathbb{N}_0^3$. Here $\top$ stands for matrix transpose. In the above equation, $P(\bm{n};t)$ denotes the probability to have the state $\bm{n}$ at time $t$. The operators $\mathbb{E}_\mathcal{J}^{\pm 1}$ are step operators that shift the copy number of TF $\mathcal{J}$ by $\pm1$. Specifically, when $\mathbb{E}_\mathcal{J}^{\pm 1}$ acts of a function $h(\bm{n})$, it gives $\mathbb{E}_\mathcal{J}^{\pm 1}h(\bm{n})=h(\bm{n} \pm \bm{e}_{\mathcal{J}})$, where $\bm{e}_{\mathcal{J}}$ is the unit vector along the $\mathcal{J}$th coordinate. The functions $f_\mathcal{J}(\bm{n})$ and $g_\mathcal{J}(\bm{n})$ denote the production and degradation propensities of TF $\mathcal{J}$, respectively. Their explicit forms depend on the type of the FFL, and are listed explicitly in Table~\ref{t1}. \\

To be specific the degradation of each TF is modeled as a first-order process, $g_{\mathcal{J}}(\bm{n})=\beta_{\mathcal{J}}n_{\mathcal{J}}$, where $\beta_{\mathcal{J}}$ is the degradation rate constant of TF $\mathcal{J}$. The production propensity $f_{\mathcal{J}}(\bm{n})$ contains a production rate, denoted by $\alpha_{\mathcal{J}}$, multiplied by a Hill-type regulatory factor (see Table~\ref{t1}). The expression of $f_{\mathcal{J}}$ contains $K_{\mathcal{U}\mathcal{U}'}$, which denotes the dissociation constant for binding of TF $\mathcal{U}$ to the promoter of TF $\mathcal{U}'$ with $(\mathcal{U},\mathcal{U}') \in \{ (X,Y), (X,Z), (Y,Z) \}$. The production rate of $X$ is constitutive, while the production rates of $Y$ and $Z$ depend on the regulatory signs and on the AND/OR logic listed in Table~\ref{t1}. As we formulate the model around a prescribed steady-state, the production rate constant $\alpha_\mathcal{J}$ is determined from the mean-field steady-state condition $f_{\mathcal{J}}(\langle\bm{n}\rangle)=g_{\mathcal{J}}(\langle\bm{n}\rangle)$ for each TF. This construction fixes the deterministic mean-field steady-state copy numbers $\langle n_\mathcal{J}\rangle$. See Table~\ref{t3} for the chosen parameter set in this study. \\

Under LNA, the CME~(\ref{eqs9}) yields the steady-state Lyapunov equation for the covariance matrix of the form \cite{Nandi2024},
\begin{equation}
    \mathbf{J} \mathbf{\Sigma} + \mathbf{\Sigma} \mathbf{J}^\top + \mathbf{D} = 0.
    \label{eqs10}
\end{equation}

\noindent We note that the detailed derivation of the Lyapunov equation from the CME is given in our previous work \cite{Nandi2024}. In this equation, $\mathbf{J}$ is the steady-state Jacobian matrix whose elements are given by,
\begin{equation}
    J_{\mathcal{J}\mathcal{J}'} = \frac{\partial}{\partial n_{\mathcal{J}'}}
    \left[ f_\mathcal{J}(\bm{n})-g_\mathcal{J}(\bm{n}) \right] \bigg|_{\bm{n}=\langle \bm{n}\rangle}
    \label{eqs11}
\end{equation}

\noindent where, $\mathcal{J},\mathcal{J}' \in \{X,Y,Z\}$ and $\langle \bm{n}\rangle=(\langle x\rangle, \langle y\rangle, \langle z\rangle)^\top$. In Eq.~(\ref{eqs10}), $\mathbf{\Sigma}$ is the covariance matrix defined as $\Sigma_{\mathcal{J}\mathcal{J}'}:=\langle \delta n_{\mathcal{J}} \delta n_{\mathcal{J}'} \rangle = \sigma_{\mathcal{J}\mathcal{J}'}$, where $\delta n_{\mathcal{J}}= n_{\mathcal{J}}-\langle n_{\mathcal{J}}\rangle$. When $\mathcal{J}=\mathcal{J}'$, $\sigma_{\mathcal{J}\mathcal{J}'}=:\sigma_{\mathcal{J}}^2$ represent the variance in $\mathcal{J}$th TF. \\

In Eq.~(\ref{eqs10}), $\mathbf{D}$ is the diffusion matrix, which accounts for the Gaussian white noise for the intrinsic fluctuations associated to each TF. As a result, the noise processes for each TF are assumed to have zero mean and are uncorrelated in time and across TFs \cite{Elf2003, Swain2004, Paulsson2004, Hayot2004, Tanase2006, deRonde2010}. This makes $\mathbf{D}$ a diagonal matrix and is given by $D_{\mathcal{J}\mathcal{J}'}=[f_\mathcal{J}(\langle \bm{n}\rangle)+g_\mathcal{J}(\langle \bm{n}\rangle)] \delta_{\mathcal{J}\mathcal{J}'}$, where $\delta_{\mathcal{J}\mathcal{J}'}$ stands for the Kronecker delta. At steady-state, $f_\mathcal{J}(\langle \bm{n}\rangle)=g_\mathcal{J}(\langle \bm{n}\rangle)$, so $D_{\mathcal{J}\mathcal{J}'}=2f_\mathcal{J}(\langle \bm{n}\rangle)=2g_\mathcal{J}(\langle \bm{n}\rangle)$. \\

Solving the Lyapunov equation [Eq.~(\ref{eqs10})] yields the second-order moments of the steady-state distribution $P(\bm{n})$, and the moments are given as,
\begin{eqnarray}
    \sigma_X^2 &=& \langle x\rangle,
    \label{eqs12} \\
    \sigma_{XY} &=& \frac{f_{YX}'}{\beta_X+\beta_Y} \sigma_X^2,
    \label{eqs13} \\
    \sigma_{XZ} &=& \frac{f_{ZX}'}{\beta_X+\beta_Z} \sigma_X^2 + \frac{f_{ZY}'}{\beta_X+\beta_Z} \sigma_{XY},
    \label{eqs14} \\
    \sigma_Y^2 &=& \langle y\rangle + \frac{f_{YX}'}{\beta_Y} \sigma_{XY},
    \label{eqs15} \\
    \sigma_{YZ} &=& \frac{f_{ZY}'}{\beta_Y+\beta_Z} \sigma_Y^2 + \frac{f_{ZX}'}{\beta_Y+\beta_Z} \sigma_{XY} 
    + \frac{f_{YX}'}{\beta_Y+\beta_Z} \sigma_{XZ},
    \label{eqs16} \\
    \sigma_Z^2 &=& \langle z\rangle + \frac{f_{ZX}'}{\beta_Z} \sigma_{XZ} + \frac{f_{ZY}'}{\beta_Z} \sigma_{YZ},
    \label{eqs17}
\end{eqnarray}

\noindent where, $f_{\mathcal{J}\mathcal{J}'}':=[\partial f_\mathcal{J}(\bm{n})/\partial n_{\mathcal{J}'}]_{\bm{n}=\langle \bm{n}\rangle}$ represent the regulatory sensitivity for the edge $\mathcal{J}' \dashrightarrow \mathcal{J}$ within the FFL. Here $\beta_\mathcal{J}$ represent the degradation rate constants of the TF $\mathcal{J}$. See Table~\ref{t1} for further details. \\

To derive the expressions of $\zeta_{XZ,d}$ and $\zeta_{XZ,ind}$ in Eq.~(\ref{eqs4}), we define the normalized covariance using Eq.~(\ref{eqs14}) and then rearrange, which yields,
\begin{eqnarray}
    \zeta_{XZ,d} &=& \frac{\mathcal{B}_1}{\langle z\rangle} f_{ZX}',
    \label{eqs18} \\
    \zeta_{XZ,ind} &=& \frac{\mathcal{E}_1}{\langle z\rangle} f_{YX}' f_{ZY}',
    \label{eqs19}
\end{eqnarray}

\noindent where, $\mathcal{B}_1=1/(\beta_X+\beta_Z)$ and $\mathcal{E}_1=1/[(\beta_X+\beta_Y)(\beta_X+\beta_Z)]$. Similarly, we derive $\eta_{Z,path}^2$ and $\eta_{Z,int}^2$ in Eq.~(\ref{eqs8}) by defining squared CV using Eq.~(\ref{eqs17}) followed by rearrangement,
\begin{eqnarray}
    \eta_{Z,path}^2 &=& 
    \underbracket{\frac{1}
    {\langle z\rangle}
    }_{=:\eta_{Z,0}^2}
    +
    \underbracket{
    \frac{\widehat{\mathcal{B}}_1 \langle x\rangle}{\langle z\rangle^2}
    (f_{ZX}')^2}_{=:\eta_{Z,d}^2} 
    +
    \underbracket{
    \frac{\mathcal{B}_2 \langle y\rangle}{\langle z\rangle^2} (f_{ZY}')^2 + \frac{\mathcal{E}_2 \langle x\rangle}{\langle z\rangle^2} (f_{YX}')^2 (f_{ZY}')^2 }_{=:\eta_{Z,ind}^2},
    \label{eqs20} \\
    \eta_{Z,int}^2 &=& 
    \frac{\mathcal{E}_3 \langle x\rangle}{\langle z\rangle^2} f_{YX}' f_{ZX}' f_{ZY}',
    \label{eqs21}
\end{eqnarray}

\noindent where, $\widehat{\mathcal{B}}_1=\mathcal{B}_1/\beta_Z$, $\mathcal{B}_2=1/[\beta_Z (\beta_Y+\beta_Z)]$, $\mathcal{E}_2=(\beta_X+\beta_Y+\beta_Z)/[\beta_Y \beta_Z (\beta_X+\beta_Y)(\beta_X+\beta_Z)(\beta_Y+\beta_Z)]$, and $\mathcal{E}_3 = 2\beta_Y \mathcal{E}_2$. \\

The coefficients $\mathcal{B}_1$, $\mathcal{E}_1$, $\widehat{\mathcal{B}}_1$, $\mathcal{B}_2$, $\mathcal{E}_2$, and $\mathcal{E}_3$ collect the dynamical filtering factors arising from the degradation rates of the nodes, whereas the regulatory sensitivities identify the corresponding path signatures. In the normalized covariance decomposition, $f_{ZX}'$ marks the direct edge $X\dashrightarrow Z$, while $f_{YX}'f_{ZY}'$ marks the indirect path $X\dashrightarrow Y\dashrightarrow Z$; hence, $\zeta_{XZ,d}$ and $\zeta_{XZ,ind}$ quantify the input-output covariance transmitted through the direct and indirect paths, respectively. In the variance decomposition, the term proportional to $(f_{ZX}')^2$ gives the direct-path noise contribution, while the terms proportional to $(f_{ZY}')^2$ and $(f_{YX}')^2(f_{ZY}')^2$ together define the indirect contribution. The former represents intrinsic fluctuations generated at $Y$ and propagated to $Z$, whereas the latter represents fluctuations originating from $X$ and propagated through the full indirect path. The interference contribution $\eta_{Z,int}^2$ is proportional to $f_{YX}'f_{ZX}'f_{ZY}'$, and therefore couples the direct and indirect path. \\

Since all prefactors in the noise decomposition are positive, the sign of each term is determined by the corresponding regulatory sensitivities. The pathway contribution $\eta_{Z,path}^2$ is always non-negative because it contains intrinsic noise and squared regulatory sensitivity terms. In contrast, the sign of $\eta_{Z,int}^2$ is determined by $f_{YX}'f_{ZX}'f_{ZY}'$, or equivalently by the relative signs of the direct-path sensitivity $f_{ZX}'$ and the indirect-path sensitivity $f_{YX}'f_{ZY}'$. Thus, coherent FFLs yield $\eta_{Z,int}^2>0$, corresponding to constructive path interference, whereas incoherent FFLs yield $\eta_{Z,int}^2<0$, corresponding to destructive path interference.

\section{Mutual information decomposition}

Based on the fluctuation decomposition, we next decompose the mutual information (MI) between $X$ and $Z$ into pathway-resolved. Since the LNA gives a Gaussian steady-state distribution, we use the Gaussian-channel expression of MI in terms of second-order moments. We begin with the Shannon definition of MI \cite{Shannon1948, Shannon1963, Cover1991},
\begin{equation}
    I(X;Z) = H(Z) - H(Z|X),
    \label{eqs22}
\end{equation}

\noindent where, $H(Z)$ is the Shannon entropy of $Z$ and $H(Z|X)$ is the conditional entropy of $Z$ given $X$. Under Gaussian channel approximation \cite{Cover1991, Barrett2015}, these entropies are determined by the variance of $Z$ and the conditional variance of $Z$ given $X$,
\begin{eqnarray}
    H(Z) &=& \frac{1}{2} \log_2 \left[2 \pi e \sigma_Z^2\right],
    \label{eqs23} \\
    H(Z|X) &=& \frac{1}{2} \log_2 \left[2 \pi e \sigma_{Z|X}^2 \right].
    \label{eqs24}
\end{eqnarray}

\noindent Here, the conditional variance is $\sigma_{Z|X}^2=\sigma_Z^2 - (\sigma_{XZ}^2/\sigma_X^2)$ \cite{Barrett2015}. We then express these quantities in normalized form using $\eta_Z^2:=\sigma_Z^2/\langle z\rangle^2$, $\eta_X^2:=\sigma_X^2/\langle x\rangle^2$, and $\zeta_{XZ}:=\sigma_{XZ}/(\langle x\rangle \langle z\rangle)$. This gives,
\begin{eqnarray}
    H(Z) &=& \frac{1}{2} \log_2 \left[2 \pi e \langle z\rangle^2 \eta_Z^2\right],
    \label{eqs25} \\
    H(Z|X) &=& \frac{1}{2} \log_2 \left[2 \pi e \langle z\rangle^2 \eta_{Z|X}^2 \right],
    \label{eqs26}
\end{eqnarray}

\noindent where, $\eta_{Z|X}^2=\eta_Z^2 - (\zeta_{XZ}^2/\eta_X^2)$ denotes the normalized conditional variance. \\

Using the decomposition of $\zeta_{XZ}$ and $\eta_Z^2$ in Eqs.~(\ref{eqs4})~and~(\ref{eqs8}), respectively, the entropy terms $H(Z)$ and $H(Z|X)$ can be expressed in pathway-resolved forms. For this purpose, we introduce the relative interference fluctuation, previously discussed in \cite{Nandi2026}, defined as $\tilde{\eta}_{Z,int}^2=\eta_{Z,int}^2/\eta_{Z,path}^2$. This dimensionless quantity compares the interference fluctuation with the path-wise output fluctuation and retains the sign of the interference term, thereby distinguishing constructive and destructive path interferences in the entropy formulation. Based on this definition, the normalized output variance can be written as $\eta_Z^2 = \eta_{Z,path}^2 (1+\tilde{\eta}_{Z,int}^2)$. Then the entropy $H(Z)$ separates naturally as,
\begin{eqnarray}
    H(Z) &=& \underbracket{\frac{1}{2} \log_2 \left[2 \pi e \langle z\rangle^2 \eta_{Z,path}^2 \right]}_{=:H_{path}(Z)} 
    +
    \underbracket{\frac{1}{2} \log_2 \left[1+\tilde{\eta}_{Z,int}^2 \right]}_{=:H_{int}(Z)},
    \label{eqs27}
\end{eqnarray}

\noindent where $H_{path}(Z)$ represents the entropy contribution arising from intrinsic and path-wise output fluctuations. The other contribution $H_{int}(Z)$ quantifies the entropy contribution at $Z$ due to direct-indirect pathway interference. \\

Similarly, $H(Z|X)$ can also be decomposed into pathway-resolved components. For this purpose, we recast the normalized conditional variance $\eta_{Z|X}^2$ using Eqs.~(\ref{eqs4})~and~(\ref{eqs8}) as,
\begin{eqnarray}
    \eta_{Z|X}^2 &=& 
    \underbracket{\eta_{Z,path}^2 - \frac{\zeta_{XZ,d}^2 + \zeta_{XZ,ind}^2}{\eta_X^2}}_{=:\eta_{Z|X,path}^2} 
    + 
    \underbracket{\eta_{Z,int}^2 - \frac{2\zeta_{XZ,d} \zeta_{XZ,ind}}{\eta_X^2}}_{=:\eta_{Z|X,int}^2},
    \label{eqs28}
\end{eqnarray}

\noindent where, $\eta_{Z|X,path}^2$ is the conditional path-wise fluctuation, while $\eta_{Z|X,int}^2$ is the conditional interference fluctuation. We then define the relative conditional interference fluctuation as $\tilde{\eta}_{Z|X,int}^2=\eta_{Z|X,int}^2/\eta_{Z|X,path}^2$, which gives, $\eta_{Z|X}^2=\eta_{Z|X,path}^2 (1+\tilde{\eta}_{Z|X,int}^2)$. This dimensionless quantity measures the conditional interference fluctuation relative to the conditional path-wise fluctuation and retains its sign. Substituting this expression into Eq.~(\ref{eqs26}) gives,
\begin{eqnarray}
    H(Z|X) &=& \underbracket{\frac{1}{2} \log_2 \left[2 \pi e \langle z\rangle^2 \eta_{Z|X,path}^2 \right]}_{=:H_{path}(Z|X)} 
    +
    \underbracket{\frac{1}{2} \log_2 \left[1+\tilde{\eta}_{Z|X,int}^2 \right]}_{=:H_{int}(Z|X)}.
    \label{eqs29}
\end{eqnarray}

\noindent Here, $H_{path}(Z|X)$ is the conditional entropy contribution arises from path-wise components and $H_{int}(Z|X)$ is the conditional entropy contribution arising from the direct-indirect pathway interference. \\

Now, substituting Eqs.~(\ref{eqs27})~and~(\ref{eqs29}) into Eq.~(\ref{eqs22}), the MI becomes,
\begin{equation}
    I(X;Z) = I_{path}(X;Z) + I_{int}(X;Z),
    \label{eqs30}
\end{equation}

\noindent where,
\begin{eqnarray}
    I_{path}(X;Z) &=& H_{path}(Z) - H_{path}(Z|X) 
    = 
    \frac{1}{2} \log_2 \left[\frac{\eta_{Z,path}^2}{\eta_{Z|X,path}^2}\right],
    \label{eqs31} \\
    I_{int}(X;Z) &=& H_{int}(Z) - H_{int}(Z|X) 
    =    \frac{1}{2} \log_2 \left[\frac{1+\tilde{\eta}_{Z,int}^2}{1+\tilde{\eta}_{Z|X,int}^2}\right].
    \label{eqs32}
\end{eqnarray}

\noindent The first term, $I_{path}(X;Z)$, represents the pathway information, which represents the joint information transmitted through the two paths in the open-loop reference. The second term, $I_{int}(X;Z)$, represents the interference information, or interference mutual information (IMI), which quantifies the information contribution generated by the direct-indirect pathway interference within the FFL. \\

Importantly, the pathway information $I_{path}(X;Z)$ is always non-negative, which directly follows from Eq.~(\ref{eqs31}) and the definition of $\eta_{Z|X,path}^2$ in Eq.~(\ref{eqs28}). The subtraction term in $\eta_{Z|X,path}^2$ is non-negative because it contains squared covariance contributions $\zeta_{XZ,d}^2$ and $\zeta_{XZ,ind}^2$. Thus, conditioning on $X$ reduces the path-wise uncertainty of $Z$, giving $\eta_{Z|X,path}^2 \le \eta_{Z,path}^2$, and therefore $I_{path}(X;Z) \ge 0$. The interference information has a different character because it is controlled by the mixed direct-indirect contribution. The sign of $I_{int}(X;Z)$ is determined by whether the relative interference fluctuation is larger before or after conditioning on $X$. \\

In coherent FFLs, the direct and indirect path sensitivities have the same sign. The mixed term $f_{YX}'f_{ZX}'f_{ZY}'$ is then positive, so the two paths constructively contribute to output fluctuations. Conditioning on $X$ removes part of the common input-driven contribution, giving $\eta_{Z,int}^2 > \eta_{Z|X,int}^2 >0$ in the present model. However, conditioning also reduces the pathway contribution, so this reduction in interference alone does not determine the sign of IMI. As in the main text, we consider operating regimes satisfying $(\eta_{Z,int}^2/\eta_{Z|X,int}^2) > (\eta_{Z,path}^2/\eta_{Z|X,path}^2)$. This condition means that conditioning removes a larger fraction of the interference contribution than of the pathway contribution. Since both interference quantities are positive, it is equivalent to $\tilde{\eta}_{Z,int}^2 > \tilde{\eta}_{Z|X,int}^2$ in Eq.~(\ref{eqs32}), which gives $I_{int}(X;Z) > 0$. \\

For incoherent FFLs, the direct and indirect path sensitivities have opposite signs. The product $f_{YX}'f_{ZX}'f_{ZY}'$ is negative, so the mixed direct-indirect contribution is destructive. In this case, the interference fluctuation reduces the output uncertainty relative to the pathway contribution. Conditioning on $X$ removes part of this input-mediated destructive interference, making the relative interference contribution less negative with $\eta_{Z,int}^2 < \eta_{Z|X,int}^2 <0$ in the present model. Again, conditioning also reduces the pathway contribution, so the change in interference alone does not determine the sign of IMI. Under the same ratio condition stated above, conditioning removes a larger fraction of the magnitude of the interference contribution than of the pathway contribution. Since both interference quantities are negative, this condition is equivalent to $\tilde{\eta}_{Z,int}^2 < \tilde{\eta}_{Z|X,int}^2$ in Eq.~(\ref{eqs32}), and hence $I_{int}(X;Z) < 0$. \\

Thus, the positive and negative IMI interpretations for coherent and incoherent FFLs apply to the specified operating regimes, rather than following from regulatory signs alone. More generally, $I_{path}(X;Z)$ quantifies the non-negative information transmitted through the individual regulatory paths, whereas $I_{int}(X;Z)$ quantifies how the two paths jointly reshape information transmission. This makes $I_{int}(X;Z)$ a pathway-interference measure in information transmission.

\section{Open-loop equivalents of FFLs as reference systems}

\begin{figure*}[!t]
\includegraphics[width=1.8\columnwidth,angle=0]{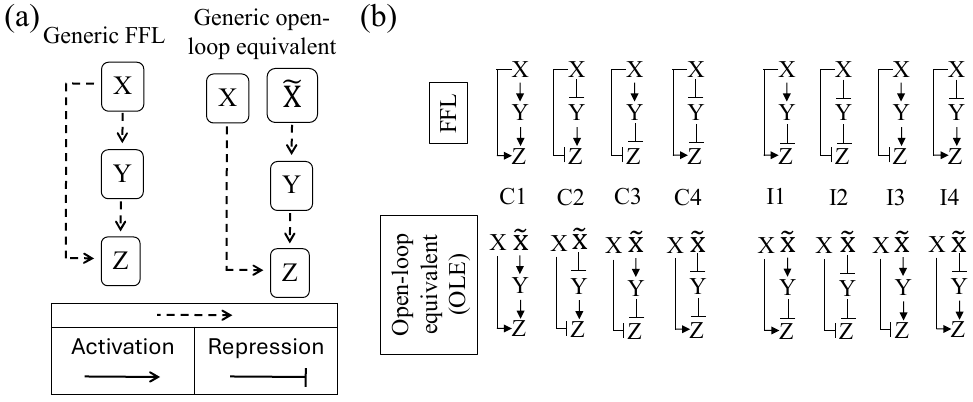}
\caption{
\textbf{Open-loop equivalent as a reference system of FFLs}.
(a) Schematic of a generic FFL and the corresponding OLE, in which the indirect path is driven by an independent but statistically equivalent input, $\tilde{X}$. 
(b) The eight FFL types and their corresponding open-loop equivalents are shown.
}
\label{sf0}
\end{figure*}

To clarify the meaning of the pathway information $I_{path}(X;Z)$, we construct an open-loop equivalent (OLE) of each FFL motif. In the original FFL, the same upstream regulator $X$ drives both the direct path $X\dashrightarrow Z$ and the indirect path $X\dashrightarrow Y\dashrightarrow Z$. This shared upstream regulator generates the direct--indirect pathway interference quantified by $I_{int}(X;Z)$. In the OLE, the direct path remains driven by $X$, whereas the indirect path is driven by an independent but statistically equivalent input $\tilde{X}$. Thus, the regulatory structure becomes $X\dashrightarrow Z$ and $\tilde{X}\dashrightarrow Y\dashrightarrow Z$ (see Fig.~\ref{sf0}). This reference construction removes the common-input coupling between the two paths while keeping the direct and indirect routes parameter-matched to the corresponding FFL. \\

The stochastic dynamics are governed by the same CME as Eq.~(\ref{eqs9}), but now with the four-dimensional copy-number vector
\begin{equation}
\bm{n}_{OLE}=(x,\tilde{x},y,z)^{\top}.
\label{eqs64}
\end{equation}

\noindent The production propensities are constructed directly from Table~S1. The only modification is that the regulation $X\dashrightarrow Y$ in the FFL is replaced by $\tilde{X}\dashrightarrow Y$ in the OLE. For example, for the C1-FFL under AND logic, the corresponding production propensities for C1-OLE are
\begin{eqnarray}
f_X^{OLE} &=& \alpha_X, \nonumber \\
f_{\tilde{X}}^{OLE} &=& \alpha_{\tilde{X}}, \nonumber \\
f_Y^{OLE} &=& \alpha_Y \frac{\tilde{x}}{K_{\tilde{X}Y}+\tilde{x}}, \nonumber \\
f_Z^{OLE} &=& \alpha_Z
\frac{x}{K_{XZ}+x}
\frac{y}{K_{YZ}+y}.
\label{eqs65}
\end{eqnarray}

\noindent For the OR-gated C1-OLE, the production propensity of $Z$ is instead
\begin{equation}
f_Z^{OLE}
=
\alpha_Z
\left[
\frac{x}{K_{XZ}+x}
+
\frac{y}{K_{YZ}+y}
\right].
\label{eqs66}
\end{equation}
The open-loop propensities for all other FFL types are obtained analogously from Table~S1 by replacing the $X$-dependent term in $f_Y$ with the corresponding $\tilde{X}$-dependent term, while keeping the $X$-dependent and $Y$-dependent terms in $f_Z$ unchanged. \\

For each optimized FFL parameter set listed in Table~\ref{t5}, the corresponding OLE is generated without any additional optimization. The independent input $\tilde{X}$ is chosen to be statistically equivalent to $X$ by setting
\begin{equation}
\beta_{\tilde{X}}=\beta_X,\qquad
\langle \tilde{x}\rangle=\langle x\rangle,\qquad
K_{\tilde{X}Y}=K_{XY}.
\label{eqs67}
\end{equation}
All remaining parameters, including $\beta_Y$, $\beta_Z$, $K_{XZ}$, $K_{YZ}$, $\langle y\rangle$, and $\langle z\rangle$, are inherited from the optimized FFL. The production rates $\alpha_X$, $\alpha_{\tilde{X}}$, $\alpha_Y$, and $\alpha_Z$ are determined from the same mean-field steady-state conditions,
\begin{equation}
f_\mathcal{J}^{OLE}(\langle\bm{n}_{OLE}\rangle)
=
g_\mathcal{J}^{OLE}(\langle\bm{n}_{OLE}\rangle),
\label{eqs68}
\end{equation}

\noindent where, $\mathcal{J}\in\{X,\tilde{X},Y,Z\}$. \\

Under LNA, the steady-state covariance matrix of the OLE satisfies the Lyapunov equation
\begin{equation}
\mathbf{J}_{OLE}\mathbf{\Sigma}_{OLE}
+
\mathbf{\Sigma}_{OLE}\mathbf{J}_{OLE}^{\top}
+
\mathbf{D}_{OLE}
=
0.
\label{eqs69}
\end{equation}

\noindent For the variable ordering $(X,\tilde{X},Y,Z)$, the Jacobian matrix is
\begin{equation}
\mathbf{J}_{OLE}
=
\begin{pmatrix}
-\beta_X & 0 & 0 & 0 \\
0 & -\beta_{\tilde{X}} & 0 & 0 \\
0 & f'_{Y\tilde{X}} & -\beta_Y & 0 \\
f'_{ZX} & 0 & f'_{ZY} & -\beta_Z
\end{pmatrix},
\label{eqs70}
\end{equation}

\noindent where the regulatory sensitivities are defined directly from the open-loop production propensities as
\begin{eqnarray}
f'_{Y\tilde{X}}
&=&
\frac{\partial f_Y^{OLE}}{\partial \tilde{x}}
\Bigg|_{\bm{n}_{OLE}=\langle\bm{n}_{OLE}\rangle},
\nonumber\\
f'_{ZX}
&=&
\frac{\partial f_Z^{OLE}}{\partial x}
\Bigg|_{\bm{n}_{OLE}=\langle\bm{n}_{OLE}\rangle},
\nonumber\\
f'_{ZY}
&=&
\frac{\partial f_Z^{OLE}}{\partial y}
\Bigg|_{\bm{n}_{OLE}=\langle\bm{n}_{OLE}\rangle}.
\label{eqs71}
\end{eqnarray}

\noindent The diffusion matrix is diagonal and is given by
\begin{equation}
\mathbf{D}_{OLE}
=
\begin{pmatrix}
2\beta_X\langle x\rangle & 0 & 0 & 0 \\
0 & 2\beta_{\tilde{X}}\langle\tilde{x}\rangle & 0 & 0 \\
0 & 0 & 2\beta_Y\langle y\rangle & 0 \\
0 & 0 & 0 & 2\beta_Z\langle z\rangle
\end{pmatrix}.
\label{eqs72}
\end{equation}

Solving Eq.~(\ref{eqs69}) gives the following second-order moments:
\begin{eqnarray}
\sigma_X^2 &=& \langle x\rangle, \quad
\sigma_{\tilde{X}}^2 = \langle \tilde{x}\rangle, \quad
\sigma_{X\tilde{X}} = 0,
\label{eqs73} \\
\sigma_{\tilde{X}Y}
&=&
\frac{f'_{Y\tilde{X}}}{\beta_{\tilde{X}}+\beta_Y}
\sigma_{\tilde{X}}^2,
\qquad
\sigma_{XY}=0,
\label{eqs74} \\
\sigma_{XZ}
&=&
\frac{f'_{ZX}}{\beta_X+\beta_Z}
\sigma_X^2,
\label{eqs75} \\
\sigma_{\tilde{X}Z}
&=&
\frac{f'_{ZY}}{\beta_{\tilde{X}}+\beta_Z}
\sigma_{\tilde{X}Y},
\label{eqs76} \\
\sigma_Y^2
&=&
\langle y\rangle
+
\frac{f'_{Y\tilde{X}}}{\beta_Y}
\sigma_{\tilde{X}Y},
\label{eqs77} \\
\sigma_{YZ}
&=&
\frac{f'_{ZY}}{\beta_Y+\beta_Z}
\sigma_Y^2
+
\frac{f'_{Y\tilde{X}}}{\beta_Y+\beta_Z}
\sigma_{\tilde{X}Z},
\label{eqs78} \\
\sigma_Z^2
&=&
\langle z\rangle
+
\frac{f'_{ZX}}{\beta_Z}
\sigma_{XZ}
+
\frac{f'_{ZY}}{\beta_Z}
\sigma_{YZ}.
\label{eqs79} 
\end{eqnarray}

\noindent These expressions show that $X$ and $\tilde{X}$ are statistically independent, while both regulate the same output $Z$ through distinct routes. \\

The normalized input-output covariances in the OLE are
\begin{eqnarray}
\zeta_{XZ}^{OLE}
&=&
\frac{\sigma_{XZ}}{\langle x\rangle\langle z\rangle}
=
\frac{1}{\langle z\rangle}
\frac{f'_{ZX}}{\beta_X+\beta_Z},
\label{eqs80} \\
\zeta_{\tilde{X}Z}^{OLE}
&=&
\frac{\sigma_{\tilde{X}Z}}{\langle \tilde{x}\rangle\langle z\rangle}
=
\frac{1}{\langle z\rangle}
\frac{f'_{Y\tilde{X}}f'_{ZY}}
{(\beta_{\tilde{X}}+\beta_Y)(\beta_{\tilde{X}}+\beta_Z)}.
\label{eqs81}
\end{eqnarray}

\noindent Using Eq.~(\ref{eqs67}), these become identical to the direct and indirect covariance contributions of the corresponding FFL,
\begin{equation}
\zeta_{XZ}^{OLE}=\zeta_{XZ,d},
\qquad
\zeta_{\tilde{X}Z}^{OLE}=\zeta_{XZ,ind}.
\label{eqs82}
\end{equation}

The squared coefficient of variation of $Z$ in the OLE is defined as
\begin{equation}
\eta_{Z,OLE}^2
=
\frac{\sigma_Z^2}{\langle z\rangle^2}.
\label{eqs83}
\end{equation}

\noindent Substituting Eqs.~(\ref{eqs75})--(\ref{eqs79}) gives
\begin{eqnarray}
\eta_{Z,OLE}^2
&=&
\frac{1}{\langle z\rangle}
+
\frac{\widehat{\mathcal{B}}_1\langle x\rangle}{\langle z\rangle^2}
(f'_{ZX})^2
+
\frac{\mathcal{B}_2\langle y\rangle}{\langle z\rangle^2}
(f'_{ZY})^2
+
\frac{\widetilde{\mathcal{E}}_2\langle\tilde{x}\rangle}{\langle z\rangle^2}
(f'_{Y\tilde{X}})^2(f'_{ZY})^2,
\label{eqs84}
\end{eqnarray}

\noindent where,
\begin{eqnarray}
\widehat{\mathcal{B}}_1
&=&
\frac{1}{\beta_Z(\beta_X+\beta_Z)},
\nonumber\\
\mathcal{B}_2
&=&
\frac{1}{\beta_Z(\beta_Y+\beta_Z)},
\nonumber\\
\widetilde{\mathcal{E}}_2
&=&
\frac{\beta_{\tilde{X}}+\beta_Y+\beta_Z}
{\beta_Y\beta_Z(\beta_{\tilde{X}}+\beta_Y)(\beta_{\tilde{X}}+\beta_Z)(\beta_Y+\beta_Z)}.
\label{eqs85}
\end{eqnarray}

\noindent In Eq.~(\ref{eqs84}), no direct--indirect interference term appears because the two routes are driven by independent upstream inputs. Under the parameter matching in Eq.~(\ref{eqs67}), we obtain,
\begin{equation}
\eta_{Z,OLE}^2
=
\eta_{Z,path}^2.
\label{eqs86}
\end{equation}

The total mutual information between the two independent inputs $(X,\tilde{X})$ and the output $Z$ is computed using the Gaussian expression,
\begin{equation}
I_{OLE}(X,\tilde{X};Z)
=
\frac{1}{2}
\log_2
\left[
\frac{\eta_{Z,OLE}^2}
{\eta_{Z|X,\tilde{X},OLE}^2}
\right],
\label{eqs87}
\end{equation}

\noindent where the normalized conditional output variance is
\begin{equation}
\eta_{Z|X,\tilde{X},OLE}^2
=
\eta_{Z,OLE}^2
-
\frac{\left(\zeta_{XZ}^{OLE}\right)^2}{\eta_X^2}
-
\frac{\left(\zeta_{\tilde{X}Z}^{OLE}\right)^2}{\eta_{\tilde{X}}^2}.
\label{eqs88}
\end{equation}

\noindent Here,
\begin{equation}
\eta_X^2=\frac{\sigma_X^2}{\langle x\rangle^2},
\qquad
\eta_{\tilde{X}}^2=\frac{\sigma_{\tilde{X}}^2}{\langle \tilde{x}\rangle^2}.
\label{eqs89}
\end{equation}

\noindent Using Eq.~(\ref{eqs82}), Eq.~(\ref{eqs88}) becomes identical to the path-wise conditional output fluctuation of the corresponding FFL,
\begin{equation}
\eta_{Z|X,\tilde{X},OLE}^2
=
\eta_{Z|X,path}^2.
\label{eqs90}
\end{equation}
Therefore,
\begin{equation}
I(X,\tilde{X};Z)
=
\frac{1}{2}
\log_2
\left[
\frac{\eta_{Z,path}^2}
{\eta_{Z|X,path}^2}
\right]
=
I_{path}(X;Z).
\label{eqs91}
\end{equation}

\noindent Thus, the pathway information of the FFL is reproduced by the total information transmission of its corresponding OLE. This comparison clarifies that $I_{path}(X;Z)$ represents the information carried by the direct and indirect routes after removing their non-additive coupling, whereas $I_{int}(X;Z)$ captures this coupling and quantifies how the shared-input FFL topology modifies information transmission. As shown in Fig.~\ref{sf9}, $I_{path}(X;Z)$ of each optimized FFL closely matches $I_{OLE}(X,\tilde{X};Z)$ of the corresponding OLE for both AND- and OR-gated architectures. Based on this, we can write the IMI as the difference between the MI content in FFL [$I_{\rm FFL}(X;Z)$] and in the corresponding OLE,
\begin{equation}
    I_{int}(X;Z) = I_{\rm FFL}(X;Z) - I_{\rm OLE}(X,\tilde{X};Z).
    \label{eqs92}
\end{equation}

\section{Optimizing interference mutual information for the abundance-like patterns}

We now ask whether the IMI [$I_{int}(X;Z)$] can mimic the observed abundance hierarchy of FFL motifs shown in Fig.~1a in the main text. For this purpose we set-up an optimization framework based on cross-entropy method (CEM). The aim of this optimization is not to infer a unique biochemical parameter set, instead, it is to test whether there exist dynamically feasible parameter regimes in which the hierarchy of IMI follows the empirical motif-abundance pattern. If such regimes are repeatedly obtained across independent optimization runs, then the IMI hierarchy can be hypothesized to explain the observed abundance pattern. \\

\noindent \textbf{Setting-up the objective.} For each organism, FFL class, and gate logic, we optimize the biochemical parameter vector
\begin{equation*}
\bm{\theta}
=
\left(\beta_X,\beta_Y,\beta_Z,K_{XY},K_{XZ},K_{YZ},\langle x\rangle,\langle y\rangle,\langle z\rangle\right).
\end{equation*}

\noindent Here, $\beta_\mathcal{J}$ is the degradation rate constant for the TF $\mathcal{J}$, and $K_{\mathcal{U}\mathcal{U}'}$ with $(\mathcal{U},\mathcal{U}')\in\{(X,Y),(X,Z),(Y,Z)\}$, denotes the dissociation constant for binding of the TF $\mathcal{U}$ to the promoter of the TF $\mathcal{U}'$. The quantities $\langle x\rangle$, $\langle y\rangle$, and $\langle z\rangle$ denote the mean copy numbers of the three TFs. Thus, the number of optimized kinetic parameters is $p=9$. For the $i$th component of $\bm{\theta}$, the lower and upper search bounds are denoted by $\theta_i^{\min}$ and $\theta_i^{\max}$, respectively. These bounds are chosen as biologically plausible order-of-magnitude search windows based on previous studies \cite{Bintu2005, Belle2006, Garcia2011, Hintsche2013, Christiano2014, Li2014, Hansen2015, Milo2015, Gupta2024}, and are listed explicitly in Table~\ref{t3}. \\

For a given organism, FFL class, and gate logic, the four motifs to be optimized are denoted by the set $\mathcal{M}=\{m_1,m_2,m_3,m_4\}$. For coherent FFLs, $\mathcal{M}=\{\rm C1,C2,C3,C4\}$, whereas for incoherent FFLs, $\mathcal{M}=\{\rm I1,I2,I3,I4\}$. For a motif $m\in\mathcal{M}$, the parameter vector $\bm{\theta}$ determines the regulatory sensitivities, the pathway and interference fluctuation terms, and hence the IMI $I_{int}^{(m)}(X;Z)$. Since $I_{int}\geq 0$ for coherent FFLs and $I_{int}\leq 0$ for incoherent FFLs, we define the IMI strength of motif $m$ as
\begin{equation}
\mathcal{I}_m(\bm{\theta})
=
\left| I_{int}^{(m)}(X;Z) \right|.
\label{eqs33}
\end{equation}

\noindent This definition allows coherent and incoherent motifs to be compared on the same positive scale. The sign of $I_{int}$ is still retained mechanistically: positive $I_{int}$ corresponds to constructive direct-indirect path interference in coherent FFLs, whereas negative $I_{int}$ corresponds to destructive direct-indirect path interference in incoherent FFLs. \\

The observed motif-abundance pattern are used only to define qualitative target ranges for the relative IMI strengths. For each organism, FFL class, and gate logic, we first identify selected motif pairs $\mathcal{Q}=\{(a,b)\}$, where motif $a$ has larger observed abundance than motif $b$, as shown in Fig.~1a in the main text. Now, for each pair $(a,b)\in\mathcal{Q}$, the corresponding IMI-strength ratio is defined as,
\begin{equation}
R_{ab}(\bm{\theta})
=
\frac{
\mathcal{I}_a(\bm{\theta})
}
{
\mathcal{I}_b(\bm{\theta})
}.
\label{eqs34}
\end{equation}

\noindent The optimization then requires this ratio to lie within a prescribed target interval,
\begin{equation}
L_{ab}
\leq
R_{ab}(\bm{\theta})
\leq
U_{ab},
\label{eqs35}
\end{equation}

\noindent where $L_{ab}$ and $U_{ab}$ are the lower and upper bounds of the target IMI-ratio window for the motif pair $(a,b)$. These bounds are chosen as broad qualitative ranges guided by the ratios of the observed abundances in Fig. 1a in the main text, rather than as exact fitted abundance values. This benchmark reflects the network reconstructions available at the time of the original abundance studies. We adopt this benchmark for comparability with the prior theoretical literature rather than as a settled ground truth. Thus, the constraints test whether the optimized IMI hierarchy can remain consistent with the observed motif-abundance hierarchy. The target IMI-ratio windows are listed in Table~\ref{t4}. \\

To quantify the violation of the target ratio constraints, we define the penalty function,
\begin{eqnarray}
\mathcal{P}(\bm{\theta})
&=&
\sum_{(a,b)\in\mathcal{Q}}
\left[
\max\left(0,\log L_{ab}-\log R_{ab}\right)^2 
+
\max\left(0,\log R_{ab}-\log U_{ab}\right)^2
\right].
\label{eqs36}
\end{eqnarray}

\noindent Thus, $\mathcal{P}(\bm{\theta})=0$ when all motif-ratio constraints are satisfied. In numerical calculations, a parameter set is defined as feasible when $\mathcal{P}(\bm{\theta})\leq \epsilon_{\mathcal{P}}$, where $\epsilon_{\mathcal{P}}$ is a small numerical tolerance. Therefore, feasibility means that the full set of ratio constraints is satisfied within numerical precision. \\

Among feasible parameter sets, we favor larger average IMI strength, defined as,
\begin{equation}
\overline{\mathcal{I}}(\bm{\theta})
=
\frac{1}{|\mathcal{M}|}
\sum_{m\in\mathcal{M}}
\mathcal{I}_m(\bm{\theta}).
\label{eqs37}
\end{equation}

\noindent Since each motif class contains four FFLs, $|\mathcal{M}|=4$. This average is important because the optimization should not reproduce the target IMI hierarchy by increasing only one motif. Instead, it should identify a dynamical regime in which the full class of coherent or incoherent FFLs displays the desired IMI hierarchy. The complete objective function is then defined as,
\begin{equation}
\mathcal{L}_{opt}(\bm{\theta})
=
\begin{cases}
-\overline{\mathcal{I}}(\bm{\theta}), 
& \mathcal{P}(\bm{\theta})\leq \epsilon_{\mathcal{P}},\\
\Omega+\mathcal{P}(\bm{\theta}), 
& \mathcal{P}(\bm{\theta})> \epsilon_{\mathcal{P}},
\end{cases}
\label{eqs38}
\end{equation}

\noindent where, $\Omega$ is a large infeasibility offset. This objective function imposes a hierarchical optimization rule. Parameter sets that satisfy the motif-ratio constraints, i.e., $\mathcal{P}(\bm{\theta})\leq \epsilon_{\mathcal{P}}$, are treated as feasible and are ranked by $-\overline{\mathcal{I}}(\bm{\theta})$, so that larger average IMI strength is favored within the feasible region. Parameter sets that violate the motif-ratio constraints are treated as infeasible and are ranked by $\Omega+\mathcal{P}(\bm{\theta})$, so that smaller violations of the target IMI hierarchy are preferred during the search. Thus, the objective first enforces the empirical motif-ratio constraints and then favors larger average IMI strength among parameter sets satisfying those constraints. The purpose is not to infer a unique globally optimized biochemical parameter set, but to identify dynamical regimes in which the IMI hierarchy remains consistent with the target motif-ratio structure. \\

\noindent \textbf{Cross-entropy method.} The optimization is performed using CEM \cite{Rubinstein1999,deBoer2005}. The search is carried out in logarithmic parameter space,
\begin{equation}
\bm{\phi}
=
\log \bm{\theta}.
\label{eqs39}
\end{equation}

\noindent This choice ensures that all kinetic parameters remain positive and that multiplicative changes in rate constants, regulatory thresholds, and mean copy numbers are treated consistently. The lower and upper parameter bounds in log space become,
\begin{equation}
\bm{\phi}^{\min}
=
\log \bm{\theta}^{\min},
\qquad
\bm{\phi}^{\max}
=
\log \bm{\theta}^{\max}.
\label{eqs40}
\end{equation}

\noindent Before the main CEM optimization, a feasibility pre-scan is performed by sampling $N_{\mathrm{scan}}$ candidate parameter vectors. The best parameter set found during this pre-scan is used as the initial center of the CEM sampling distribution. \\

For each independent run $r$, where $r=1,\ldots,N_{\mathrm{run}}$, the CEM maintains a Gaussian sampling distribution, given by,
\begin{equation}
\bm{\phi}
\sim
\mathcal{N}(\bm{\mu}_{r,t},\mathbf{C}_{r,t}).
\label{eqs41}
\end{equation}

\noindent Here, $t=1,\ldots,T$ is the CEM iteration index, $T$ is the total number of CEM iterations, $\bm{\mu}_{r,t}$ is the sampling mean in run $r$ at iteration $t$, and $\mathbf{C}_{r,t}$ is the corresponding sampling covariance matrix. We note that in all optimization cases, $T=1000$ is used as a fixed computational budget. This common stopping criterion allows different organisms, FFL classes, and gate logics to be compared under the same optimization effort. Therefore, $T=1000$ should not be interpreted as a proof that the global maximum of average IMI has been reached. \\

At iteration $t$, a population of $N$ candidate vectors is sampled within the allowed interval $[\bm{\phi}^{\min},\bm{\phi}^{\max}]$, given by, 
\begin{equation}
\left\{
\bm{\phi}_{r,t}^{(j)}
\right\}_{j=1}^{N},
\label{eqs42}
\end{equation}

\noindent where $j$ indexes the sampled parameter sets. To evaluate the objective function, each sampled log-parameter set is converted back to the original scale as $\bm{\theta}_{r,t}^{(j)} = \exp\left[ \bm{\phi}_{r,t}^{(j)} \right]$. The objective value $\mathcal{L}_{opt}(\bm{\theta}_{r,t}^{(j)})$ is then computed, and the candidates are ranked according to this value, with smaller values corresponding to better candidates. Then the elite set, $\mathcal{E}_{r,t}$, is defined as the subset of log-parameter vectors whose corresponding parameter vectors belong to the best fraction $\rho$ of this ranked population, i.e.,
\begin{eqnarray}
\mathcal{E}_{r,t}
=
\left\{
\bm{\phi}_{r,t}^{(j)}
:
\bm{\theta}_{r,t}^{(j)}
\text{ belongs to the best } \rho N
\text{ candidates}
\right\}. \nonumber \\
\label{eqs_opt16}
\end{eqnarray}

\noindent In practice, the elite size is taken as $\max(2,\lfloor \rho N\rfloor)$ to ensure that at least two samples are used in the covariance update. The CEM sampling mean and covariance matrix are then updated as,
\begin{eqnarray}
\bm{\mu}_{r,t+1}
&=&
(1-s)\bm{\mu}_{r,t}
+
s
\langle
\bm{\phi}
\rangle_{\mathcal{E}_{r,t}},
\label{eqs44}
\\
\mathbf{C}_{r,t+1}
&=&
(1-s)\mathbf{C}_{r,t}
+
s
\mathrm{Cov}_{\mathcal{E}_{r,t}}(\bm{\phi})
+
\epsilon_C\mathbb{I}.
\end{eqnarray}

\noindent Here, $\langle\bm{\phi}\rangle_{\mathcal{E}_{r,t}}$ is the mean of the elite samples, $\mathrm{Cov}_{\mathcal{E}_{r,t}}(\bm{\phi})$ is their covariance matrix, $s$ is the smoothing parameter, $\epsilon_C$ is the covariance floor, and $\mathbb{I}$ is the identity matrix. The covariance floor prevents premature collapse of the sampling distribution and maintains numerical stability. The CEM hyperparameters used in all optimization runs are listed in Table~\ref{t2}. \\

At the final CEM iteration, the best parameter set from each independent run, denoted by $\bm{\theta}_r^{\ast}=\bm{\theta}_{r,T}^{\ast}$, is reported in Table~\ref{t5}. Then, for each motif, the signed IMI value, $I_{\mathrm{int}}^{(m)}(X;Z)$, is evaluated at this parameter set. The values reported in Fig.~3a-d in the main text and in Fig.~\ref{sf6} correspond to the mean and standard deviation of these signed motif-specific IMI values across independent CEM runs. Thus, coherent FFLs are reported with positive IMI values, whereas incoherent FFLs are reported with negative IMI values. \\

\noindent \textbf{Diagnostic measures.} To evaluate whether the optimization converges and whether the optimized regimes are reproducible, we analyze several diagnostic measures, as listed below. 
\begin{enumerate}
\item For each run $r$ and iteration $t$, we define the set of all parameter vectors sampled from iteration $1$ to iteration $t$ as,
\begin{equation}
    \mathcal{S}_{r,1:t}=\left\{\bm{\theta}_{r,\tau}^{(j)}:
    1\leq \tau \leq t,\; 1\leq j\leq N\right\}.
    \label{eqs46}
\end{equation}

\noindent The best parameter set is then,
\begin{equation}
    \bm{\theta}_{r,t}^{\ast}=\arg\min_{\bm{\theta}\in\mathcal{S}_{r,1:t}}\mathcal{L}_{opt}(\bm{\theta}).
    \label{eqs47}
\end{equation}

\noindent The best IMI strength is,
\begin{equation}
    \overline{\mathcal{I}}_{r,t}^{\ast}=\overline{\mathcal{I}}
    \left(\bm{\theta}_{r,t}^{\ast}\right),
    \label{eqs48}
\end{equation}

\noindent and the corresponding best penalty is,
\begin{equation}
    \mathcal{P}_{r,t}^{\ast}=\mathcal{P}\left(
    \bm{\theta}_{r,t}^{\ast}\right),
    \label{eqs49}
\end{equation}

\noindent Because $\overline{\mathcal{I}}_{r,t}^{\ast}$ is a best quantity, it is cumulative over iterations and may continue to increase slowly even after a feasible region has been reached. Thus, saturation of $\overline{\mathcal{I}}_{r,t}^{\ast}$ is not required for interpreting the optimization. Instead, the relevant diagnostic evidence is the simultaneous reduction of $\mathcal{P}_t^{\ast}$ toward zero, the appearance of feasible sampled parameter sets, and the stabilization of the parameter-search trajectories. \\

Across independent runs, we report the averages of best IMI strength $\overline{\mathcal{I}}_t^{\ast}$ and best penalty $\mathcal{P}_t^{\ast}$ along with their standard deviations, $\mathrm{SD}(\mathcal{I}_{r,t}^{\ast})$ and $\mathrm{SD}(\mathcal{P}_{r,t}^{\ast})$, respectively. These averages and standard deviations are defined as,
\begin{eqnarray}
    \overline{\mathcal{I}}_t^{\ast} &=& \frac{1}{N_{\mathrm{run}}} \sum_{r=1}^{N_{\mathrm{run}}} \overline{\mathcal{I}}_{r,t}^{\ast},
    \label{eqs49-1} \\
    \mathrm{SD}(\mathcal{I}_{r,t}^{\ast}) &=& \sqrt{\frac{1}{N_{\mathrm{run}}-1} \sum_{r=1}^{N_{\mathrm{run}}} \left(\overline{\mathcal{I}}_{r,t}^{\ast} - \overline{\mathcal{I}}_t^{\ast}\right)^2},
    \label{eqs49-2} 
    \\
    \mathcal{P}_t^{\ast} &=& \frac{1}{N_{\mathrm{run}}} \sum_{r=1}^{N_{\mathrm{run}}} \mathcal{P}_{r,t}^{\ast},
    \label{eqs49-3} 
    \\
    \mathrm{SD}(\mathcal{P}_{r,t}^{\ast}) &=& \sqrt{\frac{1}{N_{\mathrm{run}}-1} \sum_{r=1}^{N_{\mathrm{run}}} \left(\mathcal{P}_{r,t}^{\ast} - \mathcal{P}_t^{\ast}\right)^2}.
    \label{eqs49-4} 
\end{eqnarray}

\noindent Increasing $\overline{\mathcal{I}}_t^{\ast}$ together with near-zero $\mathcal{P}_t^{\ast}$ indicates that the optimization improves IMI strength within parameter regimes satisfying the target IMI-ratio constraints.
\item We then compute the fraction of sampled parameter sets that satisfy the feasibility conditions of the optimization. For each sampled parameter vector, we define the feasibility indicator as,
\begin{equation}
\chi(\bm{\theta}) =
\begin{cases}
1, & \mathcal{P}(\bm{\theta})\leq \epsilon_{\mathcal{P}},\\
0, & \mathcal{P}(\bm{\theta}) > \epsilon_{\mathcal{P}}.
\end{cases}
\label{eqs50}
\end{equation}

\noindent Here, $\chi(\bm{\theta}) =1$ identifies parameter sets whose IMI-strength ratios lie within the target windows. For run $r$ and iteration $t$, the feasible fraction is
\begin{equation}
F_{r,t}=\frac{1}{N} \sum_{j=1}^{N} \chi \left( \bm{\theta}_{r,t}^{(j)} \right).
\label{eqs51}
\end{equation}

\noindent Across independent runs, the average of feasible fraction and the corresponding standard deviation are defined as, 
\begin{eqnarray}
F_t &=& \frac{1}{N_{\mathrm{run}}} \sum_{r=1}^{N_{\mathrm{run}}}
F_{r,t},
\label{eqs52} \\
\mathrm{SD}(F_{r,t}) &=& \sqrt{\frac{1}{N_{\mathrm{run}}-1} \sum_{r=1}^{N_{\mathrm{run}}} \left(F_{r,t} - F_t\right)^2}.
\label{eqs52-1}
\end{eqnarray}

\noindent We report these quantities for diagnosing the optimization. A rise in $F_t$ indicates that the CEM sampling distribution has moved toward a region of parameter space where many candidates satisfy the target IMI-ratio constraints.
\item We also compute a run-run distance matrix among the final optimized solutions obtained from the independent CEM runs. For each successful run, the final feasible optimized parameter vector is denoted by $\bm{\theta}_{r}^{\ast}=\bm{\theta}_{r,T}^{\ast}$ with $T$ being the final CEM iteration. Each parameter is first normalized in log space as,
\begin{equation}
q_{r,i} = \frac{\log \theta_{r,i}^{\ast}-\log \theta_i^{\min}}{\log \theta_i^{\max}-\log \theta_i^{\min}},
~~
\forall i=1,\ldots,p.
\label{eqs53}
\end{equation}

\noindent Here, $\theta_{r,i}^{\ast}$ is the $i$th component of the final feasible solution from run $r$, while $\theta_i^{\min}$ and $\theta_i^{\max}$ are the lower and upper search bounds for the same parameter, as listed in Table~\ref{t3}. The distance between two final feasible solutions from runs $r$ and $r'$ is then defined as
\begin{equation}
D_{rr'} = \left[ \frac{1}{p}\sum_{i=1}^{p} \left(
q_{r,i} - q_{r',i} \right)^2 \right]^{1/2}.
\label{eqs54}
\end{equation}

\noindent We report this root-mean-square distance $D_{rr'}$ to quantify the run-to-run similarity. Smaller values indicate that independent runs converge to similar parameter regimes, whereas larger values indicate more separated feasible regimes.
\item We next examine the trajectories of each optimized parameter along the iterations. For this purpose, we quantify the mean $\bar{\mu}_{i,t}$ and standard deviation $\mathrm{SD}(\mu_{r,t,i})$ across independent runs for parameter $\theta_i$ at iteration $t$. These quantities are defined as,
\begin{eqnarray}
\bar{\mu}_{i,t} &=& \frac{1}{N_{\mathrm{run}}} \sum_{r=1}^{N_{\mathrm{run}}}\mu_{r,t,i},
\label{eqs55} \\
\mathrm{SD}(\mu_{r,t,i}) &=& \sqrt{\frac{1}{N_{\mathrm{run}}-1} \sum_{r=1}^{N_{\mathrm{run}}} \left(\mu_{r,t,i} - \bar{\mu}_{i,t}\right)^2},
\label{eqs55-1}
\end{eqnarray}

\noindent where $\mu_{r,t,i}$ is the $i$th component of the CEM mean vector $\bm{\mu}_{r,t}$. The corresponding mean of the original parameter is,
\begin{equation}
\Theta_{i,t} = e^{\bar{\mu}_{i,t}}.
\label{eqs56}
\end{equation}

\noindent We report these mean $\Theta_{i,t}$ with corresponding standard deviation $\mathrm{SD}(\mu_{r,t,i})$ across independent runs to track the parameter evolution with CEM iterations.
\item We now compute the correlation matrix for the optimized parameters. For successful run $r$, let $\theta_{r,i}^{\ast}$ be the optimized final value of parameter $\theta_i$. The correlation between parameters $i$ and $j$ is defined as,
\begin{equation}
C_{ij} = \frac{\mathrm{Cov}_{r}\left(\theta_{r,i}^{\ast},\theta_{r,j}^{\ast}\right)}{\mathrm{SD}_{r}\left(\theta_{r,i}^{\ast}\right) \mathrm{SD}_{r} \left(\theta_{r,j}^{\ast}\right)}.
\label{eqs57}
\end{equation}

\noindent Here, the covariance $\mathrm{Cov}_{r}$ and the standard deviation $\mathrm{SD}_{r}$ are similarly computed across the final optimized parameter vectors from successful CEM runs. This correlation matrix identifies parameter dependencies among the feasible solutions selected by the optimizer. Strong positive or negative correlations indicate that two parameters co-vary across feasible solutions in order to preserve the target IMI hierarchy.
\item We compute the coefficient of variation (CV) for each final optimized parameter to quantify how much that parameter varies across successful CEM runs. For parameter $\theta_i$, the CV is defined as,
\begin{equation}
\mathrm{CV}_i = \frac{\mathrm{SD}_{r}\left(\theta_{r,i}^{\ast}\right)}{\langle \theta_{r,i}^{\ast} \rangle_{r}},
\label{eqs58}
\end{equation}

\noindent where, $\theta_{r,i}^{\ast}$ is the final optimized value of parameter $\theta_i$ in run $r$, $\langle\cdot\rangle_r$ denotes averaging across successful CEM runs, and $\mathrm{SD}_{r}(\cdot)$ denotes the corresponding standard deviation across the runs. A low $\mathrm{CV}_i$ indicates that parameter $\theta_i$ is tightly constrained by the IMI optimization. A high $\mathrm{CV}_i$ indicates that $\theta_i$ can vary across successful runs while still preserving the target IMI hierarchy.
\item Finally, we estimate how sensitive the IMI is to small relative perturbations of each parameter. A representative optimized parameter vector is first defined by averaging the final feasible optimized parameter vectors over independent CEM runs,
\begin{equation}
\bm{\theta}^{rep} = \left\langle \bm{\theta}_{r}^{\ast} \right\rangle_{r}.
\label{eqs59}
\end{equation}

\noindent Each parameter is then perturbed one at a time while keeping all other parameters fixed. Specifically, for parameter $\theta_i^{rep}$, we apply the relative perturbation,
\begin{equation}
\theta_i^{rep} \rightarrow \theta_i^{rep}(1+\delta).
\label{eqs60}
\end{equation}

\noindent Because this perturbation is multiplicative, the corresponding change in log-parameter space is $\Delta \log \theta_i = \log(1+\delta)$. For small perturbations, $\log(1+\delta) \simeq \delta$. Thus, the fractional change in IMI divided by $\delta$ measures the response of IMI to a fractional perturbation in parameter $\theta_i$. This provides a local estimate of the logarithmic sensitivity of IMI to $\theta_i$. The perturbed value is clipped to the allowed bounds $[\theta_i^{\min},\theta_i^{\max}]$ if it exceeds the search interval. Let $\bm{\theta}^{rep}_{i,\delta}$ denote the parameter vector obtained after perturbing only the $i$th parameter. For motif $m$, the finite-perturbation logarithmic sensitivity can be defined as,
\begin{eqnarray}
\mathbb{S}_{m,i} = \left\langle \left| \frac{\left[I_{int}^{(m)}\left(
\bm{\theta}^{rep}_{i,\delta}\right) - I_{int}^{(m)} \left(\bm{\theta}^{rep}\right)\right] / I_{int}^{(m)} \left(\bm{\theta}^{rep}\right)}{\delta} \right| \right\rangle_{\delta}. 
\nonumber \\
\label{eqs61}
\end{eqnarray}

\noindent Here, $\langle\cdot\rangle_{\delta}$ denotes averaging over the nonzero perturbation values used in the sensitivity analysis. This finite-difference quantity approximates
\begin{equation}
\mathbb{S}_{m,i} \simeq \left| \frac{\partial \log |I_{int}^{(m)}|}{\partial \log \theta_i} \right|.
\label{eqs62}
\end{equation}

\noindent Thus, $\mathbb{S}_{m,i}$ measures how strongly the IMI of motif $m$ responds to a relative perturbation of parameter $\theta_i$. Low sensitivity indicates local robustness of IMI to parameter perturbations, whereas high sensitivity identifies parameters that strongly control the IMI. We note that this logarithmic sensitivity is used in Figs.~\ref{sf1}-\ref{sf4} and is distinct from the absolute sensitivity used to construct the normalized robustness score in Fig.~3g of the main text.
\end{enumerate}

\noindent \textbf{Analyzing the diagnostic measures.} We first examine the AND-gated coherent FFLs in \textit{E. coli}. Across CEM iterations, the average of best IMI strength, $\overline{\mathcal{I}}_t^{\ast}$, increases while the best penalty, $\mathcal{P}_t^{\ast}$, remains near zero, indicating that the optimization enhances collective IMI strength without violating the target IMI-ratio constraints (Fig.~\ref{sf1}a). The finite fraction of feasible parameter sets, $F_t$, shows that the solutions occupy a nonzero region of parameter space (Fig.~\ref{sf1}b). The structured run-run distance matrix, $D_{rr'}$, indicates repeated convergence across independent runs rather than random scattering (Fig.~\ref{sf1}c). The profiles of mean parameter trajectories are shown in Fig.~\ref{sf1}d. The parameter-correlation matrix and CV analysis indicate compensatory but relatively stable parameter organization across successful runs (Fig.~\ref{sf1}e,f). The logarithmic sensitivity, $\mathbb{S}_{m,i}$, shows that the IMI hierarchy is controlled by a coordinated set of parameters rather than by a single dominant parameter (Fig.~\ref{sf1}g). Together, these diagnostics show repeated recovery of feasible high-IMI regimes across independent runs. \\

For AND-gated incoherent FFLs in \textit{E. coli}, Fig.~\ref{sf2}a,b shows that the optimization also reaches a feasible regime for the target IMI pattern. Compared with the coherent case, the final solutions occupy a broader but still organized region of parameter space, indicating that multiple parameter combinations can support the incoherent IMI hierarchy (Fig.~\ref{sf2}c--f). The sensitivity profile further suggests motif-specific parameter control, consistent with the need to balance opposing direct and indirect pathway contributions in incoherent FFLs (Fig.~\ref{sf2}g). \\

For AND-gated coherent FFLs in \textit{S. cerevisiae}, the optimization reaches feasible regimes, although the optimized average IMI strength is smaller than in the corresponding \textit{E. coli} coherent case (Fig.~\ref{sf3}a,b). The final solutions are more heterogeneous across runs, as reflected by broader run-run distances, more variable parameter trajectories, and larger parameter CVs (Fig.~\ref{sf3}c--f). The sensitivity profile indicates stronger motif-dependent parameter sensitivity, suggesting that the coherent yeast hierarchy requires a more delicate balance of degradation rates, regulatory thresholds, and expression levels (Fig.~\ref{sf3}g). \\

For AND-gated incoherent FFLs in \textit{S. cerevisiae}, the optimization produces the strongest increase in average IMI among the four AND-gated cases while maintaining the feasible regime for the target IMI pattern (Fig.~\ref{sf4}a,b). The run-run distances, parameter trajectories, correlations, and CVs indicate convergence toward a structured but partially flexible parameter regime (Fig.~\ref{sf4}c--f). The sensitivity analysis supports motif-specific control of IMI, consistent with the requirement to balance destructive direct-indirect pathway interference in incoherent FFLs (Fig.~\ref{sf4}g). \\

The AND-gated optimization diagnostics show that the empirical abundance-like pattern (see Fig.~1a in the main text) can be mimicked by the hierarchy of IMI in coherent and incoherent FFLs of both \textit{E. coli} and \textit{S. cerevisiae}. The diagnostic analyses indicate that the feasible regimes are repeatedly recovered from independent initial conditions. We also apply the same optimization framework to OR-gated FFLs, which can produce similar patterns as the AND-gate. Because the full convergence diagnostics are already illustrated for the AND-gated cases, we do not repeat the complete diagnostic panels for the OR-gated cases. Instead, Fig.~\ref{sf5} summarizes the IMI trajectories across CEM iterations for both AND and OR gates in \textit{E. coli} and \textit{S. cerevisiae}. This comparison tests whether the abundance-like IMI hierarchy is specific to AND-gated regulation or whether it is preserved under an alternative output logic. \\

We note that the kinetic parameter bounds used in the optimization (see Table~\ref{t3}) represent organism-specific coarse-grained search windows. The bounds used for \textit{E. coli} describe a bacterial regulatory regime with lower TF copy numbers, lower regulatory thresholds, and faster effective turnover. In contrast, the bounds for \textit{S. cerevisiae} represent a yeast regulatory regime with higher TF copy numbers, broader regulatory thresholds, and slower effective turnover. These bounds are used to restrict the optimization to biologically plausible bacterial and yeast parameter regimes while still allowing sufficient flexibility to identify dynamical regimes consistent with the target IMI-ratio hierarchy. The bounds are chosen in a way that they remain consistent with the reported order-of-magnitude differences in protein abundance, protein lifetime, and promoter regulation between bacteria and yeast \cite{Bintu2005, Belle2006, Garcia2011, Hintsche2013, Christiano2014, Li2014, Hansen2015, Milo2015, Gupta2024}. \\

To test whether the abundance-like hierarchy is specific to IMI or can be obtained from other information measures, we repeat the same CEM-based hierarchy optimization by replacing the IMI strength, $\mathcal{I}_m(\bm{\theta})=|I_{int}^{(m)}(X;Z)|$, with either the total mutual information, $\mathcal{I}_m(\bm{\theta})=I^{(m)}(X;Z)$, or the pathway information, $\mathcal{I}_m(\bm{\theta})=I_{path}^{(m)}(X;Z)$. The penalty definitions, parameter bounds, and CEM settings are kept identical to those used for IMI optimization. Thus, the information measure is changed throughout the optimization. If total MI or pathway information is sufficient to reproduce the empirical abundance-like hierarchy, the best penalty should approach zero in the same way as observed for IMI. However, Fig.~\ref{sf7} shows that the best penalties for $I(X;Z)$ and $I_{path}(X;Z)$ remain finite in several organism, class, and gate combinations, indicating that these quantities do not consistently satisfy the target hierarchy under the same optimization protocol. This control demonstrates that the abundance-like hierarchy is not a generic consequence of increasing total input-output information or path-wise information transmission, but is specifically associated with the pathway-interference contribution captured by $I_{int}(X;Z)$.

\clearpage


\onecolumngrid

\begin{center}

{\Large
SUPPLEMENTARY FIGURES
}

\end{center}

\twocolumngrid


\newpage

\begin{figure*}[!t]
\includegraphics[width=1.8\columnwidth,angle=0]{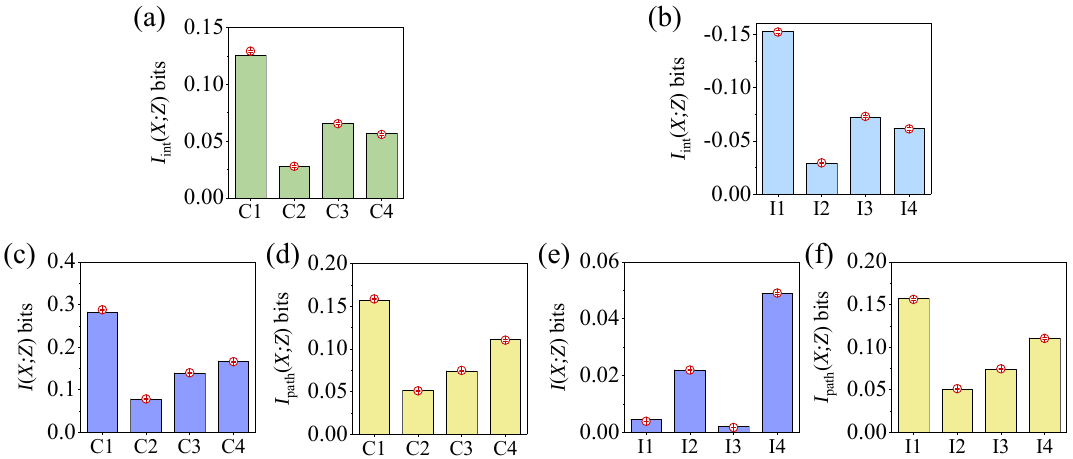}
\caption{
\textbf{Interference mutual information for OR-gate FFLs}
(a,b) IMI magnitudes for coherent and incoherent FFLs with an OR gate, respectively. These patterns closely resemble the abundance pattern of \textit{E. coli} in Fig.~1a.
(c,d) Total MI and pathway MI for coherent FFLs.
(e,f) Corresponding total MI and pathway MI for incoherent FFLs.
The total and pathway MI magnitudes for both FFL types substantially differ from any of the abundance hierarchies. Here, bars denote analytical predictions under the LNA, whereas open symbols with error bars denote numerical values generated from stochastic simulation using the Gillespie algorithm. The parameters used to generate the plots are: $\beta_X=0.1$ min$^{-1}$, $\beta_Y=1.0$ min$^{-1}$, $\beta_Z=10.0$ min$^{-1}$, $K_{XY}=100$ molecules/cell, $K_{XZ}=100$ molecules/cell, $K_{YZ}=100$ molecules/cell, $\langle x\rangle=40$ molecules/cell, $\langle y\rangle=100$ molecules/cell, and $\langle z\rangle=100$ molecules/cell. The other parameters, such as the synthesis rate constants of each TF, can be obtained from the steady-state expressions of $\langle x\rangle$, $\langle y\rangle$, and $\langle z\rangle$.
}
\label{sf01}
\end{figure*}

\newpage

\begin{figure*}[!t]
\includegraphics[width=1.8\columnwidth,angle=0]{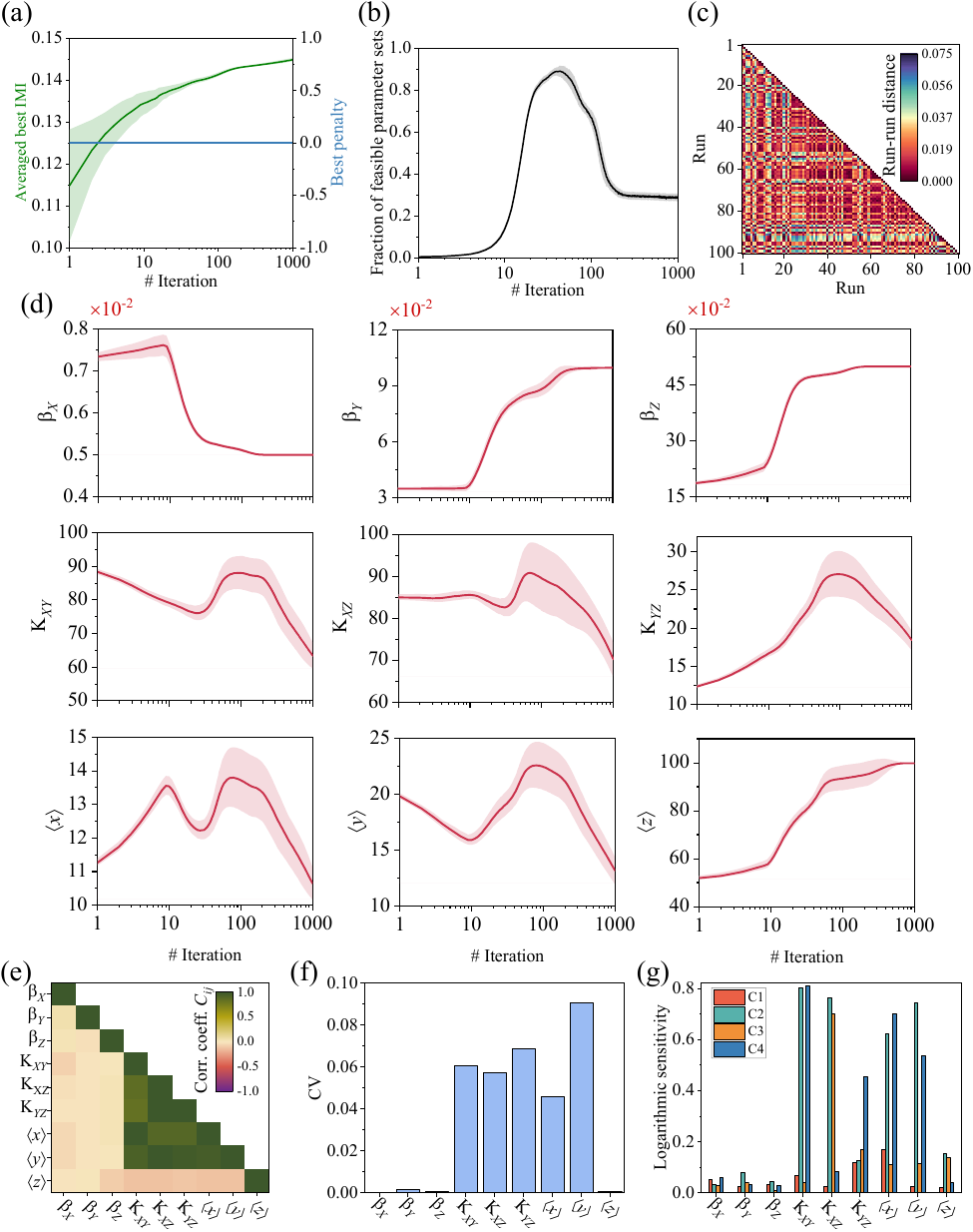}
\caption{
\textbf{CEM optimization diagnostics for AND-gated coherent FFLs in \textit{E. coli}.}
(a) Average of best IMI strengths across four motifs, $\overline{\mathcal{I}}_t^{\ast}$, and best penalty, $\mathcal{P}_t^{\ast}$, across CEM iterations.
(b) Fraction of feasible parameter sets, $F_t$, satisfying the target IMI-ratio constraints.
(c) Run--run distance matrix, $D_{rr'}$, among the final optimized solutions.
(d) Evolution of the optimized parameters across CEM iterations.
(e) Correlation matrix, $C_{ij}$, of the final optimized parameters.
(f) Coefficient of variation, $\mathrm{CV}_i$, of the final optimized parameters.
(g) Finite-perturbation logarithmic sensitivity, $\mathbb{S}_{m,i}$, of IMI to relative parameter perturbations.
Shaded regions indicate run-to-run variability across independent CEM runs where shown.
}
\label{sf1}
\end{figure*}

\newpage

\begin{figure*}[!t]
\includegraphics[width=1.8\columnwidth,angle=0]{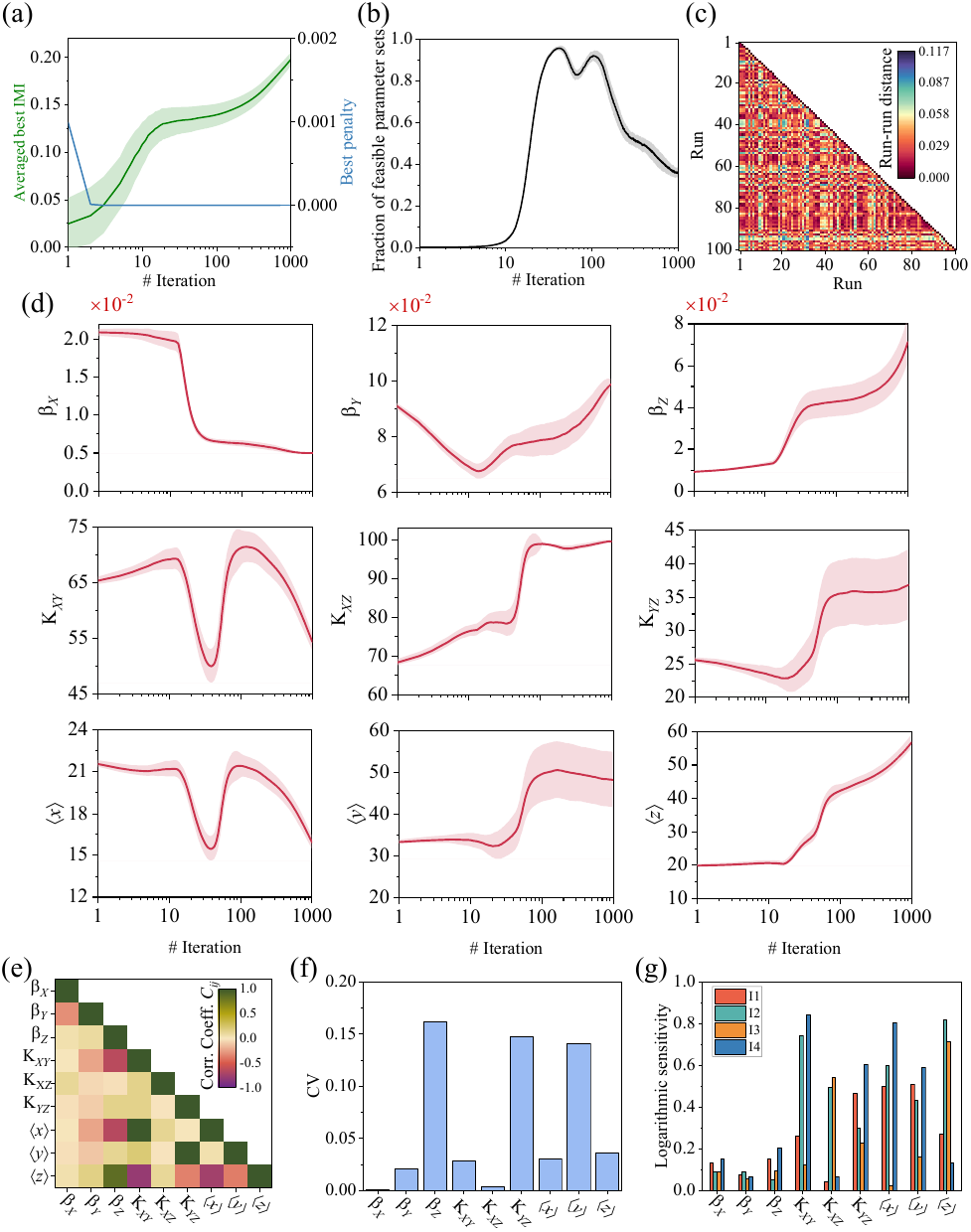}
\caption{
\textbf{CEM optimization diagnostics for AND-gated incoherent FFLs in \textit{E. coli}.}
(a) Average of best IMI strengths across four motifs, $\overline{\mathcal{I}}_t^{\ast}$, and best penalty, $\mathcal{P}_t^{\ast}$, across CEM iterations.
(b) Fraction of feasible parameter sets, $F_t$, satisfying the target IMI-ratio constraints.
(c) Run--run distance matrix, $D_{rr'}$, among the final optimized solutions.
(d) Evolution of the optimized parameters across CEM iterations.
(e) Correlation matrix, $C_{ij}$, of the final optimized parameters.
(f) Coefficient of variation, $\mathrm{CV}_i$, of the final optimized parameters.
(g) Finite-perturbation logarithmic sensitivity, $\mathbb{S}_{m,i}$, of IMI to relative parameter perturbations.
Shaded regions indicate run-to-run variability across independent CEM runs where shown.
}
\label{sf2}
\end{figure*}

\newpage

\begin{figure*}[!t]
\includegraphics[width=1.8\columnwidth,angle=0]{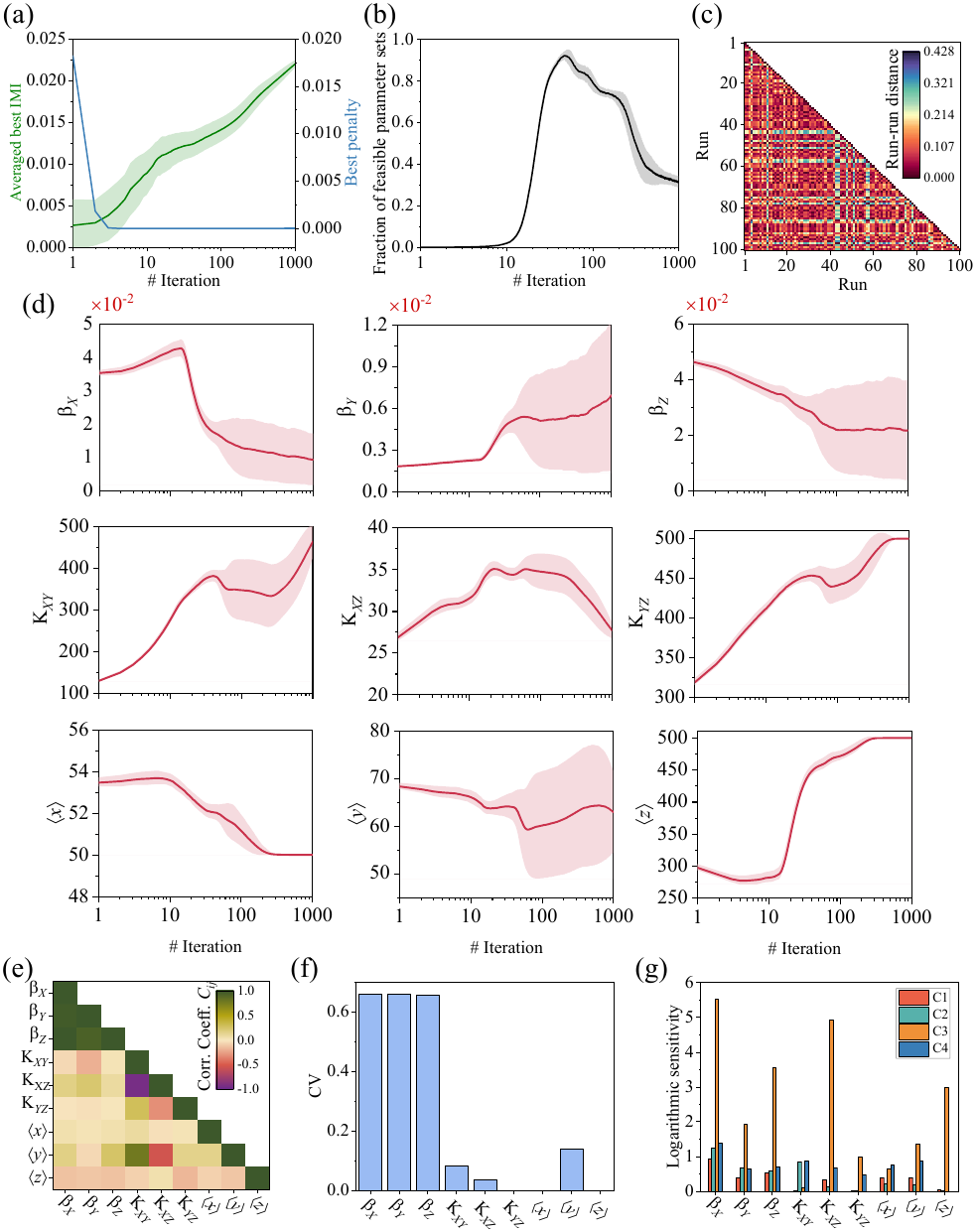}
\caption{
\textbf{CEM optimization diagnostics for AND-gated coherent FFLs in \textit{S. cerevisiae}.}
(a) Average of best IMI strengths across four motifs, $\overline{\mathcal{I}}_t^{\ast}$, and best penalty, $\mathcal{P}_t^{\ast}$, across CEM iterations.
(b) Fraction of feasible parameter sets, $F_t$, satisfying the target IMI-ratio constraints.
(c) Run--run distance matrix, $D_{rr'}$, among the final optimized solutions.
(d) Evolution of the optimized parameters across CEM iterations.
(e) Correlation matrix, $C_{ij}$, of the final optimized parameters.
(f) Coefficient of variation, $\mathrm{CV}_i$, of the final optimized parameters.
(g) Finite-perturbation logarithmic sensitivity, $\mathbb{S}_{m,i}$, of IMI to relative parameter perturbations.
Shaded regions indicate run-to-run variability across independent CEM runs where shown.
}
\label{sf3}
\end{figure*}

\newpage

\begin{figure*}[!t]
\includegraphics[width=1.8\columnwidth,angle=0]{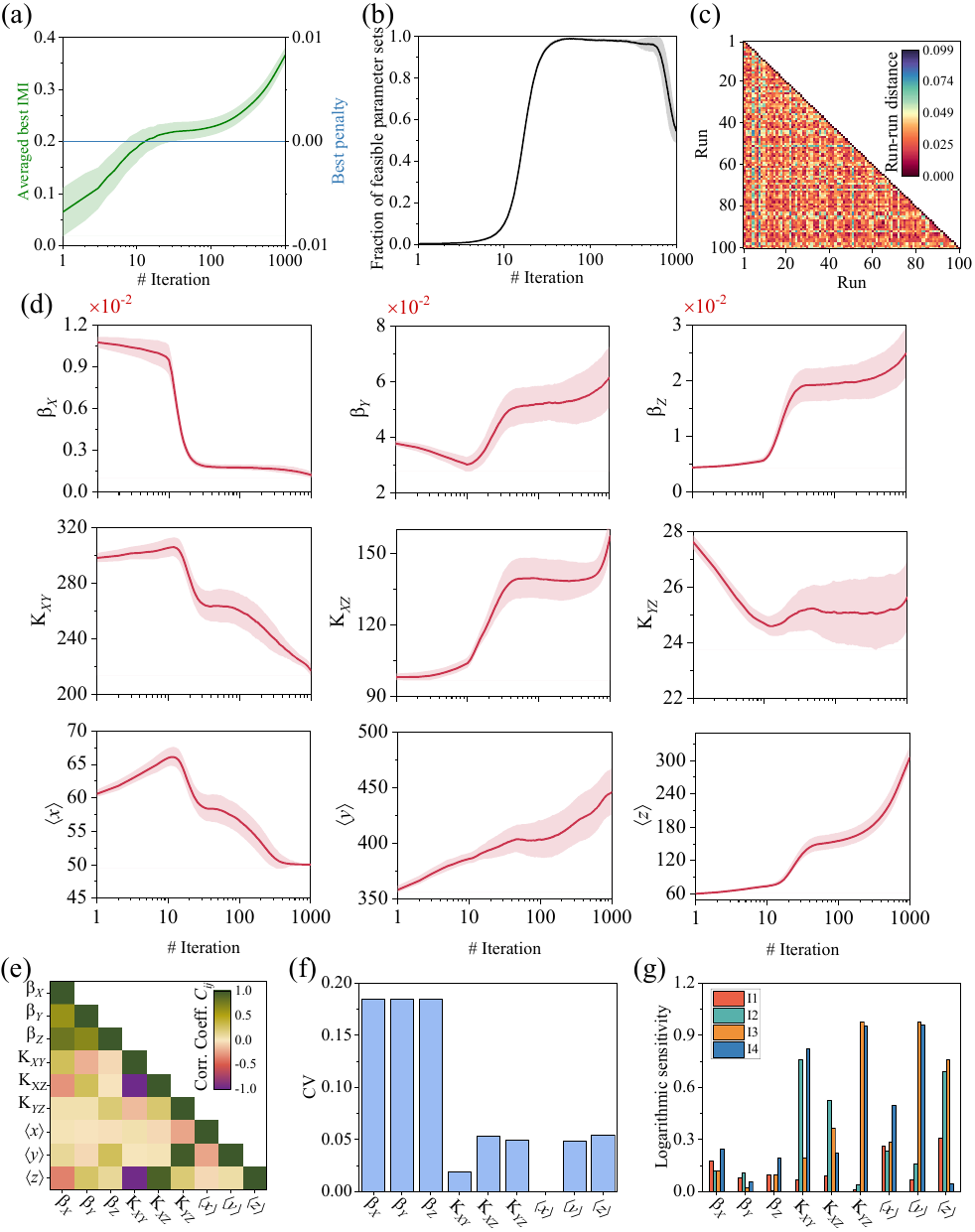}
\caption{
\textbf{CEM optimization diagnostics for AND-gated incoherent FFLs in \textit{S. cerevisiae}.}
(a) Average of best IMI strengths across four motifs, $\overline{\mathcal{I}}_t^{\ast}$, and best penalty, $\mathcal{P}_t^{\ast}$, across CEM iterations.
(b) Fraction of feasible parameter sets, $F_t$, satisfying the target IMI-ratio constraints.
(c) Run--run distance matrix, $D_{rr'}$, among the final optimized solutions.
(d) Evolution of the optimized parameters across CEM iterations.
(e) Correlation matrix, $C_{ij}$, of the final optimized parameters.
(f) Coefficient of variation, $\mathrm{CV}_i$, of the final optimized parameters.
(g) Finite-perturbation logarithmic sensitivity, $\mathbb{S}_{m,i}$, of IMI to relative parameter perturbations.
Shaded regions indicate run-to-run variability across independent CEM runs where shown.
}
\label{sf4}
\end{figure*}

\newpage

\begin{figure*}[!t]
\includegraphics[width=2.0\columnwidth,angle=0]{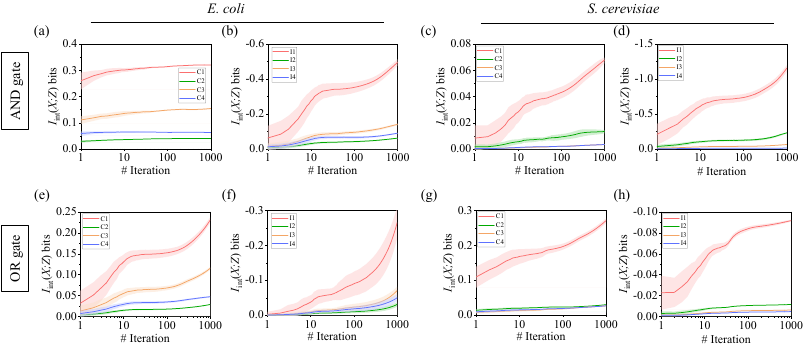}
\caption{
\textbf{Optimized IMI trajectories for AND- and OR-gated FFLs.}
Trajectories of motif-specific IMI, $I_{int}(X;Z)$, across CEM iterations for coherent and incoherent FFLs in \textit{E. coli} and \textit{S. cerevisiae}. Panels (a–d) correspond to AND-gated regulation: (a) coherent FFLs in \textit{E. coli}, (b) incoherent FFLs in \textit{E. coli}, (c) coherent FFLs in \textit{S. cerevisiae}, and (d) incoherent FFLs in  {S. cerevisiae}. Panels (e–h) show the corresponding OR-gated cases in the same order. Shaded regions indicate run-to-run variability across independent CEM runs.
}
\label{sf5}
\end{figure*}

\newpage

\begin{figure*}[!t]
\includegraphics[width=2.0\columnwidth,angle=0]{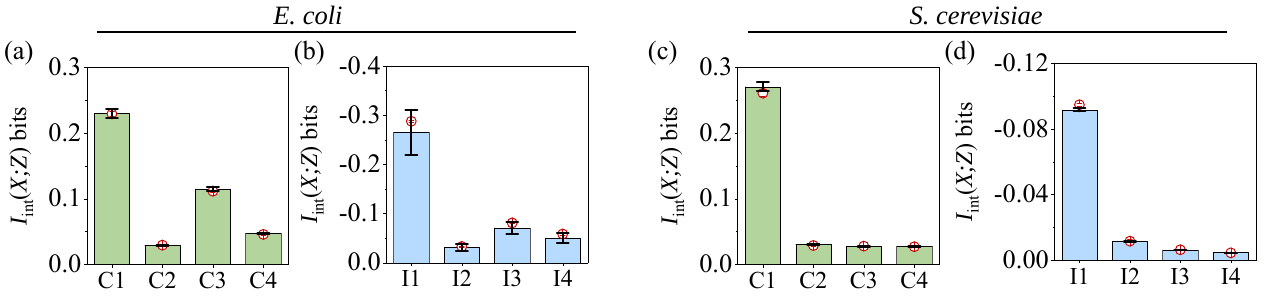}
\caption{\textbf{Optimized IMI values for OR-gate FFLs.}
(a) Coherent FFLs in \textit{E. coli}.
(b) Incoherent FFLs in \textit{E. coli}.
(c) Coherent FFLs in \textit{S. cerevisiae}.
(d) Incoherent FFLs in \textit{S. cerevisiae}.
The IMI pattern resembles the empirical abundance hierarchy shown in Fig.~1a in the main text. Here, bars denote optimized mean IMI values, with error bars indicating standard deviations across independent CEM runs. Open symbols denote mean IMI values obtained from stochastic simulations using the Gillespie algorithm, with error bars indicating standard errors of the mean across simulation replicates.
}
\label{sf6}
\end{figure*}

\newpage

\begin{figure*}[!t]
\includegraphics[width=1.5\columnwidth,angle=0]{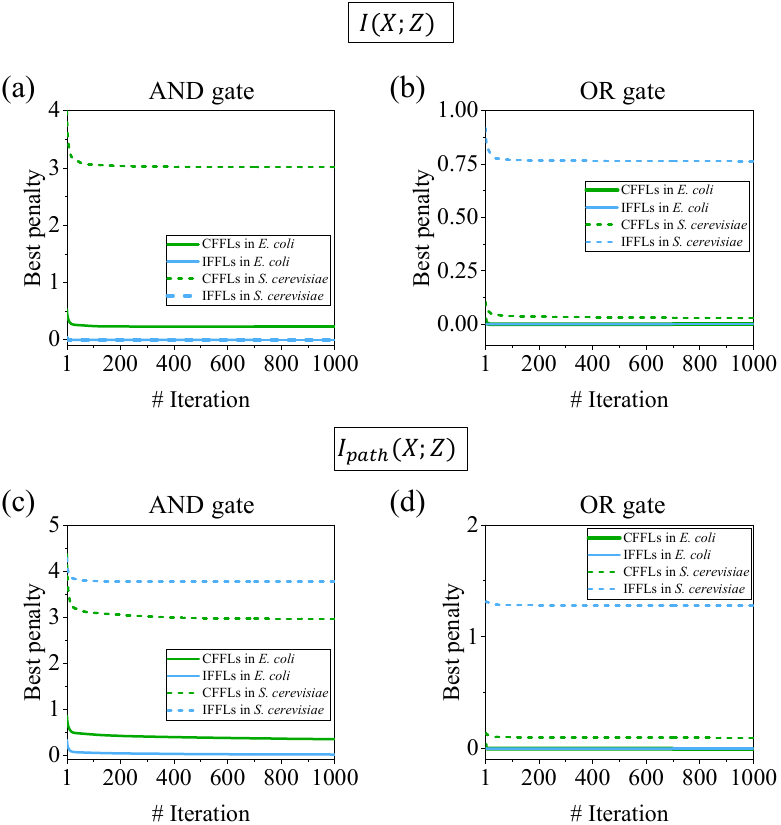}
\caption{\textbf{CEM optimization applied to total and pathway information}.
Convergence of the best penalty $\mathcal{P}_t^\ast$ obtained when the CEM optimization is applied to total mutual information, $I(X;Z)$, and pathway information, $I_{path}(X;Z)$. (a,b) Best penalty versus CEM iteration for $I(X;Z)$ in AND- and OR-gated FFLs. (c,d) Best penalty versus CEM iteration for $I_{path}(X;Z)$ in AND- and OR-gated FFLs. Plots are shown for coherent and incoherent FFL classes in \textit{E. coli} and \textit{S. cerevisiae}. Persistent nonzero penalties indicate that these metrics do not consistently satisfy the abundance-like hierarchy constraints under the same optimization setting.
}
\label{sf7}
\end{figure*}

\begin{figure*}[!t]
\includegraphics[width=1.8\columnwidth,angle=0]{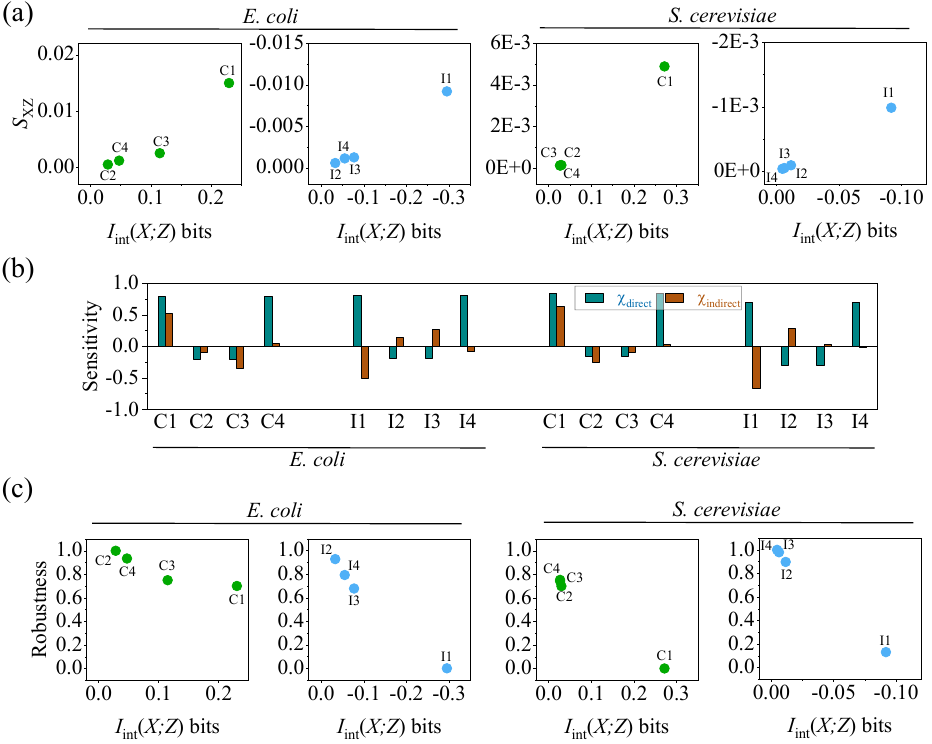}
\caption{\textbf{Biophysical origin and robustness cost of IMI hierarchy in OR-gated FFLs}.
(a) Optimized IMI $I_{int}(X;Z)$ plotted against pathway-interference strength $S_{XZ}$.
(b) Direct- and indirect-path sensitivities, $\chi_{_{direct}}$ and $\chi_{_{indirect}}$ for different FFL types.
(c) Optimized IMI plotted against robustness to parameter perturbations, $\mathcal{R}_m$. Here, we use $\varepsilon=0.01$.
In estimating $S_{XZ}$, $\chi_{_{\cdots}}$, and $\mathcal{R}_m$, we use the optimized parameter values listed in Table~\ref{t5}.
}
\label{sf8}
\end{figure*}

\begin{figure*}[!t]
\includegraphics[width=1.6\columnwidth,angle=0]{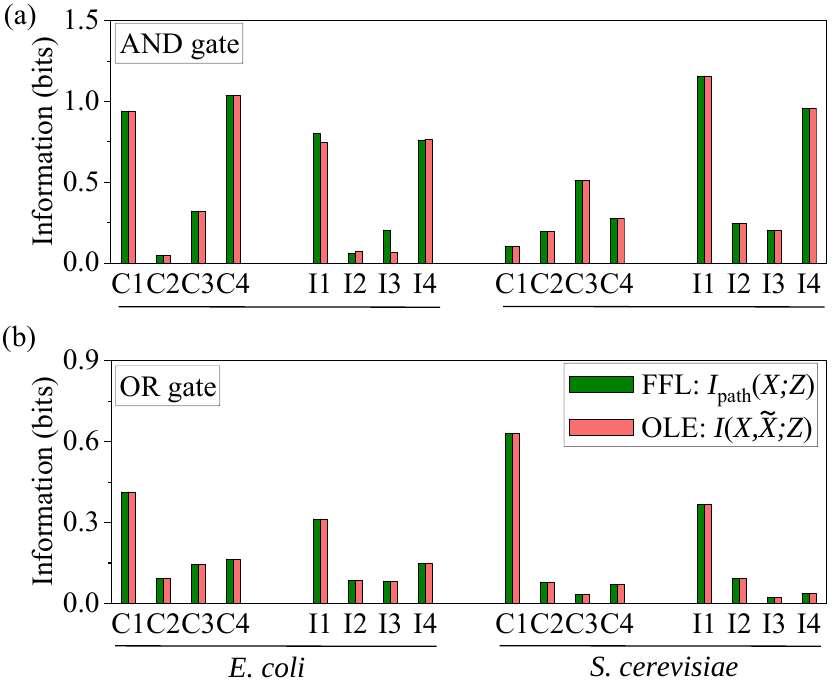}
\caption{\textbf{Comparing FFLs with open-loop equivalents}.
Comparison between the pathway information, $I_{path}(X;Z)$, of the FFL and the total information transmission, $I(X,\tilde{X};Z)$, of the corresponding OLE for (a) AND- and (b) OR-gate FFLs. The open-loop equivalents are constructed from the optimized FFL parameter sets listed in Table~\ref{t5}, without additional optimization. For each OLE, $\beta_{\tilde{X}}=\beta_X$, $\langle \tilde{x}\rangle=\langle x\rangle$, and $K_{\tilde{X}Y}=K_{XY}$, while all remaining kinetic parameters are inherited from the corresponding optimized FFL. Here, bars denote analytical predictions under the LNA.
}
\label{sf9}
\end{figure*}

\clearpage


\onecolumngrid

\begin{center}

{\Large
SUPPLEMENTARY TABLES
}

\end{center}

\twocolumngrid

\begin{table*}[!t]
    \renewcommand{\arraystretch}{1.8}
    \centering
    \caption{\textbf{Production and degradation propensities for all FFL motifs.} Functional forms of the production propensities $f_{\mathcal{J}}$ and degradation propensities $g_{\mathcal{J}}$ for the eight FFL types under AND and OR output logic. The AND and OR mechanisms affect only the output production propensity $f_Z$. Here, $\alpha_{\mathcal{J}}$ and $\beta_{\mathcal{J}}$ denote production and degradation rate constants, respectively. The parameter $K_{\mathcal{U}\mathcal{U}'}$, with $(\mathcal{U},\mathcal{U}')\in\{(X,Y),(X,Z),(Y,Z)\}$, denotes the dissociation constant for binding of the TF $\mathcal{U}$ to the promoter of the TF $\mathcal{U}'$.}
    \begin{tabular}{l|l|l|l|l}
       \hline
       \textbf{Functions}  & \textbf{C1-FFL} & \textbf{C2-FFL} & \textbf{C3-FFL} & \textbf{C4-FFL} \\
       \hline
        $f_X$ & $\alpha_X$ & $\alpha_X$ & $\alpha_X$ & $\alpha_X$ \\
        $f_Y$ & $\alpha_Y \frac{x}{K_{XY}+x}$ & $\alpha_Y \frac{K_{XY}}{K_{XY}+x}$ & $\alpha_Y \frac{x}{K_{XY}+x}$ & $\alpha_Y \frac{K_{XY}}{K_{XY}+x}$ \\
        $f_Z$ (AND) & $\alpha_Z \frac{x}{K_{XZ}+x}\frac{y}{K_{YZ}+y}$ & $\alpha_Z \frac{K_{XZ}}{K_{XZ}+x}\frac{y}{K_{YZ}+y}$ & $\alpha_Z \frac{K_{XZ}}{K_{XZ}+x}\frac{K_{YZ}}{K_{YZ}+y}$ & $\alpha_Z \frac{x}{K_{XZ}+x}\frac{K_{YZ}}{K_{YZ}+y}$ \\
        $f_Z$ (OR) & $\alpha_Z \left[\frac{x}{K_{XZ}+x}+\frac{y}{K_{YZ}+y}\right]$ & $\alpha_Z \left[\frac{K_{XZ}}{K_{XZ}+x}+\frac{y}{K_{YZ}+y}\right]$ & $\alpha_Z \left[\frac{K_{XZ}}{K_{XZ}+x}+\frac{K_{YZ}}{K_{YZ}+y}\right]$ & $\alpha_Z \left[\frac{x}{K_{XZ}+x}+\frac{K_{YZ}}{K_{YZ}+y}\right]$ \\
        \hline
        \hline
        \textbf{Functions} & \textbf{I1-FFL} & \textbf{I2-FFL} & \textbf{I3-FFL} & \textbf{I4-FFL} \\
        \hline
        $f_X$ & $\alpha_X$ & $\alpha_X$ & $\alpha_X$ & $\alpha_X$ \\
        $f_Y$ & $\alpha_Y \frac{x}{K_{XY}+x}$ & $\alpha_Y \frac{K_{XY}}{K_{XY}+x}$ & $\alpha_Y \frac{x}{K_{XY}+x}$ & $\alpha_Y \frac{K_{XY}}{K_{XY}+x}$ \\
        $f_Z$ (AND) & $\alpha_Z \frac{x}{K_{XZ}+x}\frac{K_{YZ}}{K_{YZ}+y}$ & $\alpha_Z \frac{K_{XZ}}{K_{XZ}+x}\frac{K_{YZ}}{K_{YZ}+y}$ & $\alpha_Z \frac{K_{XZ}}{K_{XZ}+x}\frac{y}{K_{YZ}+y}$ & $\alpha_Z \frac{x}{K_{XZ}+x}\frac{y}{K_{YZ}+y}$ \\
        $f_Z$ (OR) & $\alpha_Z \left[\frac{x}{K_{XZ}+x}+\frac{K_{YZ}}{K_{YZ}+y}\right]$ & $\alpha_Z \left[\frac{K_{XZ}}{K_{XZ}+x}+\frac{K_{YZ}}{K_{YZ}+y}\right]$ & $\alpha_Z \left[\frac{K_{XZ}}{K_{XZ}+x}+\frac{y}{K_{YZ}+y}\right]$ & $\alpha_Z \left[\frac{x}{K_{XZ}+x}+\frac{y}{K_{YZ}+y}\right]$ \\
        \hline
    \end{tabular}
    \label{t1}
\end{table*}

\begin{table*}[!t]
\centering
\caption{\textbf{CEM and numerical hyperparameters used in the optimization.} The hyperparameter values are kept fixed for coherent and incoherent FFLs, for AND and OR gates, and for both organisms.}
\begin{tabular}{lcc}
\hline
\textbf{Quantity} & \textbf{Symbol} & \textbf{Value} \\
\hline
Independent CEM runs & $N_{\mathrm{run}}$ & $100$ \\
Feasibility pre-scan samples & $N_{\mathrm{scan}}$ & $1000$ \\
CEM iterations & $T$ & $1000$ \\
Population size per iteration & $N$ & $1000$ \\
Elite fraction & $\rho$ & $0.1$ \\
Smoothing parameter & $s$ & $0.3$ \\
Covariance floor & $\epsilon_C$ & $10^{-6}$ \\
Infeasibility offset & $\Omega$ & $10^6$ \\
Feasibility tolerance & $\epsilon_{\mathcal{P}}$ & $10^{-12}$ \\
Number of perturbations for sensitivity & $N_{\delta}$ & $200$ \\
Relative perturbation range & $\delta$ & $[-0.1,0.1]$ \\
Pre-scan random seed & -- & $42$ \\
Run-specific random seed & -- & $100+10r$ \\
\hline
\end{tabular}
\label{t2}
\end{table*}

\begin{table*}[!t]
\centering
\caption{\textbf{Kinetic parameter bounds used in the optimization.} The \textit{E. coli} bounds represent a lower-copy-number bacterial regime, whereas the \textit{S. cerevisiae} bounds represent a higher-copy-number yeast regime with broader regulatory thresholds. The values of production rate constants $\alpha_\mathcal{J}$ are determined from the mean-field steady-state condition $f_\mathcal{J}(\langle \bm{n} \rangle)=g_\mathcal{J}(\langle \bm{n} \rangle)$ for each TF in the motifs.}
\begin{tabular}{lcccc}
\hline
\textbf{Parameter} & \textbf{\textit{E. coli} bounds} & \textbf{\textit{S. cerevisiae} bounds} & \textbf{Units}\\
\hline
$\beta_X$ & $5\times10^{-3}$ to $10^{-1}$ \cite{Hintsche2013, Gupta2024} & $10^{-3}$ to $10^{-1}$ \cite{Belle2006, Christiano2014, Milo2015} & min$^{-1}$ \\
$\beta_Y$ & $5\times10^{-3}$ to $10^{-1}$ \cite{Hintsche2013, Gupta2024} & $10^{-3}$ to $10^{-1}$ \cite{Belle2006, Christiano2014, Milo2015} & min$^{-1}$ \\
$\beta_Z$ & $5\times10^{-3}$ to $5\times10^{-1}$ \cite{Hintsche2013, Gupta2024} & $10^{-3}$ to $10^{-1}$ \cite{Belle2006, Christiano2014, Milo2015} & min$^{-1}$ \\
$K_{XY}$ & $5$ to $100$ \cite{Bintu2005, Garcia2011} & $20$ to $500$ \cite{Bintu2005, Hansen2015} & molecules/cell \\
$K_{XZ}$ & $5$ to $100$ \cite{Bintu2005, Garcia2011} & $20$ to $500$ \cite{Bintu2005, Hansen2015} & molecules/cell \\
$K_{YZ}$ & $5$ to $100$ \cite{Bintu2005, Garcia2011} & $20$ to $500$ \cite{Bintu2005, Hansen2015} & molecules/cell \\
$\langle x\rangle$ & $10$ to $100$ \cite{Li2014, Milo2015} & $50$ to $500$ \cite{Milo2015} & molecules/cell \\
$\langle y\rangle$ & $10$ to $100$ \cite{Li2014, Milo2015} & $50$ to $500$ \cite{Milo2015} & molecules/cell \\
$\langle z\rangle$ & $10$ to $100$ \cite{Li2014, Milo2015} & $50$ to $500$ \cite{Milo2015} & molecules/cell \\
\hline
\end{tabular}
\label{t3}
\end{table*}

\begin{table*}[!t]
\centering
\caption{\textbf{IMI-ratio target windows.} The ratio constraints are imposed on IMI strengths, $\mathcal{I}_m=|I_{int}^{(m)}(X;Z)|$. The notation C1/C2 denotes the IMI-strength ratio $\mathcal{I}_{\rm C1}/\mathcal{I}_{\rm C2}$, and similarly for other motif pairs. These target windows are broad qualitative ranges guided by the ratios of observed fractional motif abundances shown in Fig. 1a (see main text), rather than exact fitted abundance values.}
\begin{tabular}{ll}
\hline
\textbf{Case} & \textbf{Target windows for IMI-strength ratios} \\
\hline
\textit{E. coli}, coherent FFL, AND 
& C1/C2: 8--10, \quad C1/C3: 2--3, \quad C1/C4: 4--5 \\
\textit{E. coli}, coherent FFL, OR 
& C1/C2: 8--10, \quad C1/C3: 2--3, \quad C1/C4: 4--5 \\
\textit{E. coli}, incoherent FFL, AND 
& I1/I2: 8--10, \quad I1/I3: 3.5--4.5, \quad I1/I4: 4.5--5.5 \\
\textit{E. coli}, incoherent FFL, OR 
& I1/I2: 8--10, \quad I1/I3: 3.5--4.5, \quad I1/I4: 4.5--5.5 \\
\textit{S. cerevisiae}, coherent FFL, AND 
& C1/C2: 3--8, \quad C1/C3: 20--150, \quad C1/C4: 20--150 \\
\textit{S. cerevisiae}, coherent FFL, OR 
& C1/C2: 2--9, \quad C1/C3: 10--150, \quad C1/C4: 10--150 \\
\textit{S. cerevisiae}, incoherent FFL, AND 
& I1/I2: 5--7, \quad I1/I3: 18--22, \quad I1/I4: 50--100 \\
\textit{S. cerevisiae}, incoherent FFL, OR 
& I1/I2: 4--8, \quad I1/I3: 15--25, \quad I1/I4: 20--150 \\
\hline
\end{tabular}
\label{t4}
\end{table*}

\begin{table*}[t]
\centering
\caption{\textbf{Optimized parameter values.} The values are reported as mean $\pm$ standard deviation across runs for coherent and incoherent FFLs in \textit{E. coli} and \textit{S. cerevisiae}. From these optimized parameter values the  production rate constants $\alpha_\mathcal{J}$ can be estimated from the mean-field steady-state condition $f_\mathcal{J}(\langle \bm{n} \rangle)=g_\mathcal{J}(\langle \bm{n} \rangle)$. The optimized parameters remain within biologically plausible order-of-magnitude ranges consistent with previous studies \cite{Bintu2005, Belle2006, Garcia2011, Hintsche2013, Christiano2014, Li2014, Hansen2015, Milo2015, Gupta2024}.}
\resizebox{\textwidth}{!}{%
\begin{tabular}{l|cc|cc}
\hline
\multicolumn{5}{c}{\textbf{AND gate}} \\
\hline
\textbf{Parameter} 
& \multicolumn{2}{c|}{\textbf{\textit{E. coli}}} 
& \multicolumn{2}{c}{\textbf{\textit{S. cerevisiae}}} \\
\cline{2-5}
& \textbf{Coherent FFL} & \textbf{Incoherent FFL} & \textbf{Coherent FFL} & \textbf{Incoherent FFL} \\
\hline
$\beta_X$ 
& $(5.0007 \pm 0.0021)\times 10^{-3}$ 
& $(5.0023 \pm 0.0039)\times 10^{-3}$ 
& $(12.336 \pm 8.13)\times 10^{-3}$ 
& $(1.2494 \pm 0.231)\times 10^{-3}$ \\

$\beta_Y$ 
& $(9.9884 \pm 0.0150)\times 10^{-2}$ 
& $(9.8712 \pm 0.207)\times 10^{-2}$ 
& $(0.9040 \pm 0.595)\times 10^{-2}$ 
& $(6.2408 \pm 1.16)\times 10^{-2}$ \\

$\beta_Z$ 
& $(4.9986 \pm 0.0030)\times 10^{-1}$ 
& $(0.7257 \pm 0.117)\times 10^{-1}$ 
& $(0.2863 \pm 0.188)\times 10^{-1}$ 
& $(0.2536 \pm 0.0468)\times 10^{-1}$ \\

$K_{XY}$ 
& $63.52 \pm 3.86$ 
& $54.32 \pm 1.54$ 
& $467.23 \pm 38.39$ 
& $217.30 \pm 4.14$ \\

$K_{XZ}$ 
& $69.95 \pm 4.03$ 
& $99.55 \pm 0.41$ 
& $27.69 \pm 1.02$ 
& $157.60 \pm 8.37$ \\

$K_{YZ}$ 
& $18.40 \pm 1.27$ 
& $37.29 \pm 5.49$ 
& $499.80 \pm 0.31$ 
& $25.69 \pm 1.28$ \\

$\langle x\rangle$ 
& $10.64 \pm 0.49$ 
& $15.93 \pm 0.48$ 
& $50.00 \pm 0.005$ 
& $50.01 \pm 0.02$ \\

$\langle y\rangle$ 
& $13.17 \pm 1.20$ 
& $48.82 \pm 6.90$ 
& $63.42 \pm 8.79$ 
& $446.66 \pm 21.52$ \\

$\langle z\rangle$ 
& $99.94 \pm 0.09$ 
& $56.96 \pm 2.07$ 
& $499.96 \pm 0.12$ 
& $306.79 \pm 16.74$ \\
\hline
\multicolumn{5}{c}{\textbf{OR gate}} \\
\hline
\textbf{Parameter} 
& \multicolumn{2}{c|}{\textbf{\textit{E. coli}}} 
& \multicolumn{2}{c}{\textbf{\textit{S. cerevisiae}}} \\
\cline{2-5}
& \textbf{Coherent FFL} & \textbf{Incoherent FFL} & \textbf{Coherent FFL} & \textbf{Incoherent FFL} \\
\hline
$\beta_X$ 
& $(13.426 \pm 2.12)\times 10^{-3}$ 
& $(3.4645 \pm 2.24)\times 10^{-2}$ 
& $(1.2230 \pm 0.272)\times 10^{-3}$ 
& $(1.0933 \pm 0.0973)\times 10^{-3}$ \\

$\beta_Y$ 
& $(9.3002 \pm 0.721)\times 10^{-2}$ 
& $(6.2942 \pm 3.64)\times 10^{-2}$ 
& $(7.6122 \pm 1.61)\times 10^{-2}$ 
& $(4.4021 \pm 1.20)\times 10^{-2}$ \\

$\beta_Z$ 
& $(2.7098 \pm 0.412)\times 10^{-1}$ 
& $(2.6985 \pm 2.21)\times 10^{-1}$ 
& $(4.2529 \pm 1.23)\times 10^{-2}$ 
& $(8.8137 \pm 0.968)\times 10^{-2}$ \\

$K_{XY}$ 
& $64.61 \pm 0.68$ 
& $34.95 \pm 3.24$ 
& $134.93 \pm 8.23$ 
& $261.24 \pm 61.48$ \\

$K_{XZ}$ 
& $41.42 \pm 0.80$ 
& $44.20 \pm 3.15$ 
& $294.17 \pm 21.77$ 
& $272.19 \pm 64.10$ \\

$K_{YZ}$ 
& $26.90 \pm 1.90$ 
& $21.86 \pm 17.12$ 
& $401.08 \pm 35.53$ 
& $22.59 \pm 2.49$ \\

$\langle x\rangle$ 
& $10.00 \pm 0.01$ 
& $10.12 \pm 0.33$ 
& $51.13 \pm 2.97$ 
& $112.80 \pm 26.54$ \\

$\langle y\rangle$ 
& $17.25 \pm 1.31$ 
& $39.82 \pm 29.29$ 
& $53.04 \pm 4.87$ 
& $431.18 \pm 45.85$ \\

$\langle z\rangle$ 
& $99.95 \pm 0.07$ 
& $99.83 \pm 0.65$ 
& $357.06 \pm 38.21$ 
& $206.83 \pm 48.70$ \\
\hline
\end{tabular}%
}
\label{t5}
\end{table*}

\clearpage


\begin{thebibliography}{45}%
\makeatletter
\providecommand \@ifxundefined [1]{%
 \@ifx{#1\undefined}
}%
\providecommand \@ifnum [1]{%
 \ifnum #1\expandafter \@firstoftwo
 \else \expandafter \@secondoftwo
 \fi
}%
\providecommand \@ifx [1]{%
 \ifx #1\expandafter \@firstoftwo
 \else \expandafter \@secondoftwo
 \fi
}%
\providecommand \natexlab [1]{#1}%
\providecommand \enquote  [1]{``#1''}%
\providecommand \bibnamefont  [1]{#1}%
\providecommand \bibfnamefont [1]{#1}%
\providecommand \citenamefont [1]{#1}%
\providecommand \href@noop [0]{\@secondoftwo}%
\providecommand \href [0]{\begingroup \@sanitize@url \@href}%
\providecommand \@href[1]{\@@startlink{#1}\@@href}%
\providecommand \@@href[1]{\endgroup#1\@@endlink}%
\providecommand \@sanitize@url [0]{\catcode `\\12\catcode `\$12\catcode `\&12\catcode `\#12\catcode `\^12\catcode `\_12\catcode `\%12\relax}%
\providecommand \@@startlink[1]{}%
\providecommand \@@endlink[0]{}%
\providecommand \url  [0]{\begingroup\@sanitize@url \@url }%
\providecommand \@url [1]{\endgroup\@href {#1}{\urlprefix }}%
\providecommand \urlprefix  [0]{URL }%
\providecommand \Eprint [0]{\href }%
\providecommand \doibase [0]{https://doi.org/}%
\providecommand \selectlanguage [0]{\@gobble}%
\providecommand \bibinfo  [0]{\@secondoftwo}%
\providecommand \bibfield  [0]{\@secondoftwo}%
\providecommand \translation [1]{[#1]}%
\providecommand \BibitemOpen [0]{}%
\providecommand \bibitemStop [0]{}%
\providecommand \bibitemNoStop [0]{.\EOS\space}%
\providecommand \EOS [0]{\spacefactor3000\relax}%
\providecommand \BibitemShut  [1]{\csname bibitem#1\endcsname}%
\let\auto@bib@innerbib\@empty
\bibitem [{\citenamefont {Davidson}(2006)}]{Davidson2006}%
  \BibitemOpen
  \bibfield  {author} {\bibinfo {author} {\bibfnamefont {E.~H.}\ \bibnamefont {Davidson}},\ }\href@noop {} {\emph {\bibinfo {title} {The Regulatory Genome: Gene Regulatory Networks in Development and Evolution}}}\ (\bibinfo  {publisher} {Academic Press, Burlington, MA},\ \bibinfo {year} {2006})\BibitemShut {NoStop}%
\bibitem [{\citenamefont {Alon}(2006)}]{Alon2006}%
  \BibitemOpen
  \bibfield  {author} {\bibinfo {author} {\bibfnamefont {U.}~\bibnamefont {Alon}},\ }\href@noop {} {\emph {\bibinfo {title} {An Introduction to Systems Biology: Design Principles of Biological Circuits}}}\ (\bibinfo  {publisher} {CRC Press, Boca Raton, FL},\ \bibinfo {year} {2006})\BibitemShut {NoStop}%
\bibitem [{\citenamefont {Milo}\ \emph {et~al.}(2002)\citenamefont {Milo}, \citenamefont {Shen-Orr}, \citenamefont {Itzkovitz}, \citenamefont {Kashtan}, \citenamefont {Chklovskii},\ and\ \citenamefont {Alon}}]{Milo2002}%
  \BibitemOpen
  \bibfield  {author} {\bibinfo {author} {\bibfnamefont {R.}~\bibnamefont {Milo}}, \bibinfo {author} {\bibfnamefont {S.}~\bibnamefont {Shen-Orr}}, \bibinfo {author} {\bibfnamefont {S.}~\bibnamefont {Itzkovitz}}, \bibinfo {author} {\bibfnamefont {N.}~\bibnamefont {Kashtan}}, \bibinfo {author} {\bibfnamefont {D.}~\bibnamefont {Chklovskii}},\ and\ \bibinfo {author} {\bibfnamefont {U.}~\bibnamefont {Alon}},\ }\bibfield  {title} {\bibinfo {title} {Network motifs: simple building blocks of complex networks},\ }\href@noop {} {\bibfield  {journal} {\bibinfo  {journal} {Science}\ }\textbf {\bibinfo {volume} {298}},\ \bibinfo {pages} {824} (\bibinfo {year} {2002})}\BibitemShut {NoStop}%
\bibitem [{\citenamefont {Shen-Orr}\ \emph {et~al.}(2002)\citenamefont {Shen-Orr}, \citenamefont {Milo}, \citenamefont {Mangan},\ and\ \citenamefont {Alon}}]{Shen-Orr2002}%
  \BibitemOpen
  \bibfield  {author} {\bibinfo {author} {\bibfnamefont {S.~S.}\ \bibnamefont {Shen-Orr}}, \bibinfo {author} {\bibfnamefont {R.}~\bibnamefont {Milo}}, \bibinfo {author} {\bibfnamefont {S.}~\bibnamefont {Mangan}},\ and\ \bibinfo {author} {\bibfnamefont {U.}~\bibnamefont {Alon}},\ }\bibfield  {title} {\bibinfo {title} {{{N}etwork motifs in the transcriptional regulation network of {E}scherichia coli}},\ }\href@noop {} {\bibfield  {journal} {\bibinfo  {journal} {Nat. Genet.}\ }\textbf {\bibinfo {volume} {31}},\ \bibinfo {pages} {64} (\bibinfo {year} {2002})}\BibitemShut {NoStop}%
\bibitem [{\citenamefont {Kholodenko}\ \emph {et~al.}(2012)\citenamefont {Kholodenko}, \citenamefont {Yaffe},\ and\ \citenamefont {Kolch}}]{Kholodenko2012}%
  \BibitemOpen
  \bibfield  {author} {\bibinfo {author} {\bibfnamefont {B.}~\bibnamefont {Kholodenko}}, \bibinfo {author} {\bibfnamefont {M.~B.}\ \bibnamefont {Yaffe}},\ and\ \bibinfo {author} {\bibfnamefont {W.}~\bibnamefont {Kolch}},\ }\bibfield  {title} {\bibinfo {title} {Computational approaches for analyzing information flow in biological networks},\ }\href@noop {} {\bibfield  {journal} {\bibinfo  {journal} {Science Signaling}\ }\textbf {\bibinfo {volume} {5}},\ \bibinfo {pages} {re1} (\bibinfo {year} {2012})}\BibitemShut {NoStop}%
\bibitem [{\citenamefont {Lim}\ \emph {et~al.}(2013)\citenamefont {Lim}, \citenamefont {Lee},\ and\ \citenamefont {Tang}}]{Lim2013}%
  \BibitemOpen
  \bibfield  {author} {\bibinfo {author} {\bibfnamefont {W.~A.}\ \bibnamefont {Lim}}, \bibinfo {author} {\bibfnamefont {C.~M.}\ \bibnamefont {Lee}},\ and\ \bibinfo {author} {\bibfnamefont {C.}~\bibnamefont {Tang}},\ }\bibfield  {title} {\bibinfo {title} {Design principles of regulatory networks: Searching for the molecular algorithms of the cell},\ }\href@noop {} {\bibfield  {journal} {\bibinfo  {journal} {Mol. Cell}\ }\textbf {\bibinfo {volume} {49}},\ \bibinfo {pages} {202} (\bibinfo {year} {2013})}\BibitemShut {NoStop}%
\bibitem [{\citenamefont {Mangan}\ and\ \citenamefont {Alon}(2003)}]{Mangan2003}%
  \BibitemOpen
  \bibfield  {author} {\bibinfo {author} {\bibfnamefont {S.}~\bibnamefont {Mangan}}\ and\ \bibinfo {author} {\bibfnamefont {U.}~\bibnamefont {Alon}},\ }\bibfield  {title} {\bibinfo {title} {Structure and function of the feed-forward loop network motif},\ }\href@noop {} {\bibfield  {journal} {\bibinfo  {journal} {Proc. Natl. Acad. Sci. U.S.A.}\ }\textbf {\bibinfo {volume} {100}},\ \bibinfo {pages} {11980} (\bibinfo {year} {2003})}\BibitemShut {NoStop}%
\bibitem [{\citenamefont {Ma}\ \emph {et~al.}(2004)\citenamefont {Ma}, \citenamefont {Kumar}, \citenamefont {Ditges}, \citenamefont {Gunzer}, \citenamefont {Buer},\ and\ \citenamefont {Zeng}}]{Ma2004}%
  \BibitemOpen
  \bibfield  {author} {\bibinfo {author} {\bibfnamefont {H.~W.}\ \bibnamefont {Ma}}, \bibinfo {author} {\bibfnamefont {B.}~\bibnamefont {Kumar}}, \bibinfo {author} {\bibfnamefont {U.}~\bibnamefont {Ditges}}, \bibinfo {author} {\bibfnamefont {F.}~\bibnamefont {Gunzer}}, \bibinfo {author} {\bibfnamefont {J.}~\bibnamefont {Buer}},\ and\ \bibinfo {author} {\bibfnamefont {A.~P.}\ \bibnamefont {Zeng}},\ }\bibfield  {title} {\bibinfo {title} {{{A}n extended transcriptional regulatory network of {E}scherichia coli and analysis of its hierarchical structure and network motifs}},\ }\href@noop {} {\bibfield  {journal} {\bibinfo  {journal} {Nucleic Acids Res.}\ }\textbf {\bibinfo {volume} {32}},\ \bibinfo {pages} {6643} (\bibinfo {year} {2004})}\BibitemShut {NoStop}%
\bibitem [{\citenamefont {Mangan}\ \emph {et~al.}(2006)\citenamefont {Mangan}, \citenamefont {Itzkovitz}, \citenamefont {Zaslaver},\ and\ \citenamefont {Alon}}]{Mangan2006}%
  \BibitemOpen
  \bibfield  {author} {\bibinfo {author} {\bibfnamefont {S.}~\bibnamefont {Mangan}}, \bibinfo {author} {\bibfnamefont {S.}~\bibnamefont {Itzkovitz}}, \bibinfo {author} {\bibfnamefont {A.}~\bibnamefont {Zaslaver}},\ and\ \bibinfo {author} {\bibfnamefont {U.}~\bibnamefont {Alon}},\ }\bibfield  {title} {\bibinfo {title} {{{T}he incoherent feed-forward loop accelerates the response-time of the gal system of {E}scherichia coli}},\ }\href@noop {} {\bibfield  {journal} {\bibinfo  {journal} {J. Mol. Biol.}\ }\textbf {\bibinfo {volume} {356}},\ \bibinfo {pages} {1073} (\bibinfo {year} {2006})}\BibitemShut {NoStop}%
\bibitem [{\citenamefont {Mangan}\ \emph {et~al.}(2003)\citenamefont {Mangan}, \citenamefont {Zaslaver},\ and\ \citenamefont {Alon}}]{Mangan2003a}%
  \BibitemOpen
  \bibfield  {author} {\bibinfo {author} {\bibfnamefont {S.}~\bibnamefont {Mangan}}, \bibinfo {author} {\bibfnamefont {A.}~\bibnamefont {Zaslaver}},\ and\ \bibinfo {author} {\bibfnamefont {U.}~\bibnamefont {Alon}},\ }\bibfield  {title} {\bibinfo {title} {The coherent feedforward loop serves as a sign-sensitive delay element in transcription networks},\ }\href@noop {} {\bibfield  {journal} {\bibinfo  {journal} {J. Mol. Biol.}\ }\textbf {\bibinfo {volume} {334}},\ \bibinfo {pages} {197} (\bibinfo {year} {2003})}\BibitemShut {NoStop}%
\bibitem [{\citenamefont {Kalir}\ \emph {et~al.}(2005)\citenamefont {Kalir}, \citenamefont {Mangan},\ and\ \citenamefont {Alon}}]{Kalir2005}%
  \BibitemOpen
  \bibfield  {author} {\bibinfo {author} {\bibfnamefont {S.}~\bibnamefont {Kalir}}, \bibinfo {author} {\bibfnamefont {S.}~\bibnamefont {Mangan}},\ and\ \bibinfo {author} {\bibfnamefont {U.}~\bibnamefont {Alon}},\ }\bibfield  {title} {\bibinfo {title} {A coherent feed-forward loop with a {SUM} input function prolongs flagella expression in {E}scherichia coli},\ }\href@noop {} {\bibfield  {journal} {\bibinfo  {journal} {Mol. Syst. Biol.}\ }\textbf {\bibinfo {volume} {1}},\ \bibinfo {pages} {2005.0006} (\bibinfo {year} {2005})}\BibitemShut {NoStop}%
\bibitem [{\citenamefont {Murugan}(2012)}]{Murugan2012}%
  \BibitemOpen
  \bibfield  {author} {\bibinfo {author} {\bibfnamefont {R.}~\bibnamefont {Murugan}},\ }\bibfield  {title} {\bibinfo {title} {Theory on the dynamics of feedforward loops in the transcription factor networks},\ }\href@noop {} {\bibfield  {journal} {\bibinfo  {journal} {PLoS ONE}\ }\textbf {\bibinfo {volume} {7}},\ \bibinfo {pages} {1} (\bibinfo {year} {2012})}\BibitemShut {NoStop}%
\bibitem [{\citenamefont {Ingram}\ \emph {et~al.}(2006)\citenamefont {Ingram}, \citenamefont {Stumpf},\ and\ \citenamefont {Stark}}]{Ingram2006}%
  \BibitemOpen
  \bibfield  {author} {\bibinfo {author} {\bibfnamefont {P.~J.}\ \bibnamefont {Ingram}}, \bibinfo {author} {\bibfnamefont {M.~P.~H.}\ \bibnamefont {Stumpf}},\ and\ \bibinfo {author} {\bibfnamefont {J.}~\bibnamefont {Stark}},\ }\bibfield  {title} {\bibinfo {title} {Network motifs: structure does not determine function},\ }\href@noop {} {\bibfield  {journal} {\bibinfo  {journal} {BMC Genomics}\ }\textbf {\bibinfo {volume} {7}},\ \bibinfo {pages} {108} (\bibinfo {year} {2006})}\BibitemShut {NoStop}%
\bibitem [{\citenamefont {Widder}\ \emph {et~al.}(2012)\citenamefont {Widder}, \citenamefont {Sol{'e}},\ and\ \citenamefont {Mac{'i}a}}]{Widder2012}%
  \BibitemOpen
  \bibfield  {author} {\bibinfo {author} {\bibfnamefont {S.}~\bibnamefont {Widder}}, \bibinfo {author} {\bibfnamefont {R.}~\bibnamefont {Sol{'e}}},\ and\ \bibinfo {author} {\bibfnamefont {J.}~\bibnamefont {Mac{'i}a}},\ }\bibfield  {title} {\bibinfo {title} {Evolvability of feed-forward loop architecture biases its abundance in transcription networks},\ }\href@noop {} {\bibfield  {journal} {\bibinfo  {journal} {BMC Syst. Biol.}\ }\textbf {\bibinfo {volume} {6}},\ \bibinfo {pages} {7} (\bibinfo {year} {2012})}\BibitemShut {NoStop}%
\bibitem [{\citenamefont {Shannon}(1948)}]{Shannon1948}%
  \BibitemOpen
  \bibfield  {author} {\bibinfo {author} {\bibfnamefont {C.~E.}\ \bibnamefont {Shannon}},\ }\bibfield  {title} {\bibinfo {title} {{{T}he mathematical theory of communication}},\ }\href@noop {} {\bibfield  {journal} {\bibinfo  {journal} {Bell. Syst. Tech. J.}\ }\textbf {\bibinfo {volume} {27}},\ \bibinfo {pages} {379} (\bibinfo {year} {1948})}\BibitemShut {NoStop}%
\bibitem [{\citenamefont {Shannon}\ and\ \citenamefont {Weaver}(1963)}]{Shannon1963}%
  \BibitemOpen
  \bibfield  {author} {\bibinfo {author} {\bibfnamefont {C.~E.}\ \bibnamefont {Shannon}}\ and\ \bibinfo {author} {\bibfnamefont {W.}~\bibnamefont {Weaver}},\ }\href@noop {} {\emph {\bibinfo {title} {The Mathematical Theory of Communication}}}\ (\bibinfo  {publisher} {University of Illinois Press, Urbana},\ \bibinfo {year} {1963})\BibitemShut {NoStop}%
\bibitem [{\citenamefont {Cheong}\ \emph {et~al.}(2011)\citenamefont {Cheong}, \citenamefont {Rhee}, \citenamefont {Wang}, \citenamefont {Nemenman},\ and\ \citenamefont {Levchenko}}]{Cheong2011}%
  \BibitemOpen
  \bibfield  {author} {\bibinfo {author} {\bibfnamefont {R.}~\bibnamefont {Cheong}}, \bibinfo {author} {\bibfnamefont {A.}~\bibnamefont {Rhee}}, \bibinfo {author} {\bibfnamefont {C.~J.}\ \bibnamefont {Wang}}, \bibinfo {author} {\bibfnamefont {I.}~\bibnamefont {Nemenman}},\ and\ \bibinfo {author} {\bibfnamefont {A.}~\bibnamefont {Levchenko}},\ }\bibfield  {title} {\bibinfo {title} {{{I}nformation transduction capacity of noisy biochemical signaling networks}},\ }\href@noop {} {\bibfield  {journal} {\bibinfo  {journal} {Science}\ }\textbf {\bibinfo {volume} {334}},\ \bibinfo {pages} {354} (\bibinfo {year} {2011})}\BibitemShut {NoStop}%
\bibitem [{\citenamefont {Tka{\v c}ik}\ and\ \citenamefont {Walczak}(2011)}]{Tkacik2011}%
  \BibitemOpen
  \bibfield  {author} {\bibinfo {author} {\bibfnamefont {G.}~\bibnamefont {Tka{\v c}ik}}\ and\ \bibinfo {author} {\bibfnamefont {A.~M.}\ \bibnamefont {Walczak}},\ }\bibfield  {title} {\bibinfo {title} {{I}nformation transmission in genetic regulatory networks: a review},\ }\href@noop {} {\bibfield  {journal} {\bibinfo  {journal} {J. Phys. Condens. Matter}\ }\textbf {\bibinfo {volume} {23}},\ \bibinfo {pages} {153102} (\bibinfo {year} {2011})}\BibitemShut {NoStop}%
\bibitem [{\citenamefont {Tka{\v c}ik}\ and\ \citenamefont {Bialek}(2016)}]{Tkacik2016}%
  \BibitemOpen
  \bibfield  {author} {\bibinfo {author} {\bibfnamefont {G.}~\bibnamefont {Tka{\v c}ik}}\ and\ \bibinfo {author} {\bibfnamefont {W.}~\bibnamefont {Bialek}},\ }\bibfield  {title} {\bibinfo {title} {Information processing in living systems},\ }\href@noop {} {\bibfield  {journal} {\bibinfo  {journal} {Annu. Rev. Condens. Matter Phys.}\ }\textbf {\bibinfo {volume} {7}},\ \bibinfo {pages} {89} (\bibinfo {year} {2016})}\BibitemShut {NoStop}%
\bibitem [{\citenamefont {Tka{\v c}ik}\ and\ \citenamefont {Wolde}(2025)}]{Tkacik2025}%
  \BibitemOpen
  \bibfield  {author} {\bibinfo {author} {\bibfnamefont {G.}~\bibnamefont {Tka{\v c}ik}}\ and\ \bibinfo {author} {\bibfnamefont {P.~R.~t.}\ \bibnamefont {Wolde}},\ }\bibfield  {title} {\bibinfo {title} {Information processing in biochemical networks},\ }\href@noop {} {\bibfield  {journal} {\bibinfo  {journal} {Annu. Rev. Biophys.}\ }\textbf {\bibinfo {volume} {54}},\ \bibinfo {pages} {249} (\bibinfo {year} {2025})}\BibitemShut {NoStop}%
\bibitem [{\citenamefont {Nandi}\ \emph {et~al.}(2026)\citenamefont {Nandi}, \citenamefont {Chattopadhyay},\ and\ \citenamefont {Banik}}]{Nandi2026}%
  \BibitemOpen
  \bibfield  {author} {\bibinfo {author} {\bibfnamefont {M.}~\bibnamefont {Nandi}}, \bibinfo {author} {\bibfnamefont {S.}~\bibnamefont {Chattopadhyay}},\ and\ \bibinfo {author} {\bibfnamefont {S.~K.}\ \bibnamefont {Banik}},\ }\bibfield  {title} {\bibinfo {title} {Identifying the sources of noise synergy and redundancy in the gene expression of feed-forward loop motif},\ }\href@noop {} {\bibfield  {journal} {\bibinfo  {journal} {Phys. Biol.}\ }\textbf {\bibinfo {volume} {23}},\ \bibinfo {pages} {016003} (\bibinfo {year} {2026})}\BibitemShut {NoStop}%
\bibitem [{\citenamefont {Lipshtat}\ \emph {et~al.}(2008)\citenamefont {Lipshtat}, \citenamefont {Purushothaman}, \citenamefont {Iyengar},\ and\ \citenamefont {Ma’ayan}}]{Lipshtat2008}%
  \BibitemOpen
  \bibfield  {author} {\bibinfo {author} {\bibfnamefont {A.}~\bibnamefont {Lipshtat}}, \bibinfo {author} {\bibfnamefont {S.~P.}\ \bibnamefont {Purushothaman}}, \bibinfo {author} {\bibfnamefont {R.}~\bibnamefont {Iyengar}},\ and\ \bibinfo {author} {\bibfnamefont {A.}~\bibnamefont {Ma’ayan}},\ }\bibfield  {title} {\bibinfo {title} {Functions of bifans in context of multiple regulatory motifs in signaling networks},\ }\href@noop {} {\bibfield  {journal} {\bibinfo  {journal} {Biophys. J.}\ }\textbf {\bibinfo {volume} {94}},\ \bibinfo {pages} {2566} (\bibinfo {year} {2008})}\BibitemShut {NoStop}%
\bibitem [{\citenamefont {Bleris}\ \emph {et~al.}(2011)\citenamefont {Bleris}, \citenamefont {Xie}, \citenamefont {Glass}, \citenamefont {Adadey}, \citenamefont {Sontag},\ and\ \citenamefont {Benenson}}]{Bleris2011}%
  \BibitemOpen
  \bibfield  {author} {\bibinfo {author} {\bibfnamefont {L.}~\bibnamefont {Bleris}}, \bibinfo {author} {\bibfnamefont {Z.}~\bibnamefont {Xie}}, \bibinfo {author} {\bibfnamefont {D.}~\bibnamefont {Glass}}, \bibinfo {author} {\bibfnamefont {A.}~\bibnamefont {Adadey}}, \bibinfo {author} {\bibfnamefont {E.}~\bibnamefont {Sontag}},\ and\ \bibinfo {author} {\bibfnamefont {Y.}~\bibnamefont {Benenson}},\ }\bibfield  {title} {\bibinfo {title} {Synthetic incoherent feedforward circuits show adaptation to the amount of their genetic template},\ }\href@noop {} {\bibfield  {journal} {\bibinfo  {journal} {Mol. Syst. Biol.}\ }\textbf {\bibinfo {volume} {7}},\ \bibinfo {pages} {519} (\bibinfo {year} {2011})}\BibitemShut {NoStop}%
\bibitem [{\citenamefont {Burda}\ \emph {et~al.}(2011)\citenamefont {Burda}, \citenamefont {Krzywicki}, \citenamefont {Martin},\ and\ \citenamefont {Zagorski}}]{Burda2011}%
  \BibitemOpen
  \bibfield  {author} {\bibinfo {author} {\bibfnamefont {Z.}~\bibnamefont {Burda}}, \bibinfo {author} {\bibfnamefont {A.}~\bibnamefont {Krzywicki}}, \bibinfo {author} {\bibfnamefont {O.~C.}\ \bibnamefont {Martin}},\ and\ \bibinfo {author} {\bibfnamefont {M.}~\bibnamefont {Zagorski}},\ }\bibfield  {title} {\bibinfo {title} {Motifs emerge from function in model gene regulatory networks},\ }\href@noop {} {\bibfield  {journal} {\bibinfo  {journal} {Proc. Natl. Acad. Sci.}\ }\textbf {\bibinfo {volume} {108}},\ \bibinfo {pages} {17263} (\bibinfo {year} {2011})}\BibitemShut {NoStop}%
\bibitem [{\citenamefont {Gutierrez}\ \emph {et~al.}(2021)\citenamefont {Gutierrez}, \citenamefont {Rieke},\ and\ \citenamefont {Shea-Brown}}]{Gutierrez2021}%
  \BibitemOpen
  \bibfield  {author} {\bibinfo {author} {\bibfnamefont {G.~J.}\ \bibnamefont {Gutierrez}}, \bibinfo {author} {\bibfnamefont {F.}~\bibnamefont {Rieke}},\ and\ \bibinfo {author} {\bibfnamefont {E.~T.}\ \bibnamefont {Shea-Brown}},\ }\bibfield  {title} {\bibinfo {title} {Nonlinear convergence boosts information coding in circuits with parallel outputs},\ }\href@noop {} {\bibfield  {journal} {\bibinfo  {journal} {Proc. Natl. Acad. Sci.}\ }\textbf {\bibinfo {volume} {118}},\ \bibinfo {pages} {e1921882118} (\bibinfo {year} {2021})}\BibitemShut {NoStop}%
\bibitem [{\citenamefont {Nandi}\ \emph {et~al.}(2024)\citenamefont {Nandi}, \citenamefont {Chattopadhyay}, \citenamefont {Bandyopadhyay},\ and\ \citenamefont {Banik}}]{Nandi2024}%
  \BibitemOpen
  \bibfield  {author} {\bibinfo {author} {\bibfnamefont {M.}~\bibnamefont {Nandi}}, \bibinfo {author} {\bibfnamefont {S.}~\bibnamefont {Chattopadhyay}}, \bibinfo {author} {\bibfnamefont {S.}~\bibnamefont {Bandyopadhyay}},\ and\ \bibinfo {author} {\bibfnamefont {S.~K.}\ \bibnamefont {Banik}},\ }\bibfield  {title} {\bibinfo {title} {Channel assisted noise propagation in a two-step cascade},\ }\href@noop {} {\bibfield  {journal} {\bibinfo  {journal} {Chaos}\ }\textbf {\bibinfo {volume} {34}},\ \bibinfo {pages} {083128} (\bibinfo {year} {2024})}\BibitemShut {NoStop}%
\bibitem [{\citenamefont {Gillespie}(1976)}]{Gillespie1976}%
  \BibitemOpen
  \bibfield  {author} {\bibinfo {author} {\bibfnamefont {D.~T.}\ \bibnamefont {Gillespie}},\ }\bibfield  {title} {\bibinfo {title} {{A} general method for numerically simulating the stochastic time evolution of coupled chemical reactions},\ }\href@noop {} {\bibfield  {journal} {\bibinfo  {journal} {J. Comp. Phys.}\ }\textbf {\bibinfo {volume} {22}},\ \bibinfo {pages} {403} (\bibinfo {year} {1976})}\BibitemShut {NoStop}%
\bibitem [{\citenamefont {Gillespie}(1977)}]{Gillespie1977}%
  \BibitemOpen
  \bibfield  {author} {\bibinfo {author} {\bibfnamefont {D.~T.}\ \bibnamefont {Gillespie}},\ }\bibfield  {title} {\bibinfo {title} {{E}xact stochastic simulation of coupled chemical reactions},\ }\href@noop {} {\bibfield  {journal} {\bibinfo  {journal} {J. Phys. Chem.}\ }\textbf {\bibinfo {volume} {81}},\ \bibinfo {pages} {2340} (\bibinfo {year} {1977})}\BibitemShut {NoStop}%
\bibitem [{\citenamefont {Nandi}(2024)}]{Nandi2024_a}%
  \BibitemOpen
  \bibfield  {author} {\bibinfo {author} {\bibfnamefont {M.}~\bibnamefont {Nandi}},\ }\bibfield  {title} {\bibinfo {title} {Emergence of temporal noise hierarchy in co-regulated genes of multi-output feed-forward loop},\ }\href@noop {} {\bibfield  {journal} {\bibinfo  {journal} {Phys. Biol.}\ }\textbf {\bibinfo {volume} {22}},\ \bibinfo {pages} {016006} (\bibinfo {year} {2024})}\BibitemShut {NoStop}%
\bibitem [{\citenamefont {Maity}\ \emph {et~al.}(2015)\citenamefont {Maity}, \citenamefont {Chaudhury},\ and\ \citenamefont {Banik}}]{Maity2015}%
  \BibitemOpen
  \bibfield  {author} {\bibinfo {author} {\bibfnamefont {A.~K.}\ \bibnamefont {Maity}}, \bibinfo {author} {\bibfnamefont {P.}~\bibnamefont {Chaudhury}},\ and\ \bibinfo {author} {\bibfnamefont {S.~K.}\ \bibnamefont {Banik}},\ }\bibfield  {title} {\bibinfo {title} {{{R}ole of relaxation time scale in noisy signal transduction}},\ }\href@noop {} {\bibfield  {journal} {\bibinfo  {journal} {PLoS ONE}\ }\textbf {\bibinfo {volume} {10}},\ \bibinfo {pages} {e0123242} (\bibinfo {year} {2015})}\BibitemShut {NoStop}%
\bibitem [{\citenamefont {Bintu}\ \emph {et~al.}(2005)\citenamefont {Bintu}, \citenamefont {Buchler}, \citenamefont {Garcia}, \citenamefont {Gerland}, \citenamefont {Hwa}, \citenamefont {Kondev},\ and\ \citenamefont {Phillips}}]{Bintu2005}%
  \BibitemOpen
  \bibfield  {author} {\bibinfo {author} {\bibfnamefont {L.}~\bibnamefont {Bintu}}, \bibinfo {author} {\bibfnamefont {N.~E.}\ \bibnamefont {Buchler}}, \bibinfo {author} {\bibfnamefont {H.~G.}\ \bibnamefont {Garcia}}, \bibinfo {author} {\bibfnamefont {U.}~\bibnamefont {Gerland}}, \bibinfo {author} {\bibfnamefont {T.}~\bibnamefont {Hwa}}, \bibinfo {author} {\bibfnamefont {J.}~\bibnamefont {Kondev}},\ and\ \bibinfo {author} {\bibfnamefont {R.}~\bibnamefont {Phillips}},\ }\bibfield  {title} {\bibinfo {title} {{{T}ranscriptional regulation by the numbers: models}},\ }\href@noop {} {\bibfield  {journal} {\bibinfo  {journal} {Curr. Opin. Genet. Dev.}\ }\textbf {\bibinfo {volume} {15}},\ \bibinfo {pages} {116} (\bibinfo {year} {2005})}\BibitemShut {NoStop}%
\bibitem [{\citenamefont {Belle}\ \emph {et~al.}(2006)\citenamefont {Belle}, \citenamefont {Tanay}, \citenamefont {Bitincka}, \citenamefont {Shamir},\ and\ \citenamefont {{O'Shea}}}]{Belle2006}%
  \BibitemOpen
  \bibfield  {author} {\bibinfo {author} {\bibfnamefont {A.}~\bibnamefont {Belle}}, \bibinfo {author} {\bibfnamefont {A.}~\bibnamefont {Tanay}}, \bibinfo {author} {\bibfnamefont {L.}~\bibnamefont {Bitincka}}, \bibinfo {author} {\bibfnamefont {R.}~\bibnamefont {Shamir}},\ and\ \bibinfo {author} {\bibfnamefont {E.~K.}\ \bibnamefont {{O'Shea}}},\ }\bibfield  {title} {\bibinfo {title} {Quantification of protein half-lives in the budding yeast proteome},\ }\href@noop {} {\bibfield  {journal} {\bibinfo  {journal} {Proc. Natl. Acad. Sci.}\ }\textbf {\bibinfo {volume} {103}},\ \bibinfo {pages} {13004} (\bibinfo {year} {2006})}\BibitemShut {NoStop}%
\bibitem [{\citenamefont {Garcia}\ and\ \citenamefont {Phillips}(2011)}]{Garcia2011}%
  \BibitemOpen
  \bibfield  {author} {\bibinfo {author} {\bibfnamefont {H.~G.}\ \bibnamefont {Garcia}}\ and\ \bibinfo {author} {\bibfnamefont {R.}~\bibnamefont {Phillips}},\ }\bibfield  {title} {\bibinfo {title} {Quantitative dissection of the simple repression input--output function},\ }\href@noop {} {\bibfield  {journal} {\bibinfo  {journal} {Proc. Natl. Acad. Sci.}\ }\textbf {\bibinfo {volume} {108}},\ \bibinfo {pages} {12173} (\bibinfo {year} {2011})}\BibitemShut {NoStop}%
\bibitem [{\citenamefont {Hintsche}\ and\ \citenamefont {klumpp}(2013)}]{Hintsche2013}%
  \BibitemOpen
  \bibfield  {author} {\bibinfo {author} {\bibfnamefont {M.}~\bibnamefont {Hintsche}}\ and\ \bibinfo {author} {\bibfnamefont {S.}~\bibnamefont {klumpp}},\ }\bibfield  {title} {\bibinfo {title} {Dilution and the theoretical description of growth-rate dependent gene expression},\ }\href@noop {} {\bibfield  {journal} {\bibinfo  {journal} {J. Biol. Eng.}\ }\textbf {\bibinfo {volume} {7}},\ \bibinfo {pages} {22} (\bibinfo {year} {2013})}\BibitemShut {NoStop}%
\bibitem [{\citenamefont {Christiano}\ \emph {et~al.}(2014)\citenamefont {Christiano}, \citenamefont {Nagaraj}, \citenamefont {Fr{\"o}hlich},\ and\ \citenamefont {Walther}}]{Christiano2014}%
  \BibitemOpen
  \bibfield  {author} {\bibinfo {author} {\bibfnamefont {R.}~\bibnamefont {Christiano}}, \bibinfo {author} {\bibfnamefont {N.}~\bibnamefont {Nagaraj}}, \bibinfo {author} {\bibfnamefont {F.}~\bibnamefont {Fr{\"o}hlich}},\ and\ \bibinfo {author} {\bibfnamefont {T.~C.}\ \bibnamefont {Walther}},\ }\bibfield  {title} {\bibinfo {title} {Global proteome turnover analyses of the yeasts {S}. cerevisiae and {S}. pombe},\ }\href@noop {} {\bibfield  {journal} {\bibinfo  {journal} {Cell Rep.}\ }\textbf {\bibinfo {volume} {9}},\ \bibinfo {pages} {1959} (\bibinfo {year} {2014})}\BibitemShut {NoStop}%
\bibitem [{\citenamefont {Li}\ \emph {et~al.}(2014)\citenamefont {Li}, \citenamefont {Burkhardt}, \citenamefont {Gross},\ and\ \citenamefont {Weissman}}]{Li2014}%
  \BibitemOpen
  \bibfield  {author} {\bibinfo {author} {\bibfnamefont {G.-W.}\ \bibnamefont {Li}}, \bibinfo {author} {\bibfnamefont {D.}~\bibnamefont {Burkhardt}}, \bibinfo {author} {\bibfnamefont {C.}~\bibnamefont {Gross}},\ and\ \bibinfo {author} {\bibfnamefont {J.~S.}\ \bibnamefont {Weissman}},\ }\bibfield  {title} {\bibinfo {title} {Quantifying absolute protein synthesis rates reveals principles underlying allocation of cellular resources},\ }\href@noop {} {\bibfield  {journal} {\bibinfo  {journal} {Cell}\ }\textbf {\bibinfo {volume} {157}},\ \bibinfo {pages} {624} (\bibinfo {year} {2014})}\BibitemShut {NoStop}%
\bibitem [{\citenamefont {Hansen}\ and\ \citenamefont {{O'Shea}}(2015)}]{Hansen2015}%
  \BibitemOpen
  \bibfield  {author} {\bibinfo {author} {\bibfnamefont {A.~S.}\ \bibnamefont {Hansen}}\ and\ \bibinfo {author} {\bibfnamefont {E.~K.}\ \bibnamefont {{O'Shea}}},\ }\bibfield  {title} {\bibinfo {title} {cis {D}eterminants of promoter threshold and activation timescale},\ }\href@noop {} {\bibfield  {journal} {\bibinfo  {journal} {Cell Rep.}\ }\textbf {\bibinfo {volume} {12}},\ \bibinfo {pages} {1226} (\bibinfo {year} {2015})}\BibitemShut {NoStop}%
\bibitem [{\citenamefont {Milo}\ and\ \citenamefont {Phillips}(2015)}]{Milo2015}%
  \BibitemOpen
  \bibfield  {author} {\bibinfo {author} {\bibfnamefont {R.}~\bibnamefont {Milo}}\ and\ \bibinfo {author} {\bibfnamefont {R.}~\bibnamefont {Phillips}},\ }\href@noop {} {\emph {\bibinfo {title} {Cell Biology by the Numbers}}}\ (\bibinfo  {publisher} {Garland Science},\ \bibinfo {year} {2015})\BibitemShut {NoStop}%
\bibitem [{\citenamefont {Gupta}\ \emph {et~al.}(2024)\citenamefont {Gupta}, \citenamefont {Johnson}, \citenamefont {Cruz}, \citenamefont {Costa}, \citenamefont {Guest}, \citenamefont {Li}, \citenamefont {Hart}, \citenamefont {Nguyen}, \citenamefont {Stadlmeier}, \citenamefont {Bratton}, \citenamefont {Silhavy}, \citenamefont {Wingreen}, \citenamefont {Gitai},\ and\ \citenamefont {W{\"u}hr}}]{Gupta2024}%
  \BibitemOpen
  \bibfield  {author} {\bibinfo {author} {\bibfnamefont {M.}~\bibnamefont {Gupta}}, \bibinfo {author} {\bibfnamefont {A.~N.~T.}\ \bibnamefont {Johnson}}, \bibinfo {author} {\bibfnamefont {E.~R.}\ \bibnamefont {Cruz}}, \bibinfo {author} {\bibfnamefont {E.~J.}\ \bibnamefont {Costa}}, \bibinfo {author} {\bibfnamefont {R.~L.}\ \bibnamefont {Guest}}, \bibinfo {author} {\bibfnamefont {S.~H.-J.}\ \bibnamefont {Li}}, \bibinfo {author} {\bibfnamefont {E.~M.}\ \bibnamefont {Hart}}, \bibinfo {author} {\bibfnamefont {T.}~\bibnamefont {Nguyen}}, \bibinfo {author} {\bibfnamefont {M.}~\bibnamefont {Stadlmeier}}, \bibinfo {author} {\bibfnamefont {B.~P.}\ \bibnamefont {Bratton}}, \bibinfo {author} {\bibfnamefont {T.~J.}\ \bibnamefont {Silhavy}}, \bibinfo {author} {\bibfnamefont {N.~S.}\ \bibnamefont {Wingreen}}, \bibinfo {author} {\bibfnamefont {Z.}~\bibnamefont {Gitai}},\ and\ \bibinfo {author} {\bibfnamefont {M.}~\bibnamefont {W{\"u}hr}},\ }\bibfield  {title} {\bibinfo {title} {Global protein turnover quantification in
  {Escherichia} coli reveals cytoplasmic recycling under nitrogen limitation},\ }\href@noop {} {\bibfield  {journal} {\bibinfo  {journal} {Nat. Commun.}\ }\textbf {\bibinfo {volume} {15}},\ \bibinfo {pages} {5890} (\bibinfo {year} {2024})}\BibitemShut {NoStop}%
\bibitem [{\citenamefont {Kemmeren}\ \emph {et~al.}(2014)\citenamefont {Kemmeren}, \citenamefont {Sameith}, \citenamefont {van~de Pasch}, \citenamefont {Benschop}, \citenamefont {Lenstra}, \citenamefont {Margaritis}, \citenamefont {O'Duibhir}, \citenamefont {Apweiler}, \citenamefont {van Wageningen}, \citenamefont {Ko}, \citenamefont {van Heesch}, \citenamefont {Kashani}, \citenamefont {Ampatziadis-Michailidis}, \citenamefont {Brok}, \citenamefont {Brabers}, \citenamefont {Miles}, \citenamefont {Bouwmeester}, \citenamefont {van Hooff}, \citenamefont {van Bakel}, \citenamefont {Sluiters}, \citenamefont {Bakker}, \citenamefont {Snel}, \citenamefont {Lijnzaad}, \citenamefont {van Leenen}, \citenamefont {Groot~Koerkamp},\ and\ \citenamefont {Holstege}}]{Kemmeren2014}%
  \BibitemOpen
  \bibfield  {author} {\bibinfo {author} {\bibfnamefont {P.}~\bibnamefont {Kemmeren}}, \bibinfo {author} {\bibfnamefont {K.}~\bibnamefont {Sameith}}, \bibinfo {author} {\bibfnamefont {L.~A.~L.}\ \bibnamefont {van~de Pasch}}, \bibinfo {author} {\bibfnamefont {J.~J.}\ \bibnamefont {Benschop}}, \bibinfo {author} {\bibfnamefont {T.~L.}\ \bibnamefont {Lenstra}}, \bibinfo {author} {\bibfnamefont {T.}~\bibnamefont {Margaritis}}, \bibinfo {author} {\bibfnamefont {E.}~\bibnamefont {O'Duibhir}}, \bibinfo {author} {\bibfnamefont {E.}~\bibnamefont {Apweiler}}, \bibinfo {author} {\bibfnamefont {S.}~\bibnamefont {van Wageningen}}, \bibinfo {author} {\bibfnamefont {C.~W.}\ \bibnamefont {Ko}}, \bibinfo {author} {\bibfnamefont {S.}~\bibnamefont {van Heesch}}, \bibinfo {author} {\bibfnamefont {M.~M.}\ \bibnamefont {Kashani}}, \bibinfo {author} {\bibfnamefont {G.}~\bibnamefont {Ampatziadis-Michailidis}}, \bibinfo {author} {\bibfnamefont {M.~O.}\ \bibnamefont {Brok}}, \bibinfo {author} {\bibfnamefont {N.~A. C.~H.}\ \bibnamefont
  {Brabers}}, \bibinfo {author} {\bibfnamefont {A.~J.}\ \bibnamefont {Miles}}, \bibinfo {author} {\bibfnamefont {D.}~\bibnamefont {Bouwmeester}}, \bibinfo {author} {\bibfnamefont {S.~R.}\ \bibnamefont {van Hooff}}, \bibinfo {author} {\bibfnamefont {H.}~\bibnamefont {van Bakel}}, \bibinfo {author} {\bibfnamefont {E.}~\bibnamefont {Sluiters}}, \bibinfo {author} {\bibfnamefont {L.~V.}\ \bibnamefont {Bakker}}, \bibinfo {author} {\bibfnamefont {B.}~\bibnamefont {Snel}}, \bibinfo {author} {\bibfnamefont {P.}~\bibnamefont {Lijnzaad}}, \bibinfo {author} {\bibfnamefont {D.}~\bibnamefont {van Leenen}}, \bibinfo {author} {\bibfnamefont {M.~J.~A.}\ \bibnamefont {Groot~Koerkamp}},\ and\ \bibinfo {author} {\bibfnamefont {F.~C.~P.}\ \bibnamefont {Holstege}},\ }\bibfield  {title} {\bibinfo {title} {{{L}arge-scale genetic perturbations reveal regulatory networks and an abundance of gene-specific repressors}},\ }\href@noop {} {\bibfield  {journal} {\bibinfo  {journal} {Cell}\ }\textbf {\bibinfo {volume} {157}},\ \bibinfo
  {pages} {740} (\bibinfo {year} {2014})}\BibitemShut {NoStop}%
\bibitem [{\citenamefont {van Kampen}(2007)}]{Kampen2007}%
  \BibitemOpen
  \bibfield  {author} {\bibinfo {author} {\bibfnamefont {N.~G.}\ \bibnamefont {van Kampen}},\ }\href@noop {} {\emph {\bibinfo {title} {Stochastic Processes in Physics and Chemistry, 3rd ed.}}}\ (\bibinfo  {publisher} {North-Holland, Amsterdam},\ \bibinfo {year} {2007})\BibitemShut {NoStop}%
\bibitem [{\citenamefont {Gardiner}(2009)}]{Gardiner2009}%
  \BibitemOpen
  \bibfield  {author} {\bibinfo {author} {\bibfnamefont {C.~W.}\ \bibnamefont {Gardiner}},\ }\href@noop {} {\emph {\bibinfo {title} {Stochastic Methods: A Handbook for the Natural and Social Sciences, 4th ed.}}}\ (\bibinfo  {publisher} {Springer, Berlin},\ \bibinfo {year} {2009})\BibitemShut {NoStop}%
\bibitem [{\citenamefont {Rubinstein}(1997)}]{Rubinstein1997}%
  \BibitemOpen
  \bibfield  {author} {\bibinfo {author} {\bibfnamefont {R.~Y.}\ \bibnamefont {Rubinstein}},\ }\bibfield  {title} {\bibinfo {title} {Optimization of computer simulation models with rare events},\ }\href@noop {} {\bibfield  {journal} {\bibinfo  {journal} {Eur. J. Oper. Res.}\ }\textbf {\bibinfo {volume} {99}},\ \bibinfo {pages} {89} (\bibinfo {year} {1997})}\BibitemShut {NoStop}%
\bibitem [{\citenamefont {Rubinstein}(1999)}]{Rubinstein1999}%
  \BibitemOpen
  \bibfield  {author} {\bibinfo {author} {\bibfnamefont {R.}~\bibnamefont {Rubinstein}},\ }\bibfield  {title} {\bibinfo {title} {{The Cross-Entropy Method for Combinatorial and Continuous Optimization}},\ }\href@noop {} {\bibfield  {journal} {\bibinfo  {journal} {Methodol. Comput. Appl. Probab.}\ }\textbf {\bibinfo {volume} {1}},\ \bibinfo {pages} {127} (\bibinfo {year} {1999})}\BibitemShut {NoStop}%
\bibitem [{\citenamefont {de~Boer}\ \emph {et~al.}(2005)\citenamefont {de~Boer}, \citenamefont {Kroese}, \citenamefont {Mannor},\ and\ \citenamefont {Rubinstein}}]{deBoer2005}%
  \BibitemOpen
  \bibfield  {author} {\bibinfo {author} {\bibfnamefont {P.-T.}\ \bibnamefont {de~Boer}}, \bibinfo {author} {\bibfnamefont {D.~P.}\ \bibnamefont {Kroese}}, \bibinfo {author} {\bibfnamefont {S.}~\bibnamefont {Mannor}},\ and\ \bibinfo {author} {\bibfnamefont {R.~Y.}\ \bibnamefont {Rubinstein}},\ }\bibfield  {title} {\bibinfo {title} {A tutorial on the cross-entropy method},\ }\href@noop {} {\bibfield  {journal} {\bibinfo  {journal} {Ann. Oper. Res.}\ }\textbf {\bibinfo {volume} {134}},\ \bibinfo {pages} {19} (\bibinfo {year} {2005})}\BibitemShut {NoStop}%
\end{thebibliography}

\begin{thebibliography}{27}%
\makeatletter
\providecommand \@ifxundefined [1]{%
 \@ifx{#1\undefined}
}%
\providecommand \@ifnum [1]{%
 \ifnum #1\expandafter \@firstoftwo
 \else \expandafter \@secondoftwo
 \fi
}%
\providecommand \@ifx [1]{%
 \ifx #1\expandafter \@firstoftwo
 \else \expandafter \@secondoftwo
 \fi
}%
\providecommand \natexlab [1]{#1}%
\providecommand \enquote  [1]{``#1''}%
\providecommand \bibnamefont  [1]{#1}%
\providecommand \bibfnamefont [1]{#1}%
\providecommand \citenamefont [1]{#1}%
\providecommand \href@noop [0]{\@secondoftwo}%
\providecommand \href [0]{\begingroup \@sanitize@url \@href}%
\providecommand \@href[1]{\@@startlink{#1}\@@href}%
\providecommand \@@href[1]{\endgroup#1\@@endlink}%
\providecommand \@sanitize@url [0]{\catcode `\\12\catcode `\$12\catcode `\&12\catcode `\#12\catcode `\^12\catcode `\_12\catcode `\%12\relax}%
\providecommand \@@startlink[1]{}%
\providecommand \@@endlink[0]{}%
\providecommand \url  [0]{\begingroup\@sanitize@url \@url }%
\providecommand \@url [1]{\endgroup\@href {#1}{\urlprefix }}%
\providecommand \urlprefix  [0]{URL }%
\providecommand \Eprint [0]{\href }%
\providecommand \doibase [0]{https://doi.org/}%
\providecommand \selectlanguage [0]{\@gobble}%
\providecommand \bibinfo  [0]{\@secondoftwo}%
\providecommand \bibfield  [0]{\@secondoftwo}%
\providecommand \translation [1]{[#1]}%
\providecommand \BibitemOpen [0]{}%
\providecommand \bibitemStop [0]{}%
\providecommand \bibitemNoStop [0]{.\EOS\space}%
\providecommand \EOS [0]{\spacefactor3000\relax}%
\providecommand \BibitemShut  [1]{\csname bibitem#1\endcsname}%
\let\auto@bib@innerbib\@empty
\bibitem [{\citenamefont {Elf}\ and\ \citenamefont {Ehrenberg}(2003)}]{Elf2003}%
  \BibitemOpen
  \bibfield  {author} {\bibinfo {author} {\bibfnamefont {J.}~\bibnamefont {Elf}}\ and\ \bibinfo {author} {\bibfnamefont {M.}~\bibnamefont {Ehrenberg}},\ }\bibfield  {title} {\bibinfo {title} {{{F}ast evaluation of fluctuations in biochemical networks with the linear noise approximation}},\ }\href@noop {} {\bibfield  {journal} {\bibinfo  {journal} {Genome Res.}\ }\textbf {\bibinfo {volume} {13}},\ \bibinfo {pages} {2475} (\bibinfo {year} {2003})}\BibitemShut {NoStop}%
\bibitem [{\citenamefont {Hayot}\ and\ \citenamefont {Jayaprakash}(2004)}]{Hayot2004}%
  \BibitemOpen
  \bibfield  {author} {\bibinfo {author} {\bibfnamefont {F.}~\bibnamefont {Hayot}}\ and\ \bibinfo {author} {\bibfnamefont {C.}~\bibnamefont {Jayaprakash}},\ }\bibfield  {title} {\bibinfo {title} {{{T}he linear noise approximation for molecular fluctuations within cells}},\ }\href@noop {} {\bibfield  {journal} {\bibinfo  {journal} {Phys. Biol.}\ }\textbf {\bibinfo {volume} {1}},\ \bibinfo {pages} {205} (\bibinfo {year} {2004})}\BibitemShut {NoStop}%
\bibitem [{\citenamefont {van Kampen}(2007)}]{Kampen2007}%
  \BibitemOpen
  \bibfield  {author} {\bibinfo {author} {\bibfnamefont {N.~G.}\ \bibnamefont {van Kampen}},\ }\href@noop {} {\emph {\bibinfo {title} {Stochastic Processes in Physics and Chemistry, 3rd ed.}}}\ (\bibinfo  {publisher} {North-Holland, Amsterdam},\ \bibinfo {year} {2007})\BibitemShut {NoStop}%
\bibitem [{\citenamefont {Gardiner}(2009)}]{Gardiner2009}%
  \BibitemOpen
  \bibfield  {author} {\bibinfo {author} {\bibfnamefont {C.~W.}\ \bibnamefont {Gardiner}},\ }\href@noop {} {\emph {\bibinfo {title} {Stochastic Methods: A Handbook for the Natural and Social Sciences, 4th ed.}}}\ (\bibinfo  {publisher} {Springer, Berlin},\ \bibinfo {year} {2009})\BibitemShut {NoStop}%
\bibitem [{\citenamefont {Nandi}\ \emph {et~al.}(2024)\citenamefont {Nandi}, \citenamefont {Chattopadhyay}, \citenamefont {Bandyopadhyay},\ and\ \citenamefont {Banik}}]{Nandi2024}%
  \BibitemOpen
  \bibfield  {author} {\bibinfo {author} {\bibfnamefont {M.}~\bibnamefont {Nandi}}, \bibinfo {author} {\bibfnamefont {S.}~\bibnamefont {Chattopadhyay}}, \bibinfo {author} {\bibfnamefont {S.}~\bibnamefont {Bandyopadhyay}},\ and\ \bibinfo {author} {\bibfnamefont {S.~K.}\ \bibnamefont {Banik}},\ }\bibfield  {title} {\bibinfo {title} {Channel assisted noise propagation in a two-step cascade},\ }\href@noop {} {\bibfield  {journal} {\bibinfo  {journal} {Chaos}\ }\textbf {\bibinfo {volume} {34}},\ \bibinfo {pages} {083128} (\bibinfo {year} {2024})}\BibitemShut {NoStop}%
\bibitem [{\citenamefont {Swain}(2004)}]{Swain2004}%
  \BibitemOpen
  \bibfield  {author} {\bibinfo {author} {\bibfnamefont {P.~S.}\ \bibnamefont {Swain}},\ }\bibfield  {title} {\bibinfo {title} {{{E}fficient attenuation of stochasticity in gene expression through post-transcriptional control}},\ }\href@noop {} {\bibfield  {journal} {\bibinfo  {journal} {J. Mol. Biol.}\ }\textbf {\bibinfo {volume} {344}},\ \bibinfo {pages} {965} (\bibinfo {year} {2004})}\BibitemShut {NoStop}%
\bibitem [{\citenamefont {Paulsson}(2004)}]{Paulsson2004}%
  \BibitemOpen
  \bibfield  {author} {\bibinfo {author} {\bibfnamefont {J.}~\bibnamefont {Paulsson}},\ }\bibfield  {title} {\bibinfo {title} {Summing up the noise in gene networks},\ }\href@noop {} {\bibfield  {journal} {\bibinfo  {journal} {Nature}\ }\textbf {\bibinfo {volume} {427}},\ \bibinfo {pages} {415} (\bibinfo {year} {2004})}\BibitemShut {NoStop}%
\bibitem [{\citenamefont {T{\u a}nase-Nicola}\ \emph {et~al.}(2006)\citenamefont {T{\u a}nase-Nicola}, \citenamefont {Warren},\ and\ \citenamefont {ten Wolde}}]{Tanase2006}%
  \BibitemOpen
  \bibfield  {author} {\bibinfo {author} {\bibfnamefont {S.}~\bibnamefont {T{\u a}nase-Nicola}}, \bibinfo {author} {\bibfnamefont {P.~B.}\ \bibnamefont {Warren}},\ and\ \bibinfo {author} {\bibfnamefont {P.~R.}\ \bibnamefont {ten Wolde}},\ }\bibfield  {title} {\bibinfo {title} {{{S}ignal detection, modularity, and the correlation between extrinsic and intrinsic noise in biochemical networks}},\ }\href@noop {} {\bibfield  {journal} {\bibinfo  {journal} {Phys. Rev. Lett.}\ }\textbf {\bibinfo {volume} {97}},\ \bibinfo {pages} {068102} (\bibinfo {year} {2006})}\BibitemShut {NoStop}%
\bibitem [{\citenamefont {de~Ronde}\ \emph {et~al.}(2010)\citenamefont {de~Ronde}, \citenamefont {Tostevin},\ and\ \citenamefont {ten Wolde}}]{deRonde2010}%
  \BibitemOpen
  \bibfield  {author} {\bibinfo {author} {\bibfnamefont {W.~H.}\ \bibnamefont {de~Ronde}}, \bibinfo {author} {\bibfnamefont {F.}~\bibnamefont {Tostevin}},\ and\ \bibinfo {author} {\bibfnamefont {P.~R.}\ \bibnamefont {ten Wolde}},\ }\bibfield  {title} {\bibinfo {title} {{{E}ffect of feedback on the fidelity of information transmission of time-varying signals}},\ }\href@noop {} {\bibfield  {journal} {\bibinfo  {journal} {Phys. Rev. E}\ }\textbf {\bibinfo {volume} {82}},\ \bibinfo {pages} {031914} (\bibinfo {year} {2010})}\BibitemShut {NoStop}%
\bibitem [{\citenamefont {Shannon}(1948)}]{Shannon1948}%
  \BibitemOpen
  \bibfield  {author} {\bibinfo {author} {\bibfnamefont {C.~E.}\ \bibnamefont {Shannon}},\ }\bibfield  {title} {\bibinfo {title} {{{T}he mathematical theory of communication}},\ }\href@noop {} {\bibfield  {journal} {\bibinfo  {journal} {Bell. Syst. Tech. J.}\ }\textbf {\bibinfo {volume} {27}},\ \bibinfo {pages} {379} (\bibinfo {year} {1948})}\BibitemShut {NoStop}%
\bibitem [{\citenamefont {Shannon}\ and\ \citenamefont {Weaver}(1963)}]{Shannon1963}%
  \BibitemOpen
  \bibfield  {author} {\bibinfo {author} {\bibfnamefont {C.~E.}\ \bibnamefont {Shannon}}\ and\ \bibinfo {author} {\bibfnamefont {W.}~\bibnamefont {Weaver}},\ }\href@noop {} {\emph {\bibinfo {title} {The Mathematical Theory of Communication}}}\ (\bibinfo  {publisher} {University of Illinois Press, Urbana},\ \bibinfo {year} {1963})\BibitemShut {NoStop}%
\bibitem [{\citenamefont {Cover}\ and\ \citenamefont {Thomas}(1991)}]{Cover1991}%
  \BibitemOpen
  \bibfield  {author} {\bibinfo {author} {\bibfnamefont {T.~M.}\ \bibnamefont {Cover}}\ and\ \bibinfo {author} {\bibfnamefont {J.~A.}\ \bibnamefont {Thomas}},\ }\href@noop {} {\emph {\bibinfo {title} {{E}lements of {I}nformation {T}heory}}}\ (\bibinfo  {publisher} {Wiley-Interscience, New York},\ \bibinfo {year} {1991})\BibitemShut {NoStop}%
\bibitem [{\citenamefont {Barrett}(2015)}]{Barrett2015}%
  \BibitemOpen
  \bibfield  {author} {\bibinfo {author} {\bibfnamefont {A.~B.}\ \bibnamefont {Barrett}},\ }\bibfield  {title} {\bibinfo {title} {{{E}xploration of synergistic and redundant information sharing in static and dynamical {G}aussian systems}},\ }\href@noop {} {\bibfield  {journal} {\bibinfo  {journal} {Phys. Rev. E}\ }\textbf {\bibinfo {volume} {91}},\ \bibinfo {pages} {052802} (\bibinfo {year} {2015})}\BibitemShut {NoStop}%
\bibitem [{\citenamefont {Nandi}\ \emph {et~al.}(2026)\citenamefont {Nandi}, \citenamefont {Chattopadhyay},\ and\ \citenamefont {Banik}}]{Nandi2026}%
  \BibitemOpen
  \bibfield  {author} {\bibinfo {author} {\bibfnamefont {M.}~\bibnamefont {Nandi}}, \bibinfo {author} {\bibfnamefont {S.}~\bibnamefont {Chattopadhyay}},\ and\ \bibinfo {author} {\bibfnamefont {S.~K.}\ \bibnamefont {Banik}},\ }\bibfield  {title} {\bibinfo {title} {Identifying the sources of noise synergy and redundancy in the gene expression of feed-forward loop motif},\ }\href@noop {} {\bibfield  {journal} {\bibinfo  {journal} {Phys. Biol.}\ }\textbf {\bibinfo {volume} {23}},\ \bibinfo {pages} {016003} (\bibinfo {year} {2026})}\BibitemShut {NoStop}%
\bibitem [{\citenamefont {Bintu}\ \emph {et~al.}(2005)\citenamefont {Bintu}, \citenamefont {Buchler}, \citenamefont {Garcia}, \citenamefont {Gerland}, \citenamefont {Hwa}, \citenamefont {Kondev},\ and\ \citenamefont {Phillips}}]{Bintu2005}%
  \BibitemOpen
  \bibfield  {author} {\bibinfo {author} {\bibfnamefont {L.}~\bibnamefont {Bintu}}, \bibinfo {author} {\bibfnamefont {N.~E.}\ \bibnamefont {Buchler}}, \bibinfo {author} {\bibfnamefont {H.~G.}\ \bibnamefont {Garcia}}, \bibinfo {author} {\bibfnamefont {U.}~\bibnamefont {Gerland}}, \bibinfo {author} {\bibfnamefont {T.}~\bibnamefont {Hwa}}, \bibinfo {author} {\bibfnamefont {J.}~\bibnamefont {Kondev}},\ and\ \bibinfo {author} {\bibfnamefont {R.}~\bibnamefont {Phillips}},\ }\bibfield  {title} {\bibinfo {title} {{{T}ranscriptional regulation by the numbers: models}},\ }\href@noop {} {\bibfield  {journal} {\bibinfo  {journal} {Curr. Opin. Genet. Dev.}\ }\textbf {\bibinfo {volume} {15}},\ \bibinfo {pages} {116} (\bibinfo {year} {2005})}\BibitemShut {NoStop}%
\bibitem [{\citenamefont {Belle}\ \emph {et~al.}(2006)\citenamefont {Belle}, \citenamefont {Tanay}, \citenamefont {Bitincka}, \citenamefont {Shamir},\ and\ \citenamefont {{O'Shea}}}]{Belle2006}%
  \BibitemOpen
  \bibfield  {author} {\bibinfo {author} {\bibfnamefont {A.}~\bibnamefont {Belle}}, \bibinfo {author} {\bibfnamefont {A.}~\bibnamefont {Tanay}}, \bibinfo {author} {\bibfnamefont {L.}~\bibnamefont {Bitincka}}, \bibinfo {author} {\bibfnamefont {R.}~\bibnamefont {Shamir}},\ and\ \bibinfo {author} {\bibfnamefont {E.~K.}\ \bibnamefont {{O'Shea}}},\ }\bibfield  {title} {\bibinfo {title} {Quantification of protein half-lives in the budding yeast proteome},\ }\href@noop {} {\bibfield  {journal} {\bibinfo  {journal} {Proc. Natl. Acad. Sci.}\ }\textbf {\bibinfo {volume} {103}},\ \bibinfo {pages} {13004} (\bibinfo {year} {2006})}\BibitemShut {NoStop}%
\bibitem [{\citenamefont {Garcia}\ and\ \citenamefont {Phillips}(2011)}]{Garcia2011}%
  \BibitemOpen
  \bibfield  {author} {\bibinfo {author} {\bibfnamefont {H.~G.}\ \bibnamefont {Garcia}}\ and\ \bibinfo {author} {\bibfnamefont {R.}~\bibnamefont {Phillips}},\ }\bibfield  {title} {\bibinfo {title} {Quantitative dissection of the simple repression input--output function},\ }\href@noop {} {\bibfield  {journal} {\bibinfo  {journal} {Proc. Natl. Acad. Sci.}\ }\textbf {\bibinfo {volume} {108}},\ \bibinfo {pages} {12173} (\bibinfo {year} {2011})}\BibitemShut {NoStop}%
\bibitem [{\citenamefont {Hintsche}\ and\ \citenamefont {klumpp}(2013)}]{Hintsche2013}%
  \BibitemOpen
  \bibfield  {author} {\bibinfo {author} {\bibfnamefont {M.}~\bibnamefont {Hintsche}}\ and\ \bibinfo {author} {\bibfnamefont {S.}~\bibnamefont {klumpp}},\ }\bibfield  {title} {\bibinfo {title} {Dilution and the theoretical description of growth-rate dependent gene expression},\ }\href@noop {} {\bibfield  {journal} {\bibinfo  {journal} {J. Biol. Eng.}\ }\textbf {\bibinfo {volume} {7}},\ \bibinfo {pages} {22} (\bibinfo {year} {2013})}\BibitemShut {NoStop}%
\bibitem [{\citenamefont {Christiano}\ \emph {et~al.}(2014)\citenamefont {Christiano}, \citenamefont {Nagaraj}, \citenamefont {Fr{\"o}hlich},\ and\ \citenamefont {Walther}}]{Christiano2014}%
  \BibitemOpen
  \bibfield  {author} {\bibinfo {author} {\bibfnamefont {R.}~\bibnamefont {Christiano}}, \bibinfo {author} {\bibfnamefont {N.}~\bibnamefont {Nagaraj}}, \bibinfo {author} {\bibfnamefont {F.}~\bibnamefont {Fr{\"o}hlich}},\ and\ \bibinfo {author} {\bibfnamefont {T.~C.}\ \bibnamefont {Walther}},\ }\bibfield  {title} {\bibinfo {title} {Global proteome turnover analyses of the yeasts {S}. cerevisiae and {S}. pombe},\ }\href@noop {} {\bibfield  {journal} {\bibinfo  {journal} {Cell Rep.}\ }\textbf {\bibinfo {volume} {9}},\ \bibinfo {pages} {1959} (\bibinfo {year} {2014})}\BibitemShut {NoStop}%
\bibitem [{\citenamefont {Li}\ \emph {et~al.}(2014)\citenamefont {Li}, \citenamefont {Burkhardt}, \citenamefont {Gross},\ and\ \citenamefont {Weissman}}]{Li2014}%
  \BibitemOpen
  \bibfield  {author} {\bibinfo {author} {\bibfnamefont {G.-W.}\ \bibnamefont {Li}}, \bibinfo {author} {\bibfnamefont {D.}~\bibnamefont {Burkhardt}}, \bibinfo {author} {\bibfnamefont {C.}~\bibnamefont {Gross}},\ and\ \bibinfo {author} {\bibfnamefont {J.~S.}\ \bibnamefont {Weissman}},\ }\bibfield  {title} {\bibinfo {title} {Quantifying absolute protein synthesis rates reveals principles underlying allocation of cellular resources},\ }\href@noop {} {\bibfield  {journal} {\bibinfo  {journal} {Cell}\ }\textbf {\bibinfo {volume} {157}},\ \bibinfo {pages} {624} (\bibinfo {year} {2014})}\BibitemShut {NoStop}%
\bibitem [{\citenamefont {Hansen}\ and\ \citenamefont {{O'Shea}}(2015)}]{Hansen2015}%
  \BibitemOpen
  \bibfield  {author} {\bibinfo {author} {\bibfnamefont {A.~S.}\ \bibnamefont {Hansen}}\ and\ \bibinfo {author} {\bibfnamefont {E.~K.}\ \bibnamefont {{O'Shea}}},\ }\bibfield  {title} {\bibinfo {title} {cis {D}eterminants of promoter threshold and activation timescale},\ }\href@noop {} {\bibfield  {journal} {\bibinfo  {journal} {Cell Rep.}\ }\textbf {\bibinfo {volume} {12}},\ \bibinfo {pages} {1226} (\bibinfo {year} {2015})}\BibitemShut {NoStop}%
\bibitem [{\citenamefont {Milo}\ and\ \citenamefont {Phillips}(2015)}]{Milo2015}%
  \BibitemOpen
  \bibfield  {author} {\bibinfo {author} {\bibfnamefont {R.}~\bibnamefont {Milo}}\ and\ \bibinfo {author} {\bibfnamefont {R.}~\bibnamefont {Phillips}},\ }\href@noop {} {\emph {\bibinfo {title} {Cell Biology by the Numbers}}}\ (\bibinfo  {publisher} {Garland Science},\ \bibinfo {year} {2015})\BibitemShut {NoStop}%
\bibitem [{\citenamefont {Gupta}\ \emph {et~al.}(2024)\citenamefont {Gupta}, \citenamefont {Johnson}, \citenamefont {Cruz}, \citenamefont {Costa}, \citenamefont {Guest}, \citenamefont {Li}, \citenamefont {Hart}, \citenamefont {Nguyen}, \citenamefont {Stadlmeier}, \citenamefont {Bratton}, \citenamefont {Silhavy}, \citenamefont {Wingreen}, \citenamefont {Gitai},\ and\ \citenamefont {W{\"u}hr}}]{Gupta2024}%
  \BibitemOpen
  \bibfield  {author} {\bibinfo {author} {\bibfnamefont {M.}~\bibnamefont {Gupta}}, \bibinfo {author} {\bibfnamefont {A.~N.~T.}\ \bibnamefont {Johnson}}, \bibinfo {author} {\bibfnamefont {E.~R.}\ \bibnamefont {Cruz}}, \bibinfo {author} {\bibfnamefont {E.~J.}\ \bibnamefont {Costa}}, \bibinfo {author} {\bibfnamefont {R.~L.}\ \bibnamefont {Guest}}, \bibinfo {author} {\bibfnamefont {S.~H.-J.}\ \bibnamefont {Li}}, \bibinfo {author} {\bibfnamefont {E.~M.}\ \bibnamefont {Hart}}, \bibinfo {author} {\bibfnamefont {T.}~\bibnamefont {Nguyen}}, \bibinfo {author} {\bibfnamefont {M.}~\bibnamefont {Stadlmeier}}, \bibinfo {author} {\bibfnamefont {B.~P.}\ \bibnamefont {Bratton}}, \bibinfo {author} {\bibfnamefont {T.~J.}\ \bibnamefont {Silhavy}}, \bibinfo {author} {\bibfnamefont {N.~S.}\ \bibnamefont {Wingreen}}, \bibinfo {author} {\bibfnamefont {Z.}~\bibnamefont {Gitai}},\ and\ \bibinfo {author} {\bibfnamefont {M.}~\bibnamefont {W{\"u}hr}},\ }\bibfield  {title} {\bibinfo {title} {Global protein turnover quantification in
  {Escherichia} coli reveals cytoplasmic recycling under nitrogen limitation},\ }\href@noop {} {\bibfield  {journal} {\bibinfo  {journal} {Nat. Commun.}\ }\textbf {\bibinfo {volume} {15}},\ \bibinfo {pages} {5890} (\bibinfo {year} {2024})}\BibitemShut {NoStop}%
\bibitem [{\citenamefont {Rubinstein}(1999)}]{Rubinstein1999}%
  \BibitemOpen
  \bibfield  {author} {\bibinfo {author} {\bibfnamefont {R.}~\bibnamefont {Rubinstein}},\ }\bibfield  {title} {\bibinfo {title} {{The Cross-Entropy Method for Combinatorial and Continuous Optimization}},\ }\href@noop {} {\bibfield  {journal} {\bibinfo  {journal} {Methodol. Comput. Appl. Probab.}\ }\textbf {\bibinfo {volume} {1}},\ \bibinfo {pages} {127} (\bibinfo {year} {1999})}\BibitemShut {NoStop}%
\bibitem [{\citenamefont {de~Boer}\ \emph {et~al.}(2005)\citenamefont {de~Boer}, \citenamefont {Kroese}, \citenamefont {Mannor},\ and\ \citenamefont {Rubinstein}}]{deBoer2005}%
  \BibitemOpen
  \bibfield  {author} {\bibinfo {author} {\bibfnamefont {P.-T.}\ \bibnamefont {de~Boer}}, \bibinfo {author} {\bibfnamefont {D.~P.}\ \bibnamefont {Kroese}}, \bibinfo {author} {\bibfnamefont {S.}~\bibnamefont {Mannor}},\ and\ \bibinfo {author} {\bibfnamefont {R.~Y.}\ \bibnamefont {Rubinstein}},\ }\bibfield  {title} {\bibinfo {title} {A tutorial on the cross-entropy method},\ }\href@noop {} {\bibfield  {journal} {\bibinfo  {journal} {Ann. Oper. Res.}\ }\textbf {\bibinfo {volume} {134}},\ \bibinfo {pages} {19} (\bibinfo {year} {2005})}\BibitemShut {NoStop}%
\bibitem [{\citenamefont {Gillespie}(1976)}]{Gillespie1976}%
  \BibitemOpen
  \bibfield  {author} {\bibinfo {author} {\bibfnamefont {D.~T.}\ \bibnamefont {Gillespie}},\ }\bibfield  {title} {\bibinfo {title} {{A} general method for numerically simulating the stochastic time evolution of coupled chemical reactions},\ }\href@noop {} {\bibfield  {journal} {\bibinfo  {journal} {J. Comp. Phys.}\ }\textbf {\bibinfo {volume} {22}},\ \bibinfo {pages} {403} (\bibinfo {year} {1976})}\BibitemShut {NoStop}%
\bibitem [{\citenamefont {Gillespie}(1977)}]{Gillespie1977}%
  \BibitemOpen
  \bibfield  {author} {\bibinfo {author} {\bibfnamefont {D.~T.}\ \bibnamefont {Gillespie}},\ }\bibfield  {title} {\bibinfo {title} {{E}xact stochastic simulation of coupled chemical reactions},\ }\href@noop {} {\bibfield  {journal} {\bibinfo  {journal} {J. Phys. Chem.}\ }\textbf {\bibinfo {volume} {81}},\ \bibinfo {pages} {2340} (\bibinfo {year} {1977})}\BibitemShut {NoStop}%
\end{thebibliography}
%

\end{bibunit}
 
\end{document}